\documentclass[aps,pre,showpacs,twocolumn,superscriptaddress,floats,floatfix,notitlepage]{revtex4-1}
\usepackage{amssymb,amsmath,verbatim}
\usepackage{graphics}
\usepackage{xcolor}
\usepackage{epstopdf,inputenc}
\usepackage{ulem}
\usepackage{epsfig}
\usepackage{rotating}
\usepackage{float}
\usepackage{ulem}
\def\boxit#1{%
  \smash{\color{red}\fboxrule=1pt\relax\fboxsep=2pt\relax%
  \llap{\rlap{\fbox{\vphantom{0}\makebox[#1]{}}}~}}\ignorespaces
}

\makeatletter
\renewcommand{\p@subsection}{}
\renewcommand{\p@subsubsection}{}
\makeatother
\usepackage{bm}
\usepackage{soul}
\usepackage{hyperref}
\usepackage{url}
\usepackage{breakurl}
\newcommand{\df}{\displaystyle\frac}
\usepackage[stable,perpage]{footmisc}
\usepackage{xr}
\newcommand{\blue}[1]{\textcolor{black}{#1}}
\begin{document}
\title{Local Wavenumber Model for Inhomogeneous Two-Fluid Mixing}
\author{Nairita Pal\footnote{nairita@lanl.gov}}
\author{Ismael Boureima}
\author{Noah Braun}
\author{Susan Kurien}
\affiliation{Los Alamos National Laboratory, Los Alamos, NM 87545, USA}
\author{Praveen Ramaprabhu}
\affiliation{Mechanical Engineering and Engineering Science, University of North Carolina -- Charlotte, Charlotte, NC 28223, USA}
\author{Andrew Lawrie}
\affiliation{Hele-Shaw Laboratory, Queen's Building, University of Bristol, University Walk, Clifton BS8 1TR UK}

\date{\today}

\begin{abstract}

We analyze the local wavenumber (LWN) model, a two-point spectral closure model for turbulence, as applied to the Rayleigh-Taylor instability, the flow induced by the relaxation of a statically-unstable density stratification. Model outcomes are validated against data from 3D simulations of the RT instability.
In the first part of the study we consider the minimal model terms required to capture inhomogeneous mixing and show that this version, with suitable model coefficients,  is sufficient to capture the evolution of important mean global quantities including mix width, turbulent mass flux velocity, and Reynolds stress, if the start time is chosen such that the earliest transitions are avoided. However, this simple model does not permit the expected finite asymptote of the density-specific-volume covariance $b$. \blue{In the second part of the study, we investigate two forms for a source term for the evolution of the spectrum of density--specific-volume covariance for the LWN model. The first includes an empirically motivated calibration of the source  to achieve the final asymptotic state of constant $b$. The second form does not require calibration but, in conjunction with enhanced diffusion and drag captures the full evolution of all the dynamical quantities, namely, the mix--layer growth, turbulent mass--flux velocity, Reynolds stress, as well as the desired behavior of $b$. }


\end{abstract}

\maketitle
\section{\label{intro}Introduction}

Statistical models for turbulence with point-wise correlations as variables are known as \textit{single-point}  models and are widely used in many practical and industrial applications. Examples include those in the Reynolds-Averaged Navier-Stokes (RANS) family of models~\cite{belen1965theory,neuvazhaev1976numerical} such as the $k$-$\epsilon$~\cite{launder1983numerical}, and $k$-$\omega$ models~\cite{wilcox,pope1994relationship}, where $k$ is the energy (velocity autocorrelation) and $\epsilon$ is the energy dissipation rate, and $\omega$ is the specific dissipation rate. A similar single-point phenomenological model was introduced by Besnard, Harlow and Rauenzahn~\cite{besnard1996spectral}; such models form the basis of turbulent mixing models in many multi-physics codes widely in use for industrial and research applications \cite{gittings2008rage}. However, single-point models have difficulties with predicting phenomena such as strong transients or density variations~\cite{hanjalic2002one,cadiou2004two,schiestel1987multiple,clark1998symmetries}. This is because they do not have  information on the multiple scales generated by nonlinearities that are intrinsic to turbulence. Thus, certain flow properties may be better described using \textit{two-point} statistical models which by definition have variables depending on two points in space, and hence on the scale defined by their separation distance. Indeed as computational power increases, there are efforts towards building more accurate turbulence models beyond the well-understood workhorse RANS single-point models~\cite{tennekes1972first}. \blue{A recent comprehensive review on the status of turbulence modeling is given in \cite{ZHOU2021}. }

In this paper we examine one such model to study two-point statistics of turbulence, and we refer to this model as the Local Wavenumber (LWN) model.  This model is based on the paper by \cite{besnard1996spectral} which described a spectral model for single-fluid turbulence. This model was extended to describe two-fluid variable density turbulence in ~\cite{Steinkamp1999a,Steinkamp1999b,clark1995two}. In particular we focus on the Rayleigh-Taylor (RT) instability generated at a perturbed interface between a heavy and a light fluid, subjected to an acceleration opposing the mean density-gradient~\cite{taylor1950instability},~\cite{sharp1983overview,kull1991theory,inogamov1999role,youngs1984numerical,andronov1979effect,morgan2018two}. 
In such a configuration, the flow exhibits several interesting properties that are in general difficult to model -- first, the variable-density turbulence has large density fluctuations relative to the mean, second, the mix layer grows in thickness as the flow evolves, and third, the flow is statistically inhomogeneous and anisotropic.

The study of constant density (single fluid) homogeneous isotropic turbulence via practical (realizable) spectral models began with Eddy Damped Quasi Normal Closure (EDQNM) introduced in \cite{orszag1968formulation,orszag1970analytical}, with further developments for homogeneous flows in \cite{cambon1979modelisation, cambon1981spectral}. The modeling of anisotropic contributions in homogeneous turbulence has also been amenable to the EDQNM framework \cite{cambon2013role,cambon2017anisotropic, cambon2006anisotropic,rubinstein2015scalar} with more recent extensions to strongly anisotropic, homogeneous flow, with Unstably Stratified Homogeneous Turbulence (USHT) \cite{briard2017anisotropic}, and shear-driven and buoyancy-driven turbulent flows \cite{cambon2017anisotropic}. EDQNM models of buoyancy-driven Boussinesq flows have been studied in \cite{burlot2015spectral, burlot2015large, soulard2019permanence} for the USHT system. 

EDQNM is a more elaborate model than LWN. The former includes non-local interactions in the wavenumber space in the closure of the nonlinear terms while the latter is strictly local. While EDQNM is a more complex mathematical framework than LWN, it nevertheless does not lend itself to extension to the more general variable-density case \cite{besnard1996spectral}. Following \cite{clark1995two} we studied homogeneous, variable-density turbulence in previous work~\cite{pal2018two} using the LWN model. In that work, it was shown that the LWN model captures the time evolution of the statistics of variable-density homogeneous isotropic turbulence across large variation of  density ratios between the participating fluids. 
The LWN  model has also been shown to produce good agreement with experimental data in the case of homogeneous sheared and strained turbulence~\cite{clark1995spectral} and in anisotropic flows \cite{clarkthesis}.  Non-stationary inhomogeneous turbulence using the shear-free mixing layer (SFML) has been studied using the two-point spectral closure model developed for the purpose \cite{besnard1996spectral,bragg2017model}.  Thus for the particular considerations in RTI of variable-density, non-stationarity and inhomogeneity, the LWN model, though simpler, offers some advantages to EDQNM. The constraint of locality of triadic interactions in LWN has not presented significant drawbacks to practical implementation of the model and indeed makes it a more computable choice when compared to EDQNM.

We follow the approach of~\cite{Steinkamp1999a,Steinkamp1999b}, which formulates the evolution of an RT layer using three time-varying quantities, a Reynolds stress tensor $\hat{R}_{ij}(y,k,t)$, a velocity associated with mass flux, $\hat{a}_i(y,k,t)$ ($i$ and $j$ denote the Cartesian component components), and the covariance of density and specific volume, $\hat{b}(y,k,t)$. Here $y$ is the vertical height of the domain and $k$ is the wavenumber in the direction perpendicular to the vertical, and $t$ is the time. On average, the
flow is isotropic in the horizontal plane. Therefore, the functions do not depend on
the wave vector ${\bm k}$, but on its modulus $|{\bm k}|$. This covariance parameter can be understood as a measure of mixedness in the system, i.e., $\hat{b}(y,k)$ has a high value in a segregated domain, and gradually decreases as the fluids mix.

The work presented here is a first effort at a minimal augmentation of the model in~\cite{pal2018two} in order to capture inhomogeneous flow physics. \blue{In particular we are interested in statistical properties of the flows and do not attempt to correlate to physical features such as bubble-spike dynamics which form a complementary approach \cite{oron2001dimensionality,hecht1994potential,zufiria1988bubble}.}
In the first part of the paper we consider terms corresponding to a Leith-type \cite{leith67} spatial diffusion for each of the dynamical variables. Other additional terms relative to the homogeneous case, like the spatial advection terms, are also retained because they are exact. Although this version is quite successful at recovering aspects of the physics, some deficits appear particularly in the fidelity to the density-specific-volume covariance. Therefore, in the latter part of the paper, we include options for a kinematic source term in the evolution equation for the covariance of density and specific volume. 

The objective of this work is to highlight the roles of the different terms (derived and heuristic) in the LWN equations for inhomogeneous mixing and motivate further refinement based on the insights gathered. 
The numerical simulations data against which the model predictions are compared, are generated using MOBILE \cite{lawrie2010rayleigh,lawrie2011turbulent,ramaprabhu2013rayleigh}, following the implicit large-eddy simulation (ILES) methodology of \cite{margolin2006modeling,snider1996simulation}. MOBILE integrates the incompressible variable-density miscible equations of fluid motion and has been successfully used to study systems with Atwood numbers up to $0.9$~\cite{ramaprabhu2013rayleigh}.
Throughout this study the resolution of the LWN model calculation is identical (in the number of modes and in the grid resolution) to that of the ILES simulations. This permits a valid comparison between the two without needing to resort to very high resolution simulations. 


The outline of the paper is as follows: in section~\ref{implement} we introduce the model, and explain the physical significance of the various terms involved. 
In section~\ref{iles} we give details of our numerical setup and details of the MOBILE simulations. In section~\ref{results} we show results from comparison against MOBILE data for the simplest model, and establish the importance of various terms in the different flow regimes. 
We discuss the model results for the RT system across three different density-ratios between the participating fluids. In section~\ref{source_term} we show results for a modified LWN model with a source term for the density--specific--volume--covariance{\blue{\sout{$b$}}}. Finally, in section~\ref{conclusions} we provide a discussion and summary and point towards future model development.

\section{Model equations and implementation\label{implement}}
We will follow the development proposed for single-fluid incompressible flow by Besnard et al~\cite{besnard1996spectral}, and subsequently adapted for variable-density flow by~\cite{Steinkamp1999a,Steinkamp1999b}. We first decompose the flow-field variables, i.e., density $\rho$, velocity ${\bm u}$, and pressure $p$ into their mean and fluctuating parts as follows:
\begin{eqnarray}
\rho & = & \overline{\rho} + \rho^{\prime} \\
{\bm u} & = & \overline{\bm u} +{\bm u}^{\prime}\\
p & = & \overline{p} + p^{\prime}
\end{eqnarray}
where the overbar denotes the mean, and the primes the fluctuations about the mean.
In the case of variable-density flows, it is useful to work with the mass-weighted averages introduced
by Favre. The Favre-averaged velocity $\tilde{{\bm u}}$ is 
\begin{equation}
\tilde{{\bm u}} = \frac{\overline{\rho {\bm u}}}{\overline{\rho}}.
\label{tildeu}
\end{equation}
Let ${\bm u}^{\prime\prime}$ denote the fluctuation about this Favre averaged velocity $\tilde{\bm u}$. Then we have
\begin{equation}
{\bm u} = \tilde{{\bm u}} + {\bm u}^{\prime\prime}.
\end{equation}
If we apply the standard Reynolds decomposition to $\overline{\rho {\bm u}}$ we get,
\begin{eqnarray}
\overline{\rho {\bm u}} &= & {\overline \rho}~{\overline{\bm u}}+\overline{\rho^{\prime}{\bm u}^{\prime}} 
\end{eqnarray}
since $\overline{{\bm u}^{\prime}}=0$ and $\overline{\rho^{\prime}}=0$.
Using Eq.~(\ref{tildeu}) we then obtain,
\begin{eqnarray}
\overline{\rho}\bm{\tilde{u}}&=&{\overline \rho}~{\overline {\bm u}}+\overline{\rho^{\prime}{\bm u}^{\prime}}\nonumber\\
\tilde{{\bm u}}&=&\overline{{\bm u}}+\frac{\overline{\rho^{\prime}{\bm u}^{\prime}}}{\overline{\rho}}
\label{defa}
\end{eqnarray}
We define a velocity ${\bm a}$ associated with the net turbulent mass flux as follows : 
\begin{eqnarray}
{\bm a} &= & \frac{\overline{\rho^{\prime}{\bm u}^{\prime}}}{\overline{\rho}} \\
{\rm So,} \quad\tilde{{\bm u}}&=&\overline{{\bm u}}+\bm{a}
\end{eqnarray}
From Eq. (\ref{defa}) then, we can define ${\bm a}$ as the flux of mass relative to $\tilde{{\bm u}}$.

For two arbitrary points ${\bm x}_1$ and ${\bm x}_2$ in space, the mass-weighted Reynolds stress tensor is defined as,
\begin{equation}
R_{ij}({\bm x}_1,{\bm x}_2)= \frac{1}{2}\overline{[\rho({\bm x}_1)+\rho({\bm x}_2)]u_i^{\prime\prime}({\bm x}_1)u_j^{\prime\prime}({\bm x}_2)}.
\label{beg_r}
\end{equation}
Defining the specific volume as $\displaystyle\upsilon({\bm x})=\frac{1}{\rho({\bm x})}$ and its fluctuations $\upsilon'(\bm{x})$  defined with respect to the mean specific-volume, the velocity associated with the turbulent mass-flux is defined as
\begin{equation}
a_i({\bm x}_1,{\bm x}_2)= -\overline{u_i^{\prime\prime}({\bm x}_1)\rho({\bm x}_1)\upsilon({\bm x}_2)}\label{a_def},
\end{equation}
and the covariance of the density and specific-volume is defined as
\begin{equation}
b({\bm x}_1,{\bm x}_2)=-\overline{\rho^{\prime}({\bm x}_1)\upsilon^{\prime}({\bm x}_2)}.
\end{equation} 

Alternatively, these two points can be expressed in terms of a center of mass
$\bm{x}=\frac{1}{2}(\bm{x}_1+\bm{x}_2)$, and separation $\bm{r}=\bm{x}_1-\bm{x}_2$ vectors. The corresponding 
Fourier transform, in terms of the wavevector, $\bm{k}$, associated with scale $\bm{r}$, 
\begin{eqnarray}
 \tilde{R}_{ij}({\bm x},{\bm k})&=&\int R_{ij}({\bm x},{\bm r})e^{-i{\bm k}\cdot{\bm r}}{\rm d}{\bm r},\\
\tilde{a}_i({\bm x},{\bm k})&=&\int a_i({\bm x},{\bm r})e^{-i{\bf k}\cdot{\bm r}}{\rm d}{\bm r},\\
\tilde{b}({\bm x},{\bm k})&=&\int b({\bm x},{\bm r})e^{-i{\bm k}\cdot{\bf r}}{\rm d}{\bm r}
\end{eqnarray}

To simplify further, we average over the sphere in $\bm{k}$-space to obtain
\blue{
\begin{eqnarray}
\hat{R}_{ij}({\bm x},k)&=&\int \tilde{R}_{ij}({\bm x},{\bm k})\frac{k^2{\rm d}\Omega_k}{4\pi},\\
\hat{a}_i({\bm x},k)&=&\int \tilde{a}_i({\bm x},{\bm k})\frac{k^2{\rm d}\Omega_k}{4\pi}, \\
\hat{b}({\bm x},k)&=&\int \tilde{b}({\bm x},{\bm k})\frac{k^2{\rm d}\Omega_k}{4\pi}.
\end{eqnarray}}

where ${\rm d}\Omega_k=\sin\theta~{\rm d}\theta~{\rm d}\phi$ for $0\leq \theta \leq \pi$; $0\leq \phi \leq 2\pi$.
Henceforth we will use \blue{$\hat{R}_{ij}$}, \blue{$\hat{a}_i$} and \blue{$\hat{b}$} to denote the spectral quantities at a particular time $t$, and will omit their respective arguments. 
Following Steinkamp {\textit {et.~al}}~\cite{Steinkamp1999a} we write the mass and momentum conservation equations for variable-density flows driven by gravity in the $y$-direction as follows:
\begin{eqnarray}
\frac{\partial\overline{\rho}}{\partial t} + \frac{\partial \overline{\rho} \tilde{u}_y}{\partial y} & = & 0\label{drhodt}\\
\frac{\partial \overline{\rho}\tilde{u}_y }{\partial t}+\frac{\partial {\overline\rho} \tilde{u}_y\tilde{u}_y}{\partial y} & = &
-\frac{\partial \overline{p}}{\partial y}+\overline{\rho}g-\frac{\partial R_{yy}}{\partial y}.
\label{dpdy}
\end{eqnarray}

Here, \blue{$R_{yy}(y,t)=\int \hat{R}_{yy} (y,k,t)dk$} is the vertical component of the Reynolds stress tensor.
The equations we use for our comparison studies are obtained by multiplying quantities such as $\overline{\upsilon}$ with Eq.~\eqref{drhodt} or $\tilde{u}_y$ with Eq.~\eqref{dpdy}, and integrating to get the ensemble-averaged correlation variables defined in Eqs.~\eqref{beg_r} and~\eqref{a_def}. The equations for the correlation variables can then be simplified by taking Fourier transforms across each homogeneous plane in the RT system. In the three dimensional RT system, the direction of gravity is $y$, and the interface lies on an $x-z$ plane. For practical purposes, we consider the $x-z$ plane to be homogeneous, and thus take Fourier transforms across those planes. The vertical ($y$) direction is, however, inhomogeneous, and in this direction we retain a physical-space representation (see Fig.~\ref{visual} for details).
The detailed derivation of the LWN system is presented in~\cite{Steinkamp1999a, Steinkamp1999b}, and here we simply summarize the governing equations, removing the source term when considering the RT system. Defining $\hat{R}_{nn}(y,k,t)$ as the trace of $\hat{R}_{ij}(y,k,t)$, the final set of evolution equations for the correlation variables are as follows:
\begin{widetext}
\begin{eqnarray}
\frac{\partial \blue{\hat{R}_{nn}}(y,k,t)}{\partial t} & = &-\frac{\partial \hat{R}_{nn}\tilde{u}_y}{\partial y}+\int \limits_{-\infty}^{+\infty}2\hat{a}_y \df{\partial \overline{p}}{\partial y}(k \exp{(-2k|y^{\prime}-y|)})dy^{\prime}+\frac{\partial}{\partial k}\left[k\Theta^{-1}\left[-C_{r1}\hat{R}_{nn}+C_{r2}k\frac{\partial \hat{R}_{nn}}{\partial k}\right]\right]\nonumber\\&&-2\hat{R}_{yy}\frac{\partial \tilde{u}_y}{\partial y}+C_d\frac{\partial }{\partial y}\left(\upsilon_t\frac{\partial \hat{R}_{nn}}{\partial y}\right)\label{main_rnn}\\
\frac{\partial \blue{\hat{R}_{yy}}(y,k,t)}{\partial t} & = &-\frac{\partial \hat{R}_{yy}\tilde{u}_y}{\partial y}+\int \limits_{-\infty}^{+\infty}2\hat{a}_y \df{\partial \overline{p}}{\partial y}(k \exp{(-2k|y^{\prime}-y|)})dy^{\prime}+\frac{\partial}{\partial k}\left[k\Theta^{-1}\left[-C_{r1}\hat{R}_{yy}+C_{r2}k\frac{\partial \hat{R}_{yy}}{\partial k}\right]\right]\nonumber\\&&-2\hat{R}_{yy}\frac{\partial \tilde{u}_y}{\partial y}+C_d\frac{\partial }{\partial y}\left(\upsilon_t\frac{\partial \hat{R}_{yy}}{\partial y}\right)+C_m\Theta^{-1}\left(\df{\delta_{ij}}{3}\hat{R}_{nn}-\hat{R}_{yy}\right)\label{main_ryy}\\
\frac{\partial \blue{\hat{a}}_y(y,k,t)}{\partial t}&=&-\tilde{u}_y\frac{\partial \hat{a}_y}{\partial y}+\frac{\hat{b}}{\overline{\rho}}
\frac{\partial \overline{p}}{\partial y}-\left[C_{rp1}k^2\sqrt{\hat{a}_{\hat{n}}\hat{a}_{\hat{n}}}+C_{rp2}\Theta^{-1}\right]\hat{a}_y-\frac{\hat{R}_{yy}}{\overline{\rho}^2}\frac{\partial \overline{\rho}}{\partial y}+C_d\frac{\partial }{\partial y}\left(\upsilon_t\frac{\partial \hat{a}_y}{\partial y}\right)
\label{press}\nonumber\\
&&+\frac{\partial}{\partial k}\left[k\Theta^{-1}\left[-C_{a1}\hat{a}_y+C_{a2}k\frac{\partial \hat{a}_y}{\partial k}\right]\right]
\label{main_a}\\
\displaystyle\frac{\partial \blue{\hat{b}}(y,k,t)}{\partial t}&=&\frac{\partial}{\partial k}\left[k\Theta^{-1}\left[-C_{b1}\hat{b}+C_{b2}k\frac{\partial \hat{b}}{\partial k}\right]\right]
+C_d\frac{\partial }{\partial y}\left(\upsilon_t\frac{\partial \hat{b}}{\partial y}\right)\label{main_b}
\end{eqnarray}
\end{widetext}
where the turbulence frequency $\Theta^{-1}=\sqrt{\int_0^{k}\frac{k^2\hat{R}_{nn}}{\overline{\rho}}dk}$, and the turbulent viscosity $\upsilon_t= \int \limits_0^{\infty}\sqrt{\df{k\hat{R}_{nn}}{\overline{\rho}}}\df{dk}{k^2}$. In the equations (\ref{main_rnn}-\ref{main_b}) the respective dynamical variables $\hat{R}_{nn}$, $\hat{R}_{yy}$, $\hat{a}_y$ and $\hat{b}$ are functions of the vertical height $y$, the horizontal wave number $k$ and the time $t$. We drop the explicit arguments for brevity. In Eqs.~\eqref{main_rnn}--\eqref{main_b}, the variables are functions of time, but the time argument is dropped for brevity. Here and in what follows, the arguments $y$ and $k$ are implicit unless otherwise specified.
$C_d$ is the spatial diffusion coefficient.

\blue{In Eq.~\eqref{main_rnn}, $\hat{R}_{nn}(y,k,t)$ may be integrated to obtain planar averaged values
$R_{nn}(y,t) = \int \hat{R}_{nn}(y,k,t) dk$. Similarly, $R_{yy}(y,t) = \int \hat{R}_{yy}(y,k,t) dk$, $a_{y}(y,t) = \int \hat{a}_{y}(y,k,t) dk$ and $b(y,t) = \int \hat{b}(y,k,t) dk$. Here $t$ is the time.} $R_{nn}(y,t)$ is related to the turbulent kinetic energy $E(y,t)$ in the following way:
\begin{equation}
E(y,t)=\frac{1}{2\overline{\rho}}R_{nn}(y,t)
\label{const}
\end{equation} 
The first term on the RHS of Eq.~\eqref{main_rnn}, i.e., $-\df{\partial \hat{R}_{nn}\tilde{u}_y}{\partial y}$ is the advection term. The second term is the pressure-velocity transport term, and is responsible for the onset of instability and turbulence. In this study we use the ``nonlocal'' formulation of the pressure gradient term, $\int \limits_{-\infty}^{+\infty}2\hat{a}_y \df{\partial \overline{p}}{\partial y}(k \exp{(-2k|y^{\prime}-y|)})dy^{\prime}$, which couples $\hat{a}_y$ with the mean pressure gradient and is the principal driving term in the equation for $\hat{R}_{nn}$. The ``nonlocal'' or integral formulation in physical space helps characterize instantaneous propagation of pressure waves from one physical location to another. The non-local integral is evaluated as a function of the vertical coordinate $y$ in our code. It is obtained by summing over the integrand from $0$ to $y$ for each $y$. 
The third term $\frac{\partial}{\partial k}\left[k\Theta^{-1}\left[-C_{r1}\hat{R}_{nn}+C_{r2}k\frac{\partial \hat{R}_{nn}}{\partial k}\right]\right]$ accounts for the energy cascade in $k$-space. The term with the coefficient $C_{r1}$ has a ``wave-like'' contribution to the cascade, whereas the term with the $C_{r2}$ coefficient makes a diffusive contribution. The $C_{r1}$ term is deemed ``wave-like'' because by retaining only the $C_{r1}$ term on the RHS, we obtain a wave equation (a hyperbolic equation) after taking a second derivative of $\hat{R}_{nn}$ with respect to time. $C_{r1}> 0$ gives rise to a forward cascade in $k$ space, and $C_{r2} > 0$ results in both forward and reverse cascades. The fourth term, $2\hat{R}_{yy}\frac{\partial \tilde{u}_y}{\partial y}$, is also a driving term which accounts for the coupling of $R_{yy}$ with the gradients in velocity. The final term, $C_d\displaystyle\frac{\partial }{\partial y}\left(\upsilon_t\displaystyle\frac{\partial \hat{R}_{nn}}{\partial y}\right)$, accounts for spatial diffusion and is derived from the velocity triple correlation term in the equation of motion of the Reynolds stress tensor components $R_{ij}$.

Eq.~\eqref{main_ryy} is the equation for the vertical component, $\hat{R}_{yy}$, of the Reynolds stress tensor and shares its form with Eq.~\eqref{main_rnn}. The first term on the RHS, $-\displaystyle\frac{\partial \hat{R}_{yy}\tilde{u}_y}{\partial y}$, represents advection and the second, $$\int \limits_{-\infty}^{+\infty}2\hat{a}_y \df{\partial \overline{p}}{\partial y}(k \exp{(-2k|y^{\prime}-y|)})dy^{\prime},$$, is a principal drive term. The third term represents the energy cascade and has an equivalent in $\hat{R}_{nn}(y,k,t)$. The fourth term $2\hat{R}_{yy}\displaystyle\frac{\partial \tilde{u}_y}{\partial y}$ is another drive term, and the fifth, $C_d\displaystyle\frac{\partial }{\partial y}\left(\upsilon_t\frac{\partial \hat{R}_{yy}}{\partial y}\right)$, represents spatial diffusion of $\hat{R}_{yy}$.

The final term in Eq.~\eqref{main_ryy} describes the rate of return to isotropy. The main contribution of the coefficient $C_m$ is a redistribution of energy between components of the $R_{ij}$ tensor. A high value of $C_m$ draws the distribution closer to equality amongst the three diagonal components, whereas a low value of $C_m$ biases the energy towards $\hat{R}_{yy}(y,k)$. Though we note that $C_m$ may not be constant in all circumstances, possibly varying with Atwood number or with the rate at which isotropy is restored in the flow, in the present study we set $C_m=1$, following previous literature~\cite{Steinkamp1999a}.

The equation for the turbulent mass flux velocity, $\hat{a}_y$, (Eq.~\eqref{main_a}) has a similar form to both Eqs.~\eqref{main_rnn} and~\eqref{main_ryy}. Advection in $\hat{a}_y$ by the velocity field $\tilde{u}_y$ is represented by $-\tilde{u}_y\df{\partial \hat{a}_y}{\partial y}$, while $\df{\hat{b}}{\overline{\rho}}
\df{\partial \overline{p}}{\partial y}$ is the production term for $\hat{a}_y(y,k,t)$. Here, $\hat{b}(y,k,t)$ couples directly to the pressure-gradient to produce $\hat{a}_y(y,k,t)$. The term $\left[C_{rp1}k^2\sqrt{\hat{a}_{\hat{n}}\hat{a}_{\hat{n}}}+C_{rp2}\Theta^{-1}\right]\hat{a}_y$ applies drag to $\hat{a}_y(y,k,t)$. This is a modeled term introduced in \cite{Steinkamp1999a} to account for drag perpendicular to the interface, denoted as $C_{rp1}$, and correspondingly $C_{rp2}$ denotes drag parallel to the interface. The other term driving $\hat{a}_y(y,k,t)$ is $-\df{\hat{R}_{yy}}{\overline{\rho}^2}\frac{\partial \overline{\rho}}{\partial y}$, and provides a flux opposite to the density gradient. The final two terms on the RHS of Eq.~\eqref{main_a} describe spatial diffusion and energy cascade respectively.

Finally Eq.~\eqref{main_b} describes the evolution of the spectrum of the covariance of density and specific volume, $\hat{b}$. The 
cascade and spatial diffusion terms are similar in form to the previous equations.

The spectral model calculations presented in this paper are performed with a code using a MacCormack scheme~\cite{pletcher2012computational} which is second order accurate in both space and time, and employs a two-step methodology. Each of these steps uses single-sided differences for the first order derivatives, a left-biased stencil for the first step and a right-biased one for the second. The code uses a logarithmic grid for a modified wavenumber $$z = z_s \ln\left\{\frac{k}{k_0}\right\}$$ where, following~\cite{besnard1996spectral}, we take $k_0$ and $z_s$ to be unit scale factors. 
We employ a specific choice of variables in the cascade terms, $\{k\hat{R}_{nn},k\hat{R}_{yy},k\hat{a}_i,k\hat{b}\}$, that retains a conservation form when expressed in terms of $z$ rather than $k$. It follows that the values of the integrals of the spectral quantities are easily determined, e.g.;
\begin{widetext} 
\begin{equation}
 R_{nn}\left(y, t\right) = \int_0^{+\infty} \hat{R}_{nn}\left(y,k,t\right) dk 
= \int_{-\infty}^{+\infty} \hat{R}_{nn}\left(y,z,t\right) \frac{k_0}{z_s}\exp\left\{\frac{z}{z_s}\right\}dz. 
\end{equation}
\end{widetext}
\blue{with similar definitions of integral quantities $R_{yy}(y,t)$, $b(y,t)$ and $a_y(y,t)$.}
Setting $k_0=1$ and $z_s=1$ gives 
$$ R_{nn}\left(y,t\right) = \int_{-\infty}^{+\infty} \exp\left(z\right)\hat{R}_{nn}\left(y,z,t\right) dz, $$
where $\exp\left(z\right)\hat{R}_{nn}\left(y,z,t\right)  = k\hat{R}_{nn}\left(y,k,t\right)$. The boundary conditions at $k=1$ and $k = k_{max}$ are Neumann (zero flux) as are those at $y=0$ and $y=L_y$, where $L_y$ is the height of the domain.

The LWN model for inhomogeneous RT fluid mixing can as written here be thought of as the minimal necessary augmentation of the corresponding equations for homogeneous turbulence to take account of inhomogeneity. It includes a Leith-type diffusion term \cite{leith67} that accounts for inhomogeneous growth and the spreading of $\hat{b}(y,k)$ in a manner analogous to that for $\hat{R}_{nn}(y,k)$ and $\hat{a}_y(y,k)$. 

In presenting results we will use integral quantities for analysis. These quantities are $b(y,t) = \int \hat{b}(y,k,t) dk$, $a_y(y,t) = \int \hat{a}_y(y,k,t) dk$, and $R_{nn}(y,t) = \int \hat{R}_{nn}(y,k,t) dk$. Here $t$ is the simulation time.
In our study, the Atwood number of the system is defined as $A=\df{\rho_2-\rho_1}{\rho_2+\rho_1}$, where $\rho_1$ is the density of the light fluid and $\rho_2$ is the density of the heavy fluid.
One important metric in our study is the mix-width of the RT system.  The mix-width $W(t)$ is obtained from the position of contours of volume fraction, and for robustness is here taken to be 
\begin{equation}
W(t)= y|_{\alpha_h=95\%}-y|_{\alpha_h=5\%},
\label{mixeq}
\end{equation}
where $ y|_{\alpha_h=95\%}$ denotes the domain height at which the volume fraction of the heavy fluid ($\alpha_h$) is $95\%$ and $y|_{\alpha_h=5\%}$ denotes the same for a $5\%$ volume fraction.

As presented, the LWN model for inhomogeneous RT fluid mixing can be thought of as the minimal necessary augmentation of the corresponding equations for homogeneous turbulence to take account of inhomogeneity. It is the least elaborate two-point formulation for variable-density, and includes a Leith-type diffusion term \cite{leith67} that accounts for inhomogeneous growth and the spreading of $\hat{b}(y,k)$ in a manner analogous to that for $\hat{R}_{nn}(y,k)$ and $\hat{a}_y(y,k)$.

\section{\label{iles}Implicit Large Eddy Simulation: Description of the MOBILE Code}
We briefly review the numerical methods employed in the ILES of variable density turbulent flows. The Rayleigh-Taylor (RT) simulations were performed using MOBILE~\cite{lawrie2010rayleigh,lawrie2011turbulent,ramaprabhu2013rayleigh,ramaprabhu2016evolution}, a three-dimensional, hydrodynamic solver. MOBILE solves the incompressible Navier-Stokes equations, and
adjusts the pressure field to conserve volume. In MOBILE, computational expediency is achieved through decomposing the incompressible governing equations~\eqref{mob1}--\eqref{mob2} given below into hyperbolic (advective transport), and non-hyperbolic (diffusion and viscous dissipation) and elliptic (pressure and velocity correction) components.
\begin{eqnarray}
\df{\partial \rho}{\partial t}+\df{\partial }{\partial x_i}(\rho u_i) &=&0 \label{mob1}\\
\df{\partial}{\partial t}(\rho u_i)+\df{\partial}{\partial x_i}(\rho u_iu_i+p\delta_{ij})&=&\rho g_i \label{mob2}\\
\frac{\partial }{\partial x_i} u_i &=& 0
\end{eqnarray}
							             
MOBILE employs a split, high-order advection scheme using a fractional step approach comprised of a sequence of one-dimensional updates of the conserved variables (mass and momentum) along
the X, Y and Z coordinate directions. Following Strang~\cite{strang1968construction}, a sequence of sweeps [X-Y-Z-Z-Y-X] results in a net truncation error which is close to second order in time.
MOBILE has been shown to accurately predict
global flow features such as symmetry break-down of rising bubbles and spikes in the single
mode RT simulation at Atwood numbers upto $0.5$ \cite{lawrie2010rayleigh}. Further, MOBILE has been validated for several fluid mixing and transport problems including single-mode and multimode Rayleigh-Taylor flows up to Atwood $A = 0.9$ ~\cite{lawrie2010rayleigh,lawrie2011turbulent,ramaprabhu2013rayleigh,ramaprabhu2016evolution}, Kelvin-Helmholtz instability~\cite{lawrie2010rayleigh}, lock-release gravity currents~\cite{lawrie2010rayleigh}, systems with unusual geometries \cite{lawrie2011turbulent}, jet flows with background flows \cite{lawrie2011axisymmetric,atthanayake2019formation}, and systems with variable acceleration~\cite{aslangil2016numerical}. For additional details on these methods and codes, the reader is referred to~\cite{lawrie2010rayleigh,lawrie2011turbulent,ramaprabhu2013rayleigh,ramaprabhu2016evolution}.
While MOBILE may be used in both DNS and ILES modes, the simulations in this paper employ the ILES approach. When used in DNS mode, the dissipation of kinetic energy is dominated by an explicit representation of physical viscosity. In contrast, ILES exploits numerical dissipation of kinetic energy (and scalar fluctuation energy) and corresponds closely to the use of a subgrid turbulence model where the scale filter is applied at the grid scale implicitly by the numerical method. Such a simulation strategy has the additional benefit of being monotonicity preserving, and this is an essential property to faithfully represent sharp material interfaces in two-fluid mixing. It has been shown \cite{youngs2009application} that such simulations correspond to the high Reynolds number (and Schmidt number ${\rm Sc} = 1$) limit, where the flow has exceeded the Reynolds number (${\rm Re}$) threshold \cite{sharan2019turbulent} for mixing transition, beyond which several key mixing properties have been observed to lose their dependence on ${\rm Re}$. In this study, we have examined the performance of the LWN model in this high ${\rm Re}$, ${\rm Sc} = 1$ limit by comparison with the ILES calculations, while the extension to finite Re (and non-unity Sc) will be pursued in follow-up studies. 

Parameters for three MOBILE computations used to test and validate the LWN model are tabulated in Table \ref{table1}.  The Atwood numbers range from low to moderate and the grid resolution in all cases remains fixed. The acceleration due to gravity is fixed at $2.0 cm/s^2$ and density of the lighter fluid is fixed at $\rho_1 = 1.0 g cm^{-3}$. In each case, we eventually non-dimensionalize the time \blue{$t$} with the typical Atwood dependent timescale $\displaystyle t' = \frac{1}{\sqrt{Ag/L_x}}$ in each case, and $L_x$ is the domain length in the horizontal direction $\hat{\bf x}$. The non-dimensional time is $\tau=\df{t}{t^{\prime}}$. \blue{Thus, $\hat{R}_{nn}=\hat{R}_{nn}(y,k,t)=\hat{R}_{nn}(y,k,\tau)$, $\hat{R}_{yy}=\hat{R}_{yy}(y,k,t)=\hat{R}_{yy}(y,k,\tau)$, $\hat{a}_{y}=\hat{a}_{y}(y,k,t)=\hat{a}_{y}(y,k,\tau)$, $\hat{b}=\hat{b}(y,k,t)=\hat{b}(y,k,\tau)$. Similarly, for the integrated quantities, $R_{nn}=R_{nn}(y,\tau)$, $R_{yy}=R_{yy}(y,\tau)$, $a_y=a_y(y,\tau)$ and $b=b(y,\tau)$ unless otherwise mentioned.}

\begin{figure*}[h!]
\begin{center}
\includegraphics[width=.11\linewidth]{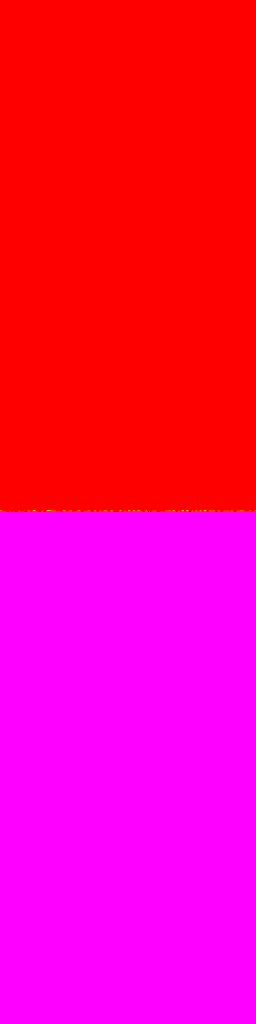}
\put(-50,194){\bf \scriptsize (a)}
\hspace{0.08in}
\includegraphics[width=.11\linewidth]{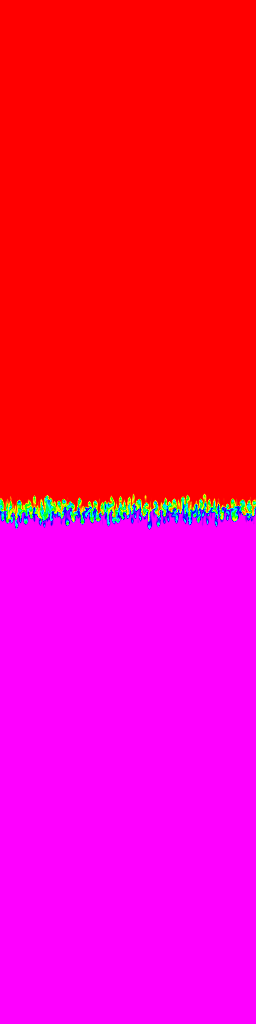}
\put(-50,194){\bf \scriptsize (b)}
\hspace{0.08in}
\includegraphics[width=.11\linewidth]{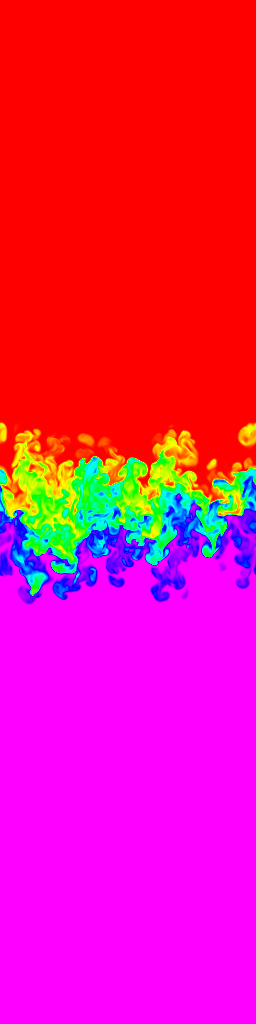}
\put(-50,194){\bf \scriptsize (c)}
\hspace{0.08in}
\includegraphics[width=.11\linewidth]{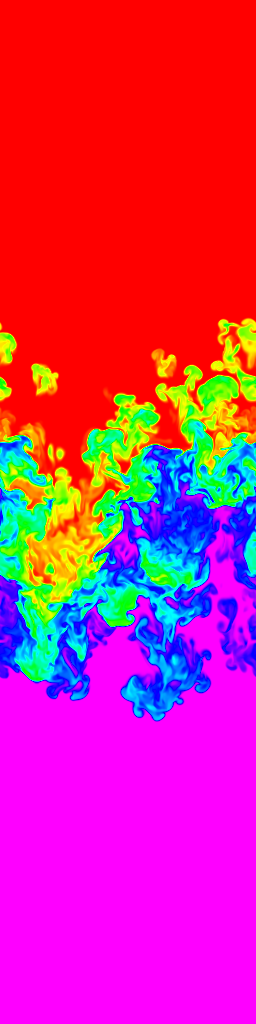}
\put(-50,194){\bf \scriptsize (d)}
\hspace{0.08in}
\includegraphics[width=.11\linewidth]{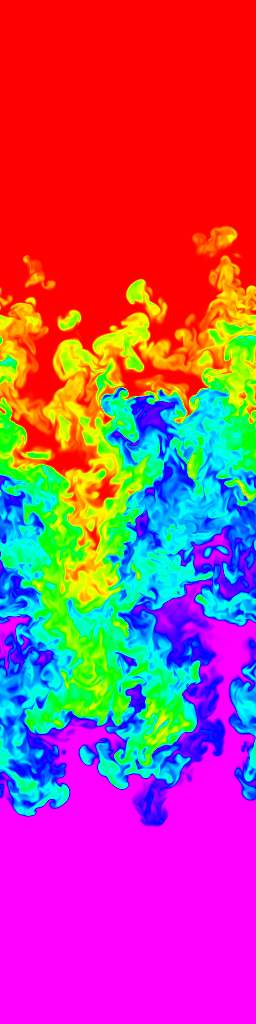}
\put(-50,194){\bf \scriptsize (e)}

\includegraphics[width=.9\linewidth]{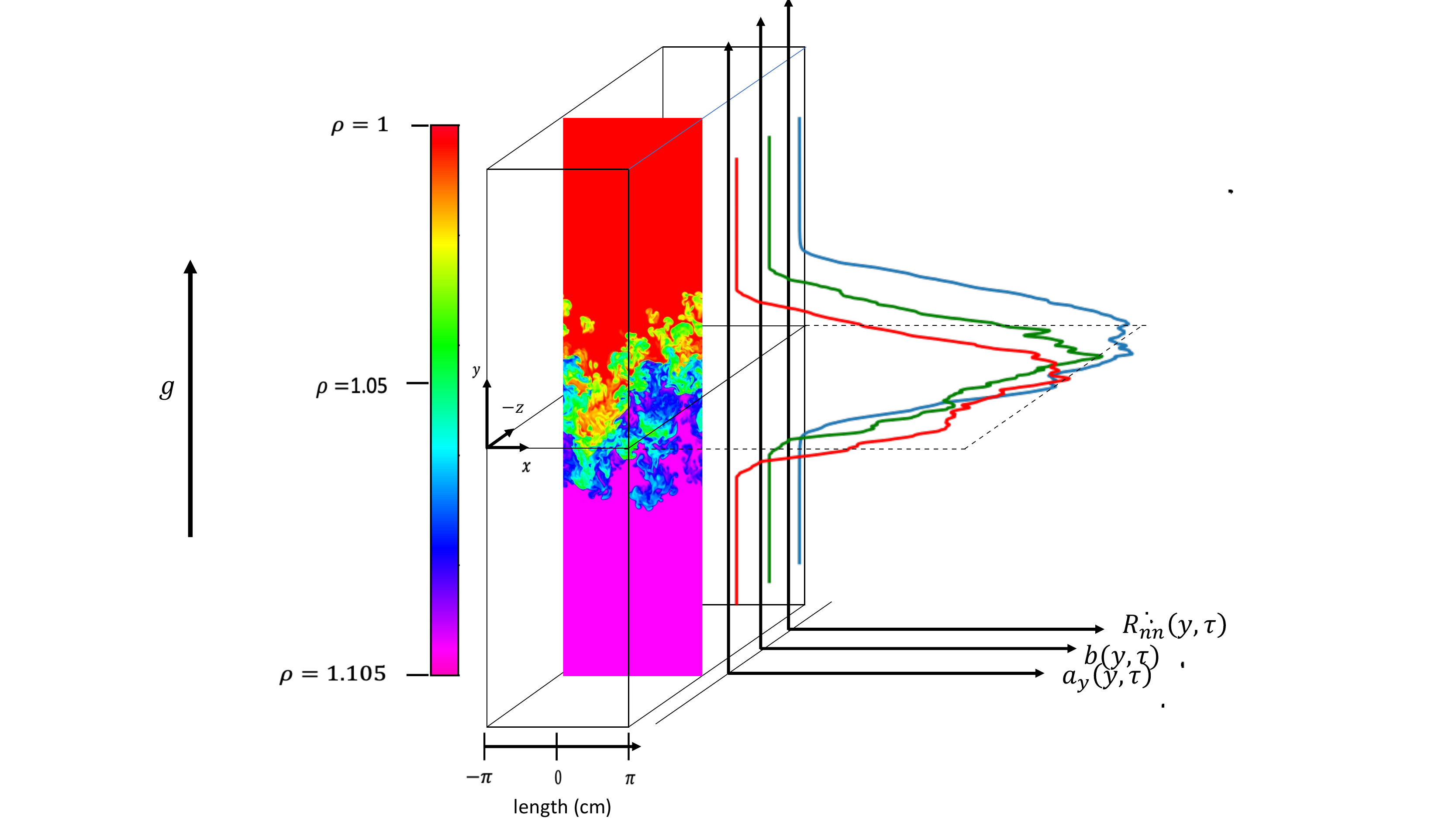}
\put(-80,194){\bf \scriptsize (f)}
\end{center}
\caption{Visualization of the density field in the MOBILE simulation run {\tt R3} at times (a) $\tau=0$; (b) $\tau=0.42$; (c) $\tau=3.164$; (d) $\tau=5.19$ and (e) $\tau=6.5$; (f) schematic of the LWN system, with a slice-through visualization from 3D simulations at $\tau=5.19$. LWN variables such as $R_{nn}(y,\tau)=\int \hat{R}_{nn}(y,k,\tau)dk, a_y(y,\tau)=\int \hat{a}_y(y,k,\tau)dk$ and $b(y,\tau)=\int \hat{b}(y,k,\tau)dk$ at $\tau=5.19$ are shown in the same plot, and these plots show that the maximum value of all these variables occur at the center-plane of the RT system. The horizontal length $L_x=2\pi$ cm and the vertical length $L_y=8\pi$ cm.}
\label{visual}
\end{figure*}

In Fig.~\ref{visual} we show visualization of the density field obtained from the MOBILE simulation over a range of times. The profiles of $a_y(y), b(y)$ and $R_{nn}(y)$ are overlaid at time $\tau=5.19$. \blue{While presenting the results, we note that $b(y,\tau)$ is a dimensionless quantity, and thus we do not provide explicit dimensions of $b(y,\tau)$. Unless otherwise mentioned, $R_{nn}(y,\tau)$ has units of ${\rm g \ cm^{-1} \ s^{-2}}$,  $a_y(y,\tau)$ has units of ${\rm cm \ s^{-1}}$, $g$ has units of ${\rm cm \ s^{-2} }$ and $\rho_1, \rho_2, \overline{\rho}$ has units of ${\rm g \ cm^{-3}}$. Note that the corresponding spectral quantities will have the following units : $\hat{b}(y,k,\tau)$ has units of ${\rm cm }$, $\hat{a}_y(y,k,\tau)$ has units of ${\rm cm^2 \ s^{-1}}$, and $\hat{R}_{nn}(y,k,\tau)$ has units of ${\rm g \ s^{-2}}$.}

\begin{table*}
{
\begin{tabular}{|c|c|c|c|c|c|c|c|c|c|c|c|}
\hline
Run&$A$&$N_{x}$ & $N_z$&$N_{v}$  & $L_x$ \blue{[cm]} & $L_z$ \blue{[cm]}&$L_y$ \blue{[cm]}  & $g [\blue{\rm cm \ s^{-2}}]$&$\rho_1 [\blue{{\rm g \ cm^{-3} }}]$ & $\rho_2 [\blue{{\rm g \ cm^{-3}}}]$ & $\displaystyle t^{\prime} [\blue{{\rm sec}}]$\\
\hline
{\tt R1}&$0.25$&$256$ & $256$ &$512$ & $2\pi$ & $2\pi$ & $4\pi$  & $2.0$ &$1.0$ & $1.667$ & $3.5$\\
\hline
{\tt R2}&$0.1$&$256$ & $256$&$1024$ & $2\pi$ & $2\pi$ &$8\pi$ & $2.0$ &$1.0$ & $1.228$ & $5.6$\\
\hline
{\tt R3}&$0.05$&$256$ & $256$&$1024$ & $2\pi$ & $2\pi$&$8\pi$ & $2.0$&$1.0$ & $1.105$& $7.9$\\
\hline
\end{tabular}
}
\caption{Table showing system parameters used in the MOBILE simulations for runs {\tt R1},{\tt R2}, {\tt R3}. $N_x,N_z$ are the two horizontal resolutions and $N_v$ is the vertical resolution. $L_x, L_z$ are the domain lengths in the two horizontal directions and $L_y$ is the domain height (vertical direction). $\rho_1$ and $\rho_2$ are the densities of the light and heavy fluids respectively. $t^{\prime}=\df{1}{\sqrt{Ag/L_x}}$ denotes a typical timescale for the flow.
}
\label{table1} 
\end{table*}

\section{\label{results}Results}
In this section, we seek to compare the results from the LWN model with MOBILE simulations. 
We first demonstrate that the LWN model code converges as numerical resolution is increased. 

\subsection{A test case\label{testcase}}
\begin{figure*}
\includegraphics[width=.32\linewidth]{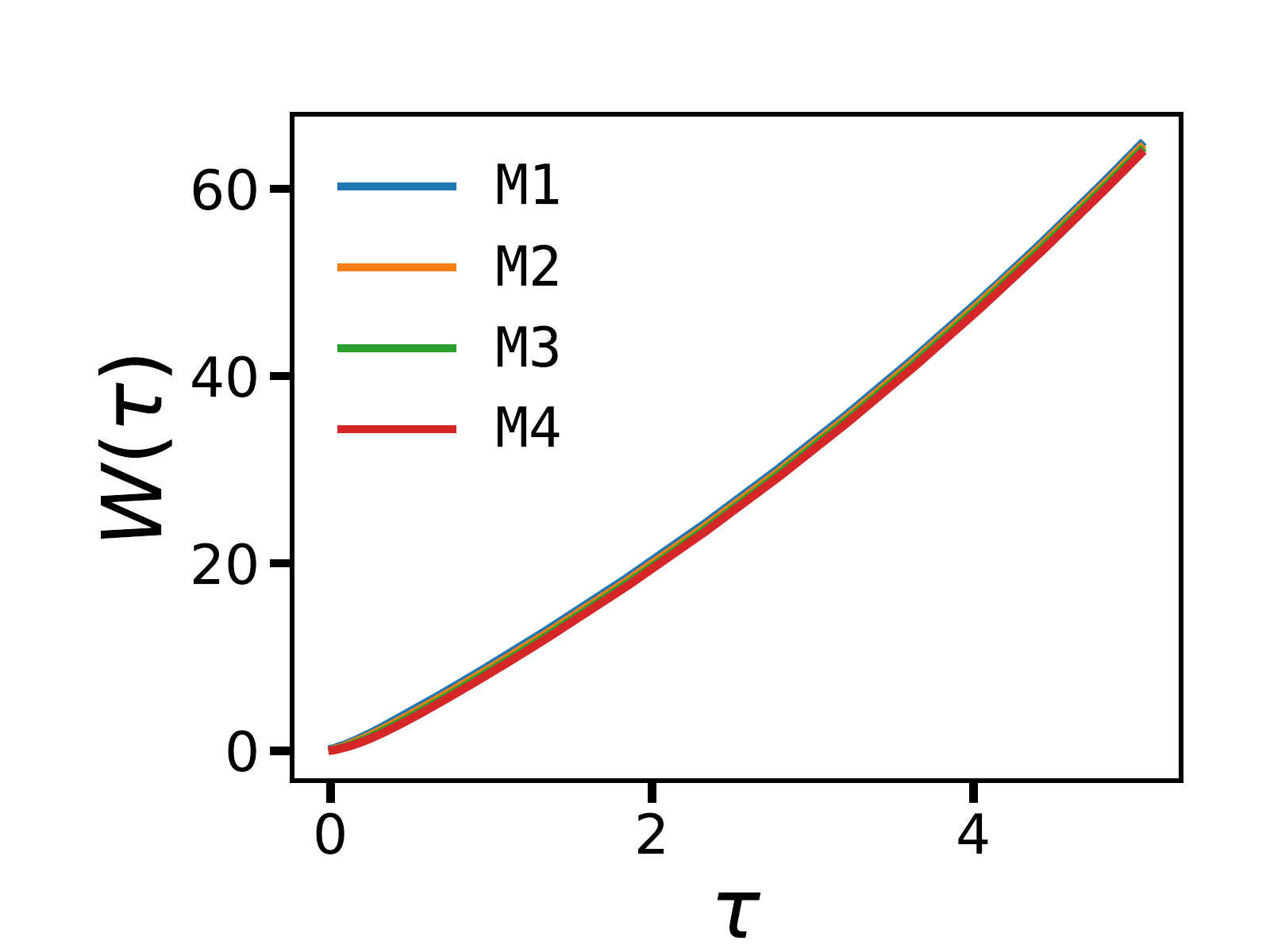}
\put(-30,40){\bf \scriptsize (a)}
\includegraphics[width=.32\linewidth]{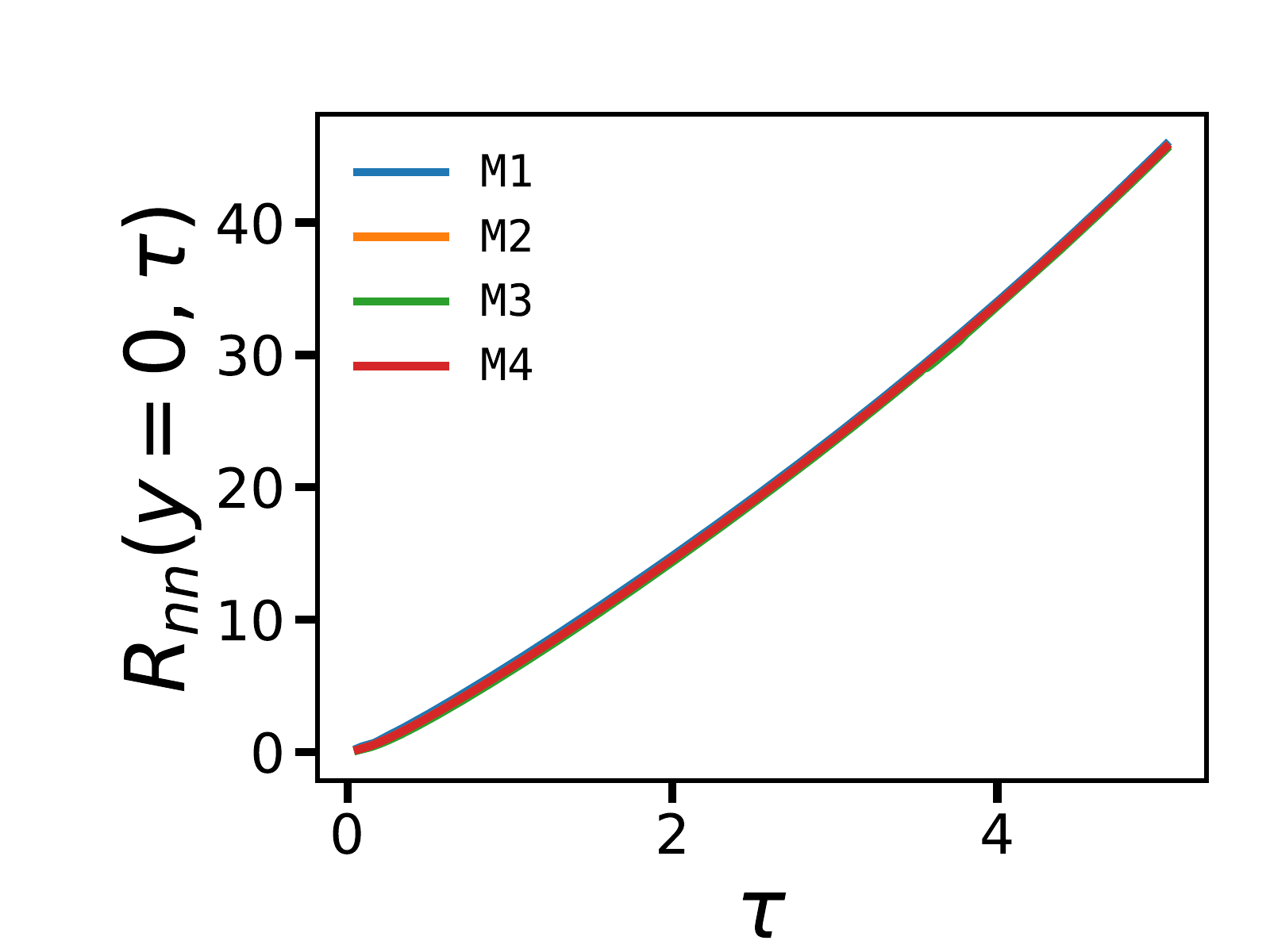}
\put(-30,40){\bf \scriptsize (b)}
\includegraphics[width=.32\linewidth]{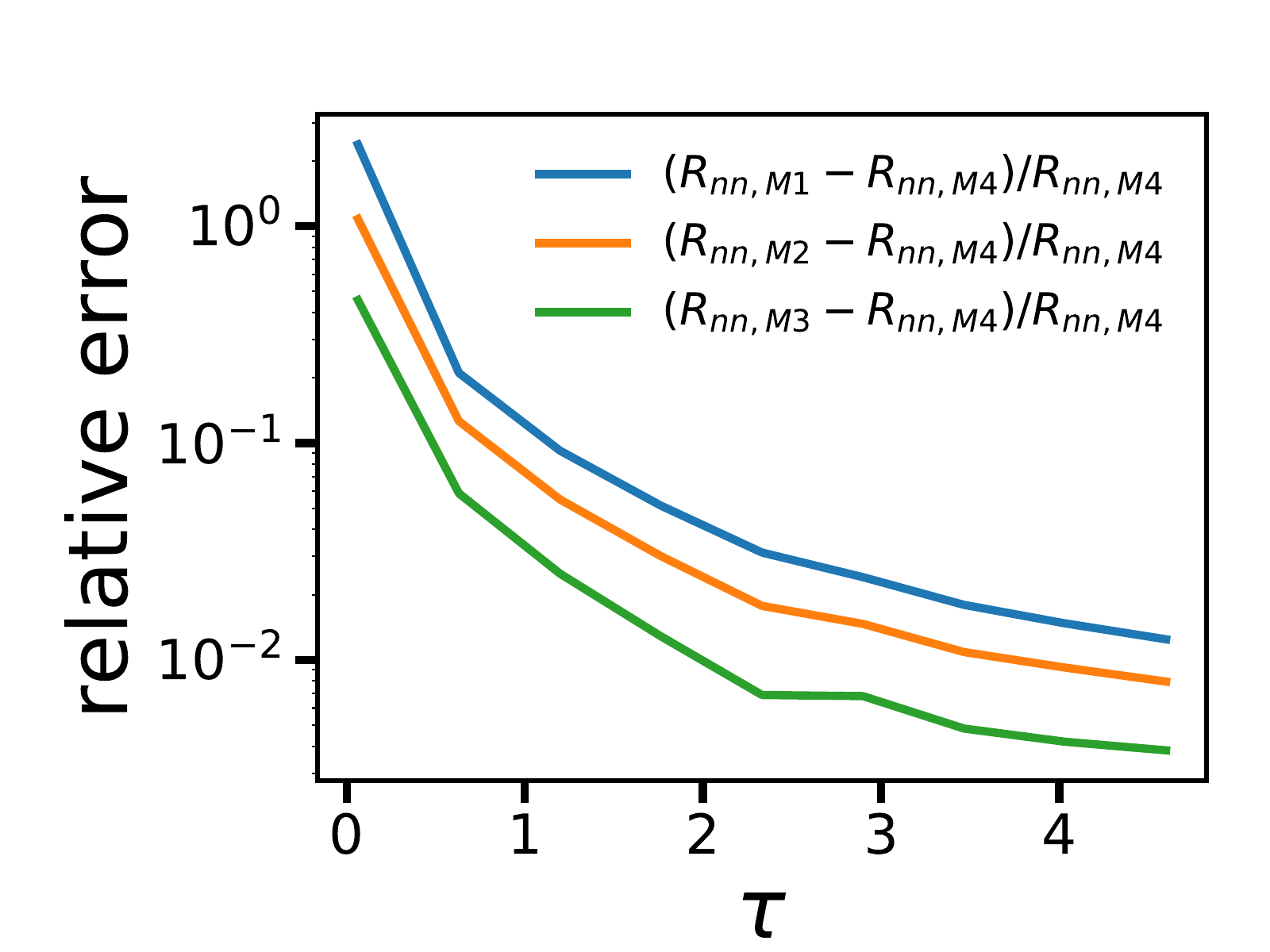}
\put(-115,40){\bf \scriptsize (c)}
\caption{(a) The mix-width $W(\tau)$ \blue{(in cm)};  (b) mean $R_{nn}(y=0,\tau)$ (in \blue{${\rm g \ cm^{-1} \ s^{-2}}$}) for {\tt M1}, {\tt M2}, {\tt M3}, and {\tt M4}; (c) relative error of $R_{nn}(y=0,\tau)$ for {\tt M1} and {\tt M2} with respect to $R_{nn}(y=0,\tau)$ for {\tt M3}. showing convergence as resolution is increased. The system coefficients are given in the first three rows of Table~\ref{table1a}.}
\label{spec_test}
\end{figure*}

We configure our LWN model for an idealized RT system starting with an analytically specified initial spectrum and demonstrate that our implementation exhibits convergence with grid refinement for the relevant metrics.
The initial $\hat{b}(y=0,k)$ has the following functional form~\cite{Steinkamp1999b} :
\begin{equation}
\hat{b}(y=0,k)=\df{\gamma_1k^m}{1+\gamma_2 k^{m+\frac{5}{3}}}
\label{init_b}
\end{equation} 
and $\hat{R}_{yy}(y,k,\tau=0)=0$ and $\hat{a}_y(y,k,\tau=0)=0$. 
The two constants $\gamma_1$ and $\gamma_2$ are chosen to ensure that the maximum of $\hat{b}$ occurs at $k=1$ and the initial spectral integral \blue{\sout{$b(y)=0.5$} $b(y)$} corresponds to equal volume fraction of the two fluids in the $y=0$ cell. 
This particular functional form of $\hat{b}(y=0,k)$ was used previously in \cite{steinkamp1996spectral} so that when $k$ is small, $\hat{b}(y=0,k) \sim k^m$, and when $k$ is large, $\hat{b}(y=0,k) \sim k^{-\frac{5}{3}}$, following the anticipated power-law scaling of the turbulent kinetic energy spectrum within the inertial range. In Fig. \ref{spec_test}(a) and (b) we demonstrate convergence of the mix-width $W(\tau)$ and $R_{nn}(y=0,\tau)$ with grid refinement in physical space. In Fig.~\ref{spec_test}(c) we plot the relative error in $R_{nn}(y=0,\tau)$ for the runs {\tt M1} and {\tt M2} with respect to $R_{nn}(y=0,\tau)$ of the most refined grid, i.e., {\tt M3}. As the plot shows, the relative error goes to $0$ asymptotically with time. The system coefficients are given in the first two rows of Table~\ref{table1a}.

\begin{table*}
{
\begin{tabular}{|c|c|c|c|c|c|c|c|c|c|c|c|c|c|c|c|}
\hline
Test case &$A$& $g [\blue{{\rm cm \ s^{-2}}}]$ &$L_x$ \blue{[cm]} &$L_y$ \blue{[cm]} &$N_k$&$dk_{max} [\blue{{\rm cm^{-1}}}]$&$N_v$ &$C_{r1}$&$C_{r2}$&$C_d$&$C_{rp1}$ & $C_{rp2}$ & $t^{\prime} [\blue{{\rm sec}}]$ \\
\hline
{\tt M1}&$0.5$&$2.0$&$62.8$&$30.0$&$140$& $0.1$ &$120$&$0.12$&$0.06$ &$0.03$ &$1.0$ & $1.0$ & $7.92$ \\
\hline
{\tt M2}&$0.5$&$2.0$&$62.8$&$30.0$&$140$& $0.1$ &$240$&$0.12$&$0.06$ &$0.03$ &$1.0$ & $1.0$ & $7.92$ \\
\hline
{\tt M3}&$0.5$&$2.0$&$62.8$&$30.0$&$140$& $0.1$ &$480$&$0.12$&$0.06$ &$0.03$ &$1.0$ & $1.0$ & $7.92$ \\
\hline
{\tt M4}&$0.5$&$2.0$&$62.8$&$30.0$&$140$& $0.1$ &$960$&$0.12$&$0.06$ &$0.03$ &$1.0$ & $1.0$ & $7.92$ \\
\hline
Homogeneous&$0.05$&$1.0$&$2\pi$&--&$1024$&--&&$0.12$& $0.06$ &$0.0$ &$1.0$ & $1.0$ & $1.0$ \\
\hline
\end{tabular}
}
\caption{
Table summarizing parameters for test case at three resolutions, and the coefficients used. The previously studied homogeneous case is also tabulated to indicate the coefficients optimized for that study \cite{pal2018two}. $N_k$ and $N_v$ are the horizontal (spectral) and vertical resolutions respectively. $A$ is the Atwood number and $g$ denotes the acceleration due to gravity. $dk_{max}$ is the maximum value of the spectral discretization. We use a logarithmic discretization in the spectral space. $t^{\prime}=\df{1}{\sqrt{Ag/L_x}}$ denotes a typical timescale for the flow.}
\label{table1a} 
\end{table*}

\subsection{Comparison with MOBILE simulations}

An extensive literature on the topic, encompassing the full breadth of theory, experiments and numerical simulations, e.g.~\cite{dalziel1999self,dimonte2004comparative, ZHOU20171,zhou2017rayleigh,zhou2019time}, has shown that RT instability is a complex mixing phenomenon with three stages of evolution, a linear, a weakly nonlinear and a fully nonlinear turbulent stage. Any small perturbations present at the initial condition grow exponentially in the linear stage, after which they interact with each other in the nonlinear stage, with a final transition to turbulence. The mix-layer grows exponentially in the first stage and later grows as $\tau^2$, where $\tau = t/t^{\prime}$ is the non-dimensionalized time, in the final turbulent stage. In this subsection we specify the stage of the flow evolution in MOBILE that we take as an initial condition for the LWN model calculations.

\begin{figure*}
\includegraphics[scale=0.26]{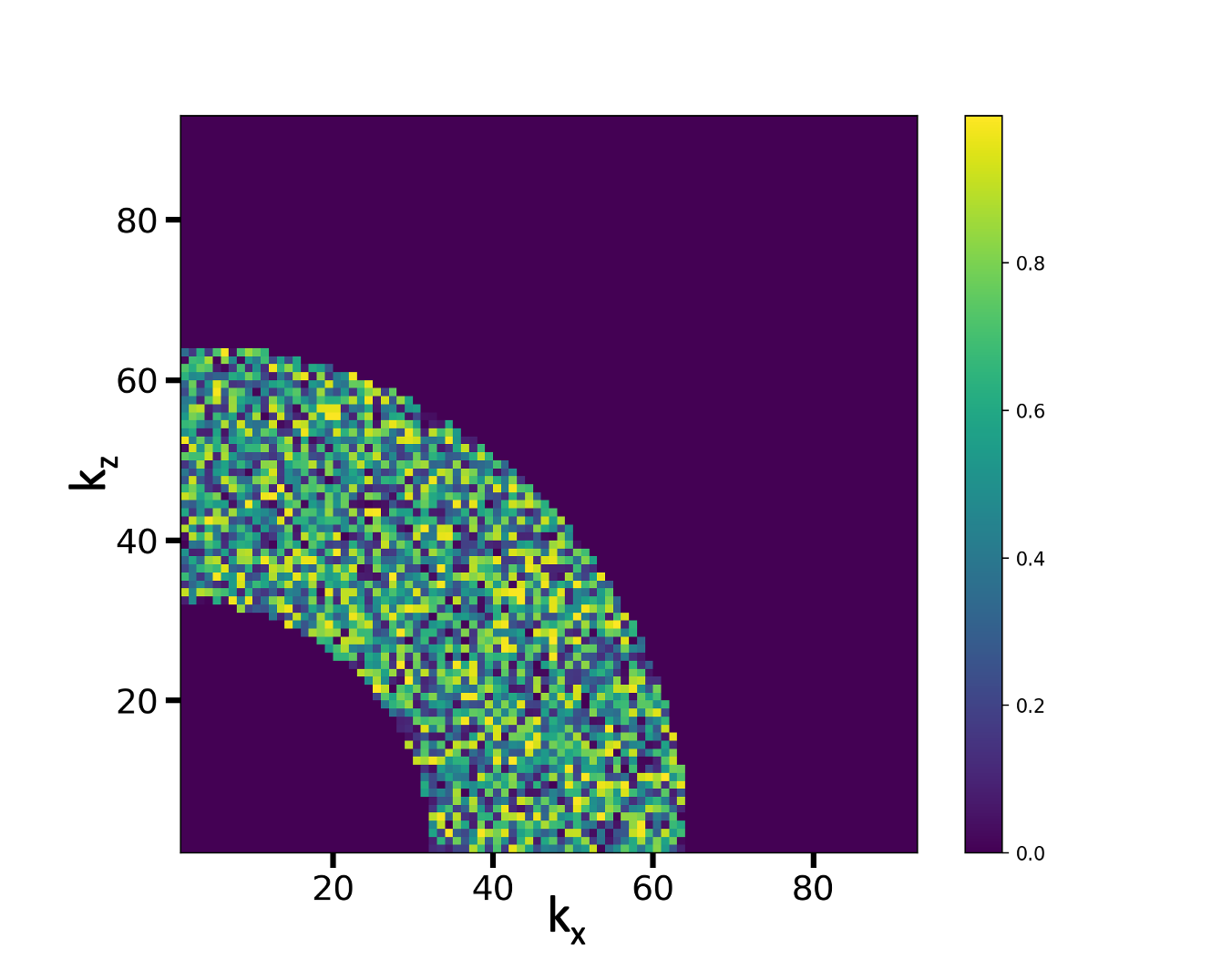}
\put(-60,100){\bf \scriptsize \textcolor{white}{(a)}}
\includegraphics[scale=0.265]{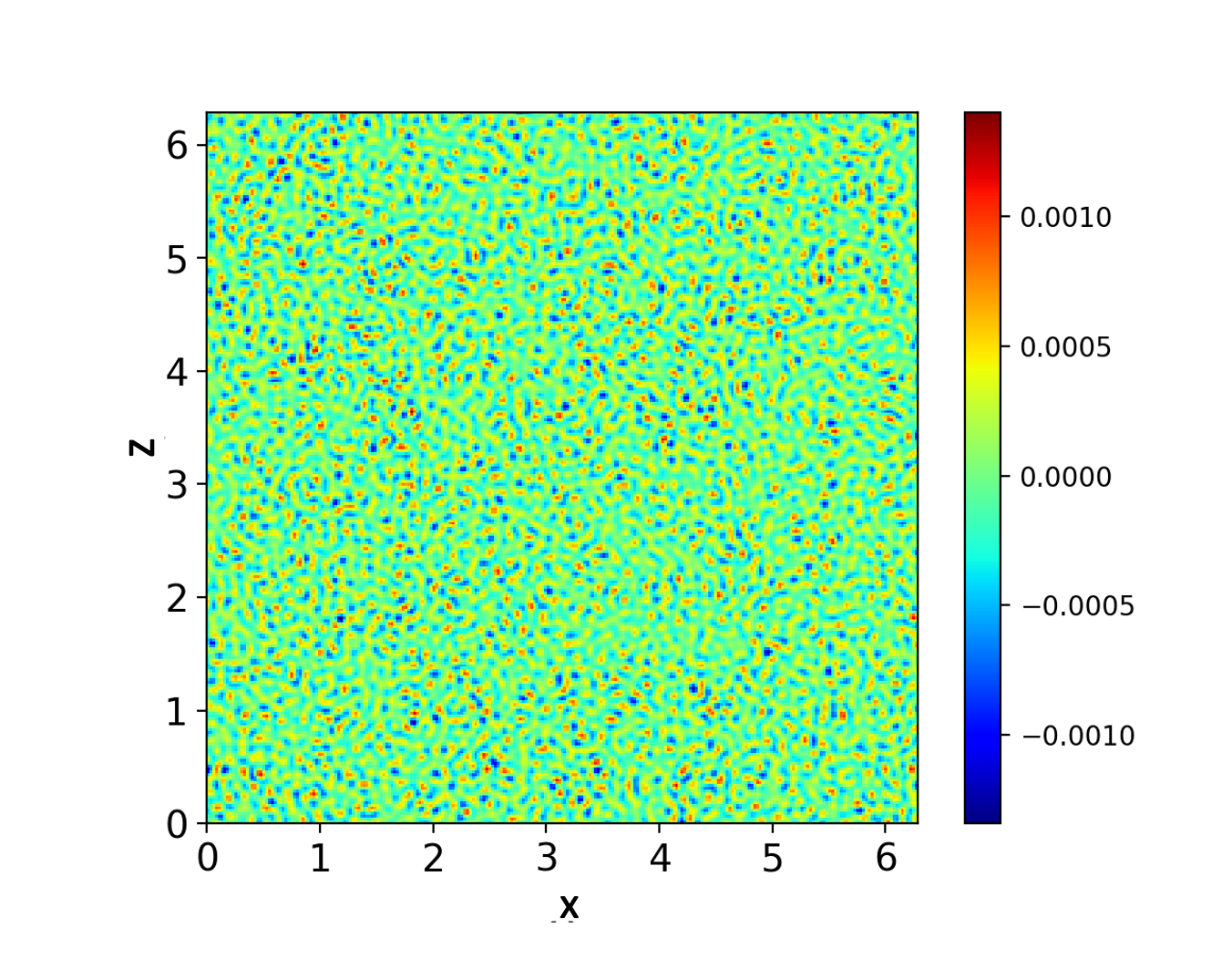}
\put(-60,100){\bf \scriptsize \textcolor{black}{(b)}}
\includegraphics[scale=0.38]{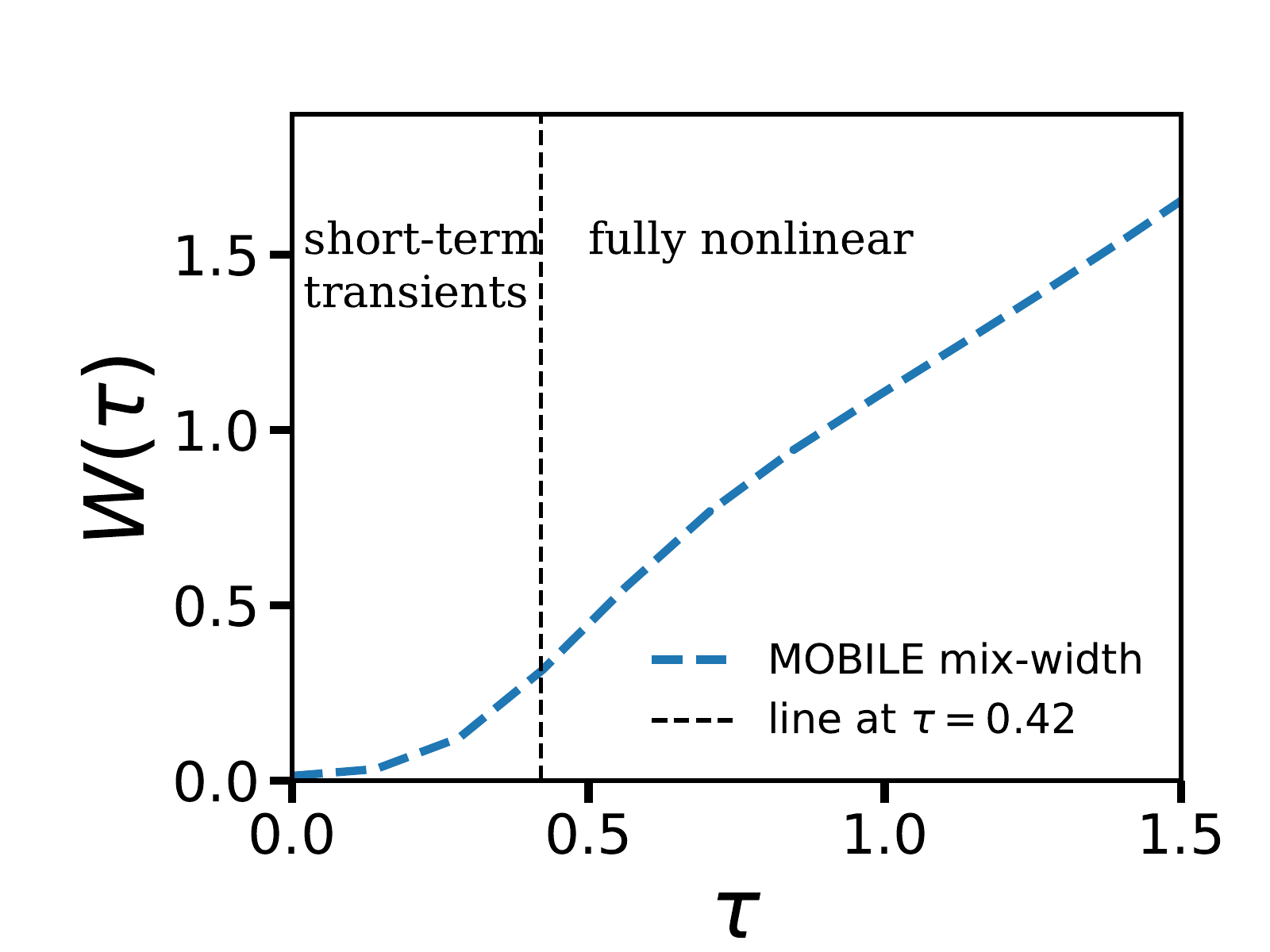}
\put(-35,100){\bf (c)}
\caption{(a) Visualization of population of 2D-spectral modes in the $k_x$-$k_z$ plane at the centerline of the domain, (b) amplitude $h_0(x,y)$ (in \blue{${\rm cm}$}) in physical space; (c) time-evolution of mix-width $W(\tau)$ \blue{(in ${\rm cm}$)} obtained from the MOBILE data showing the exponential and nonlinear growth stages; the black dotted line shows the start time for LWN calculations.}
\label{iles_init}
\end{figure*}

For all our model calculations, early-time perturbation spectra from MOBILE calculations were used to initialize the model. 
The multimode perturbation imposed at the interface separating the two fluids~ \cite{alpha_group} can be obtained by the function:
\begin{eqnarray} 
h(x,z,t=0)&=&\nonumber\\
\sum_{k_y, k_z}
\begin{bmatrix}
a_k \times cos(k_x x) \times cos(k_z z) + &&\\
b_k \times cos(k_x x) \times sin(k_z z) + &&\\
c_k \times sin(k_x x) \times cos(k_z z) + &&\\
d_k \times sin(k_x x) \times sin(k_z z)
\end{bmatrix}
\end{eqnarray}
Here $h(x,z,t)$ is the amplitude of perturbation in the horizontal $x--z$ plane.
$k_x$ and $k_z$ are wave numbers in the two horizontal directions.
 $h_{0_{RMS}} \sim 3.0 \times 10^{-4} L_x$ is the RMS amplitude, and with
the spectral amplitudes $a_k$ , $b_k$ , $c_k$ , $d_k$ chosen randomly chosen within narrow-band spectrum in wavenumber space in the range $32 \leq k \leq 64$ as shown in Fig.~\ref{iles_init}(a). The amplitude profile $h_0(x,y)$ in physical space is shown in Fig.~\ref{iles_init}(b). The initial amplitudes were converted to volume fraction perturbations.

We choose as the initial condition for evolution of the LWN model a non-dimensional time $\tau_0=0.42$, where $t^{\prime} = 1/\sqrt{Ag/L_x}$ is the characteristic timescale of the flow. At this time, the mix-layer as calculated by the data is approaching the end of the early growth stage, as shown in Fig. \ref{iles_init}(c). Immediately thereafter, the growth rate smoothly transitions from that associated with the initial interface spectrum to a fully nonlinear development. As we show later in the paper, this time is close to the position of the peak of the mean $b(y=0,\tau)$ evolution. The simple version of the LWN model does not contain a kinematic source term in the equation for $b(y,\tau)$ (see Eq.~\ref{main_b}). Since there is no mechanism in the LWN model to represent these earliest stages of growth, particularly the initial growth of $\hat{b}$, the modeled mix-layer goes directly into the quadratic nonlinear regime~\cite{inogamov2001stochastic}. Therefore we start the model calculations at a time when the MOBILE mix-layer has settled out of its early transients. As we show later in the paper, this time is also close to the position of the peak of the mean $a_y(y=0,\tau)$ evolution.
 


\begin{figure*}
\includegraphics[width=.35\linewidth]{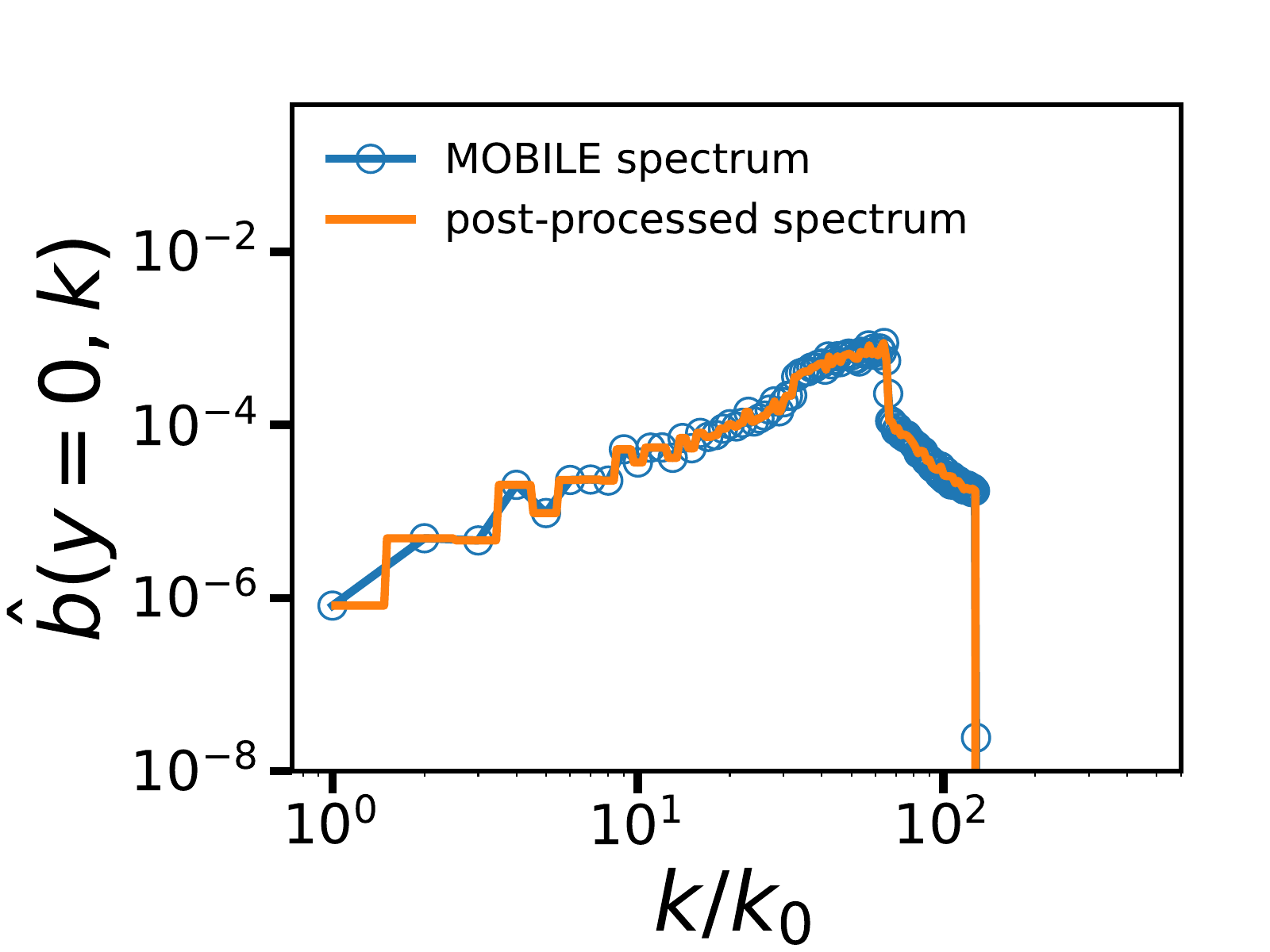}
\put(-30,100){\bf (a)}
\includegraphics[width=.35\linewidth]{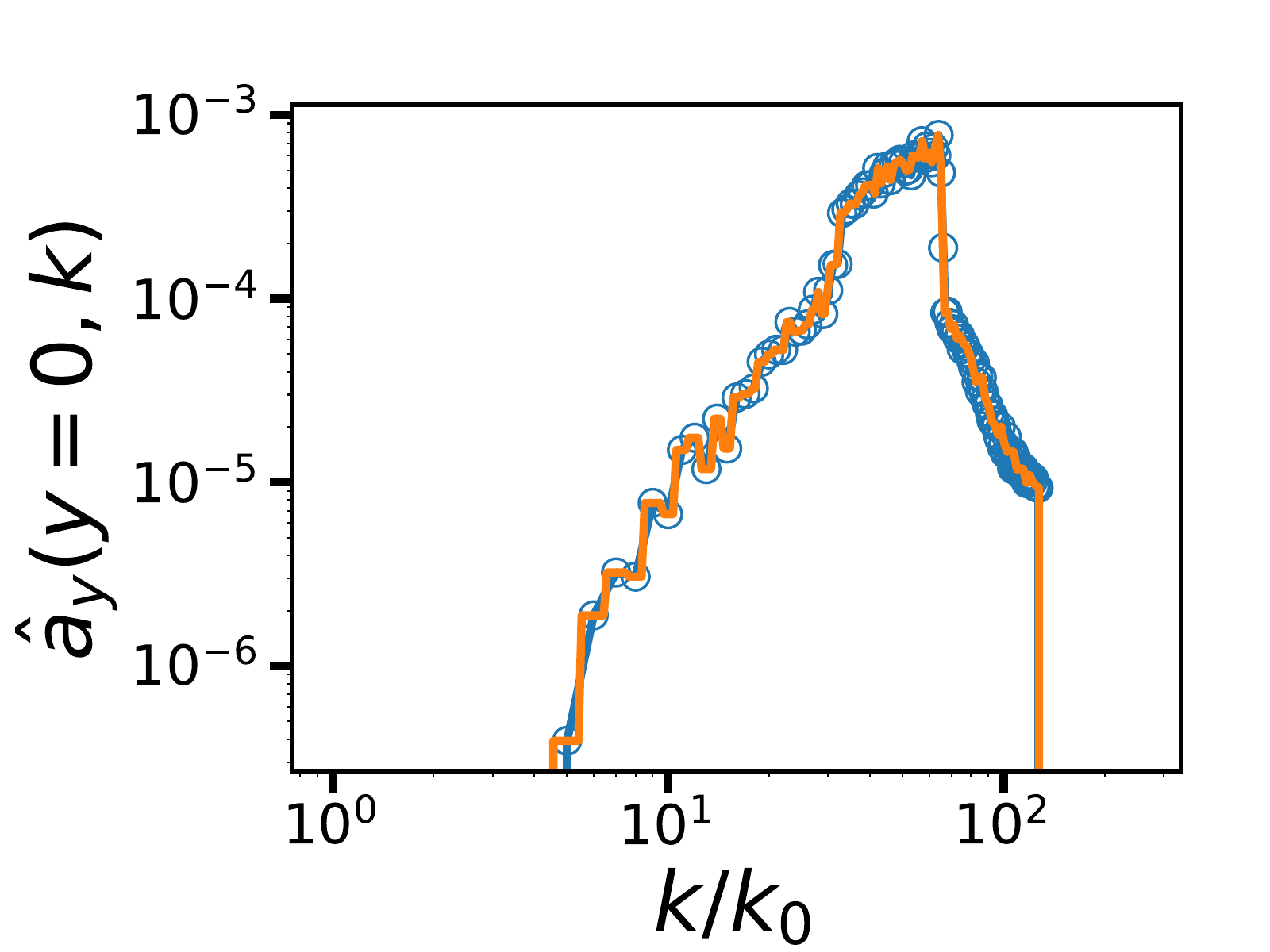}
\put(-30,100){\bf \scriptsize (b)}
\includegraphics[width=.35\linewidth]{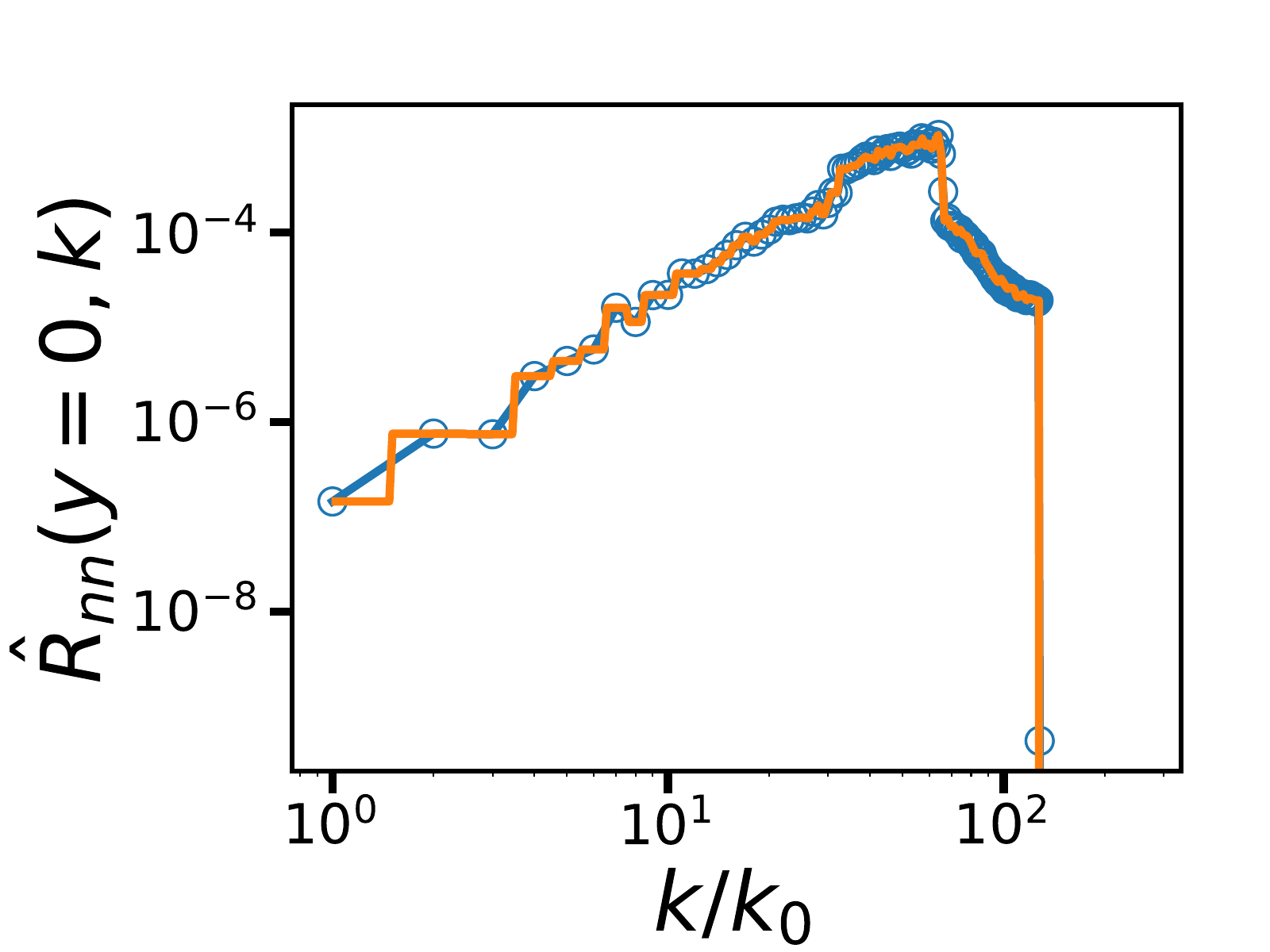}
\put(-30,100){\bf \scriptsize (c)}
\caption{
\blue{Initial spectral data for
LWN taken from MOBILE {\tt R1} at $\tau = 0.42$ (blue line): (a) $\hat{b}(y=0,k)$ [cm], (b)$\hat{a}_y(y=0,k)$ [${\rm cm}^2 {\rm \ s}^{-1}$] and (c) spectra for $\hat{R}_{nn}(y=0,k)$ [${\rm g \ s}^{-1}$]. The spectral data are taken at the
central line: $y=0$}}
\label{spec}
\end{figure*}

At the chosen time $\tau_0$, the $\hat{b}(y=0,k)$, $\hat{a}_y(y=0,k)$ and $\hat{R}_{nn}(y = 0, k)$ spectral initial conditions for the model as obtained from the MOBILE data are shown in Fig.~\ref{spec}.
Here the lowest wavenumber $k_0=\displaystyle\frac{2\pi}{L_x}$, where $L_x$ is the domain length (see Fig.~\ref{visual}).The $y\neq0$ planes are initialized from the MOBILE spectra in a similar manner. The maximum wavenumber is $k_{max}=128$ due to the de-aliasing operation~\cite{canuto2012spectral}. 

\subsubsection{Results from the LWN model: $A=0.25$}

\begin{table}
{
\begin{tabular}{|l|l|l|l|l|l|l|l|}
\hline
run &$A$ & $\tau_0$ &$C_{r1}$ & $C_{r2}$ &$C_d$&$C_{rp1}$ & $C_{rp2}$ \\
\hline
{\tt T1}&$0.25$&$0.42$&$0.12$&$0.06$ &$0.03$&$1.0$ & $1.0$ \\
\hline
{\tt T2}&$0.25$& $0.42$&$0.12$& $0.06$ &$0.1$&$1.0$ & $1.0$\\
\hline
{\tt T3}&$0.25$&$0.42$ &$0.12$& $0.06$ &$0.5$&$1.0$ & $1.0$\\
\hline
{\tt T4}&$0.25$&$0.42$ &$0.12$& $0.06$ &$1.0$&$1.0$ & $1.0$\\
\hline
{\tt T5}&$0.25$&$0.42$ &$0.12$& $0.06$ &$0.5$&$0.08$ & $0.08$\\
\hline
{\tt T6}&$0.25$& $0.42$&$0.12$& $0.06$ &$0.5$&$0.2$ & $0.2$\\
\hline
\boxit{2.22in}
{\tt T7}&$0.25$&$0.42$ &$0.12$ &$0.06$&$0.5$&$0.5$ & $0.5$\\
\hline
\end{tabular}
}
\caption{Table summarizing comparison study of LWN model against the MOBILE data (run {\tt R1}). The set of coefficients inside the red box gives the best agreement between the LWN model and MOBILE results that were obtained in this study.}
\label{table2} 
\end{table}

\begin{figure*}
\includegraphics[width=.45\linewidth]{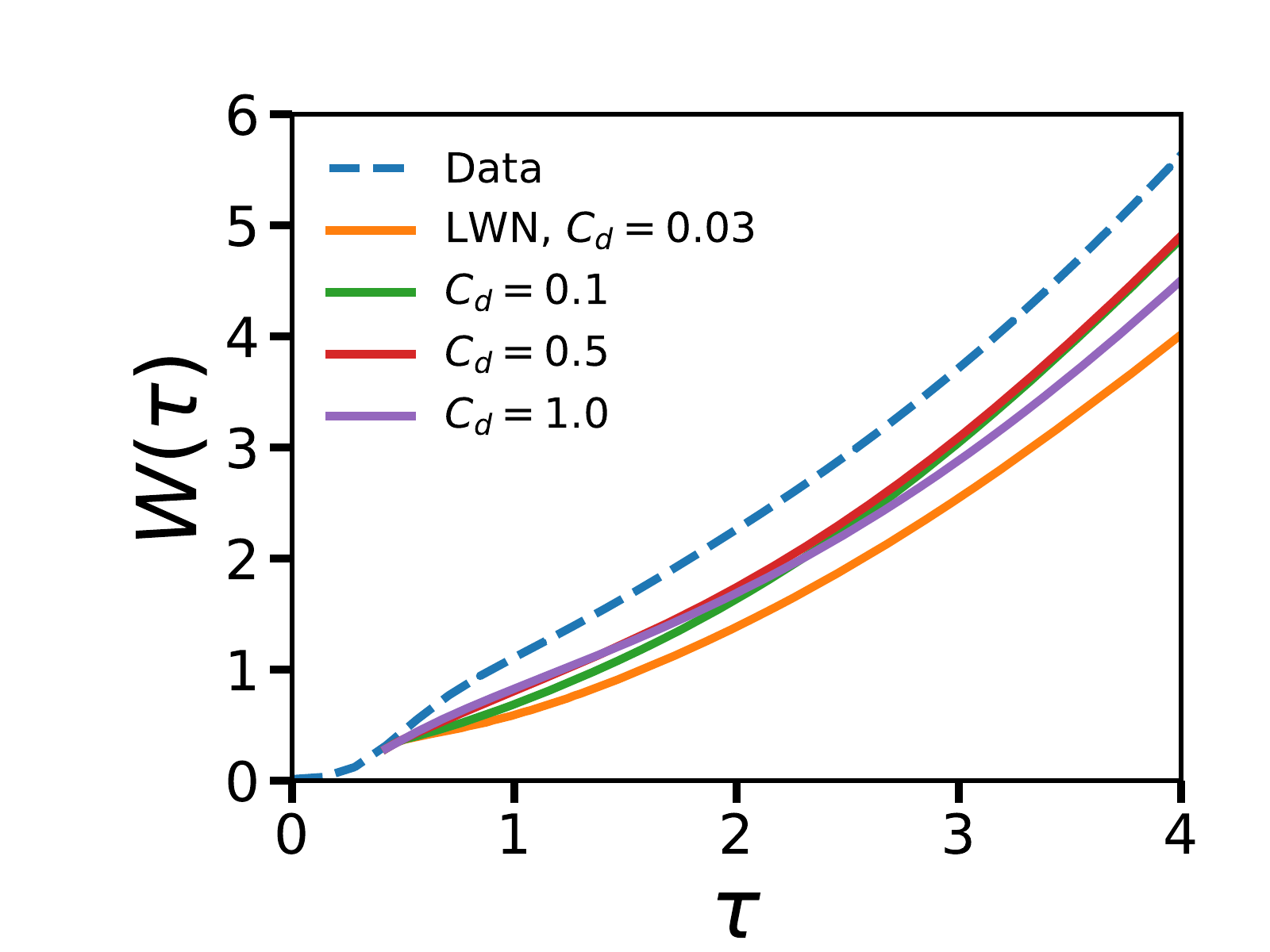}
\put(-30,40){\bf \scriptsize (a)}\\

\includegraphics[width=.32\linewidth]{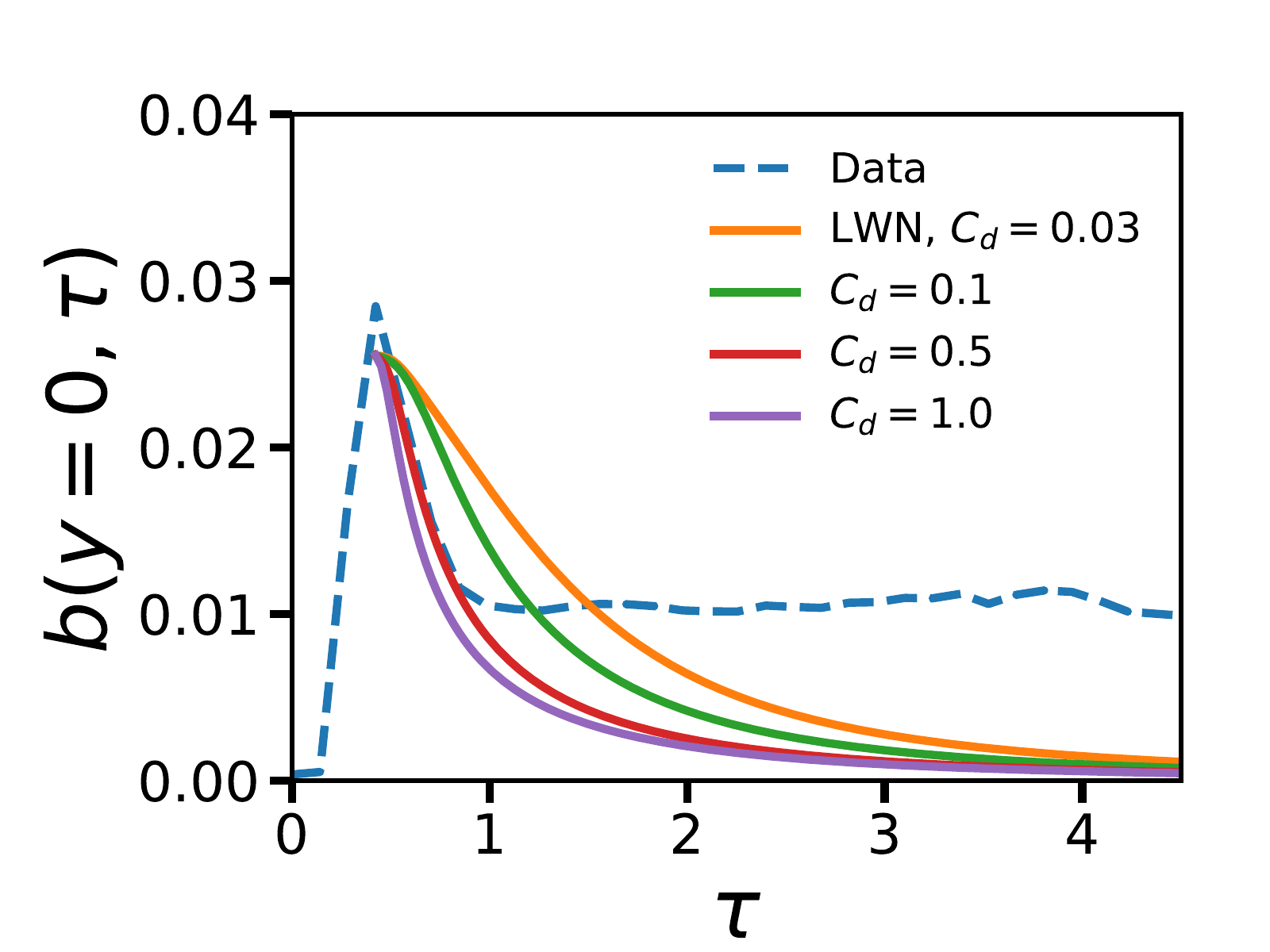}
\put(-118,90){\bf \scriptsize (b)}
\includegraphics[width=.32\linewidth]{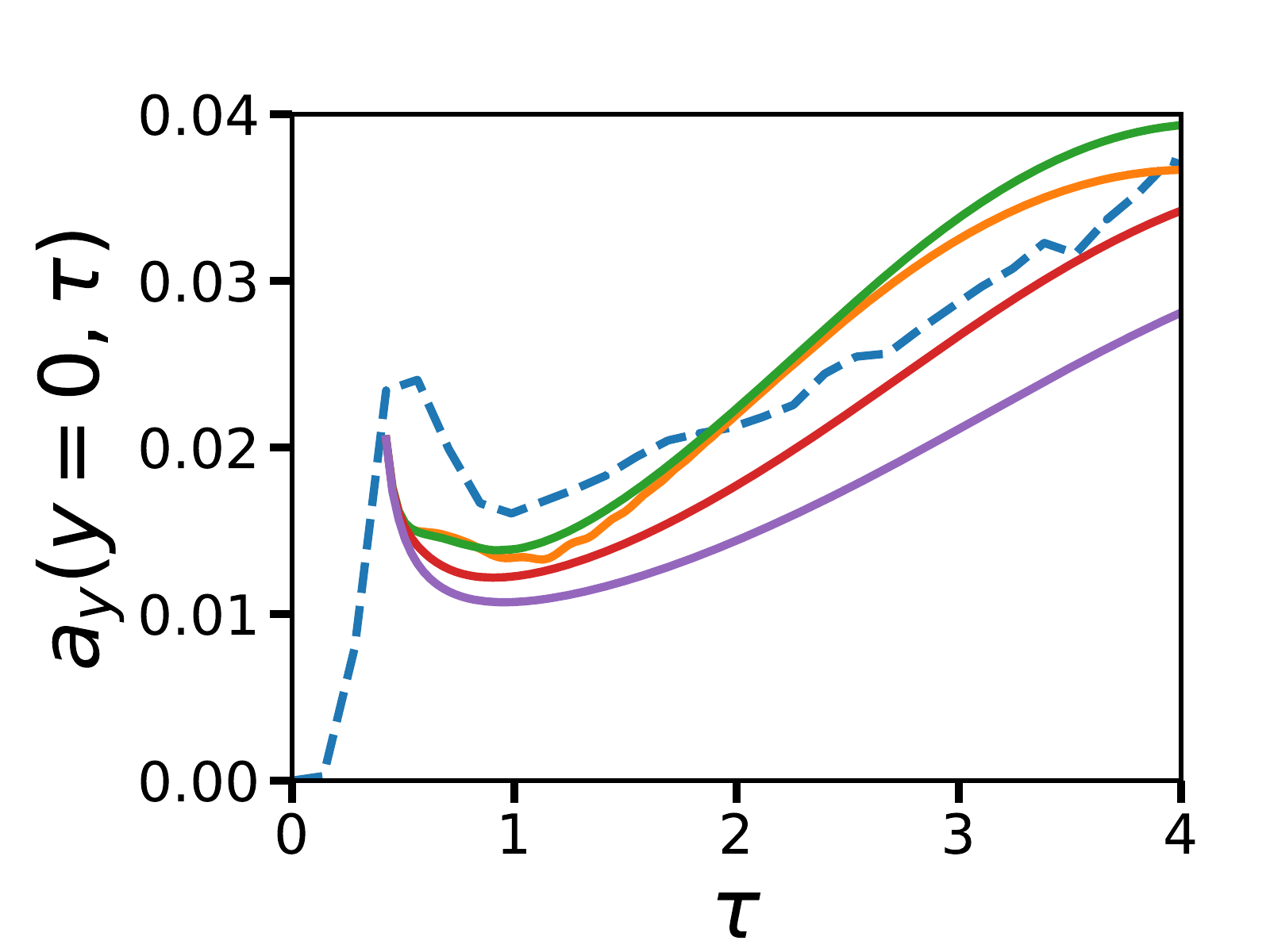}
\put(-115,90){\bf \scriptsize (c)}
\includegraphics[width=.32\linewidth]{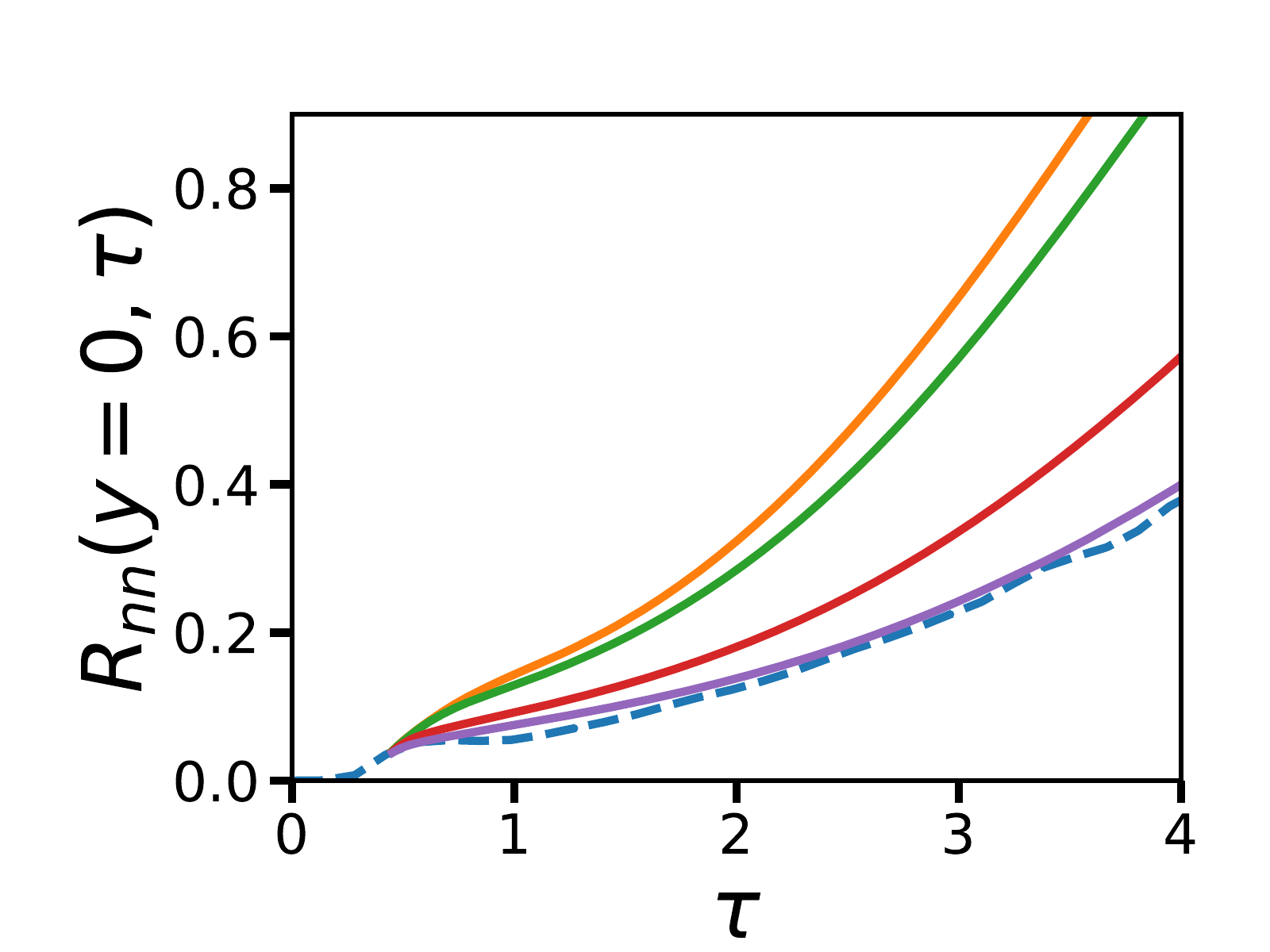}
\put(-115,90){\bf \scriptsize (d)}
\caption{
(a) Mix-width (in \blue{${\rm cm}$}) from the MOBILE data ({\tt R1}, blue dashed line) and that from the LWN model calculations ({\tt T1} orange, {\tt T2} green, {\tt T3} red, {\tt T4} purple); (b) $b(y=0,\tau)$; (c) $a_y(y=0,\tau)$ (in \blue{${\rm cm \ s^{-1} }$}); (d) $R_{nn}(y=0,\tau)$ (in \blue{${\rm g \ cm^{-1} \ s^{-2} }$}).
}
\label{mix_high_crp}
\end{figure*}

The coefficients used to compute a series of LWN runs for $A=0.25$ are listed in  Table~\ref{table2}. We keep the spectral transfer coefficients $C_r1$ and $C_{r2}$ fixed to the values obtained in the homogeneous turbulence study of \cite{pal2018two}. 
In {\tt T1}--{\tt T4} we vary the spatial diffusion coefficient $C_d$, keeping the drag coefficients $C_{rp1}$ and $C_{rp2}$ fixed at $1.0$, their value in the homogeneous variable-density case (see last line of Table~\ref{table1a}). 
Figure.~\ref{mix_high_crp}(a) shows that the mix-layer width is under-predicted compared to the MOBILE data (blue dashed line) for {\tt T1}--{\tt T4}, and relatively insensitive to large changes in $C_d$. In Figs.~\ref{mix_high_crp}(b),(c) and (d) we see that as $C_d$ is increased, $b(y=0,\tau)$ decays faster, and growth of both the magnitude of $a_y(y=0,\tau)$ and $R_{nn}(y=0,\tau)$ is suppressed at later times. It appears that $C_d$ roughly ${\cal O}(1)$ is a reasonable choice. 

\begin{figure*}
\includegraphics[width=.325\linewidth]{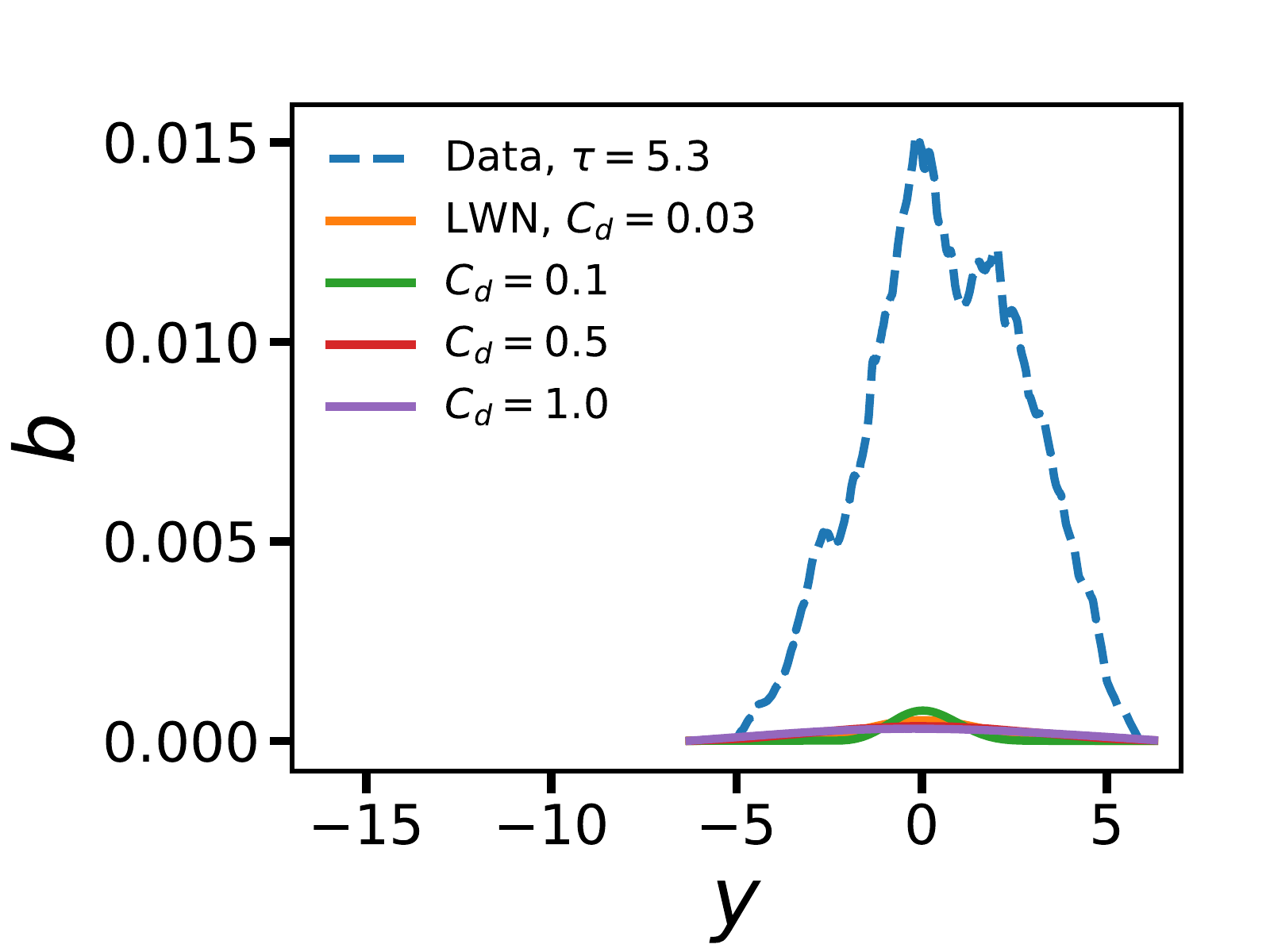}
\put(-25,95){\bf \scriptsize (a)}
\includegraphics[width=.325\linewidth]{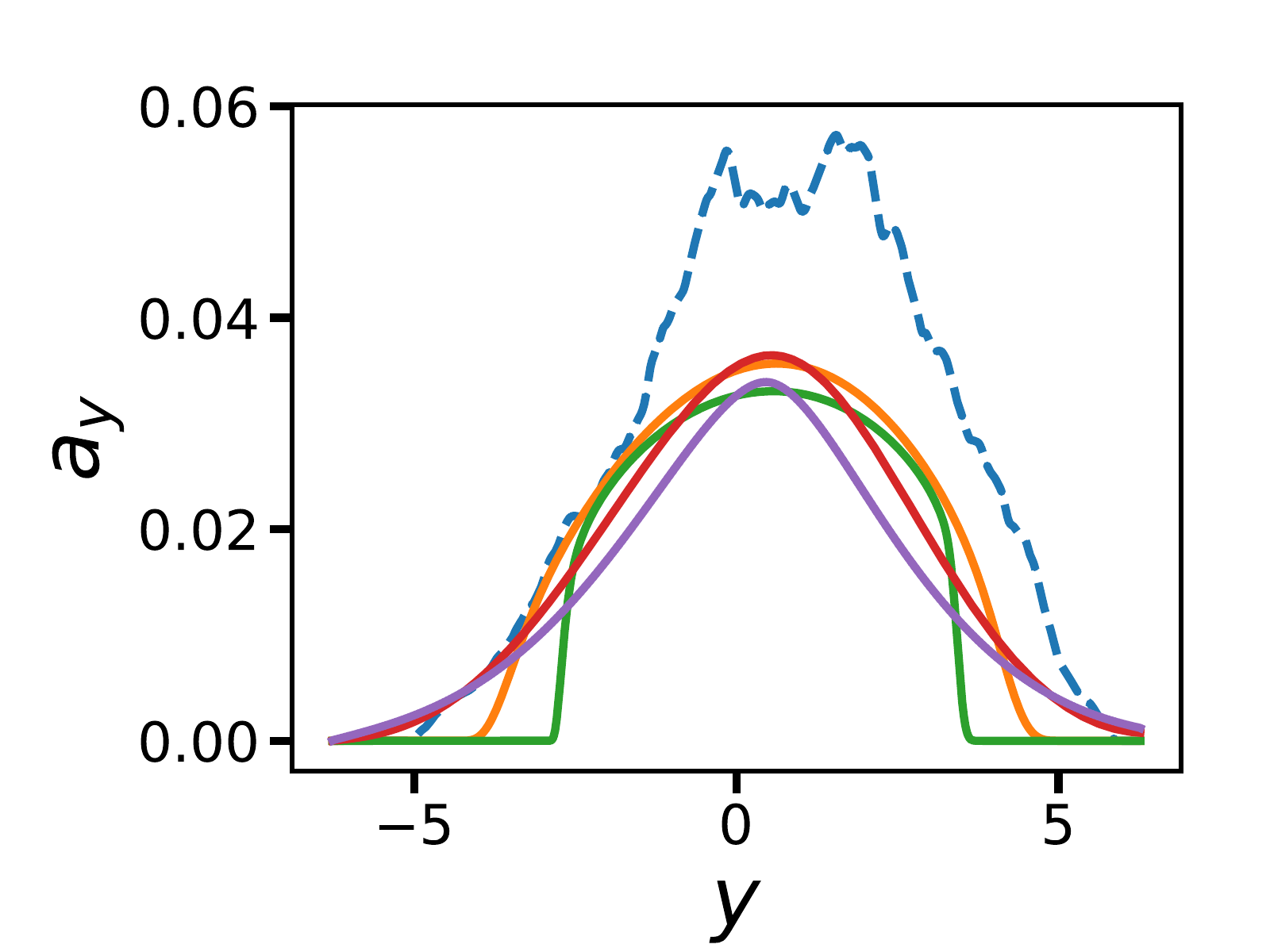}
\put(-25,95){\bf \scriptsize (b)}
\includegraphics[width=.325\linewidth]{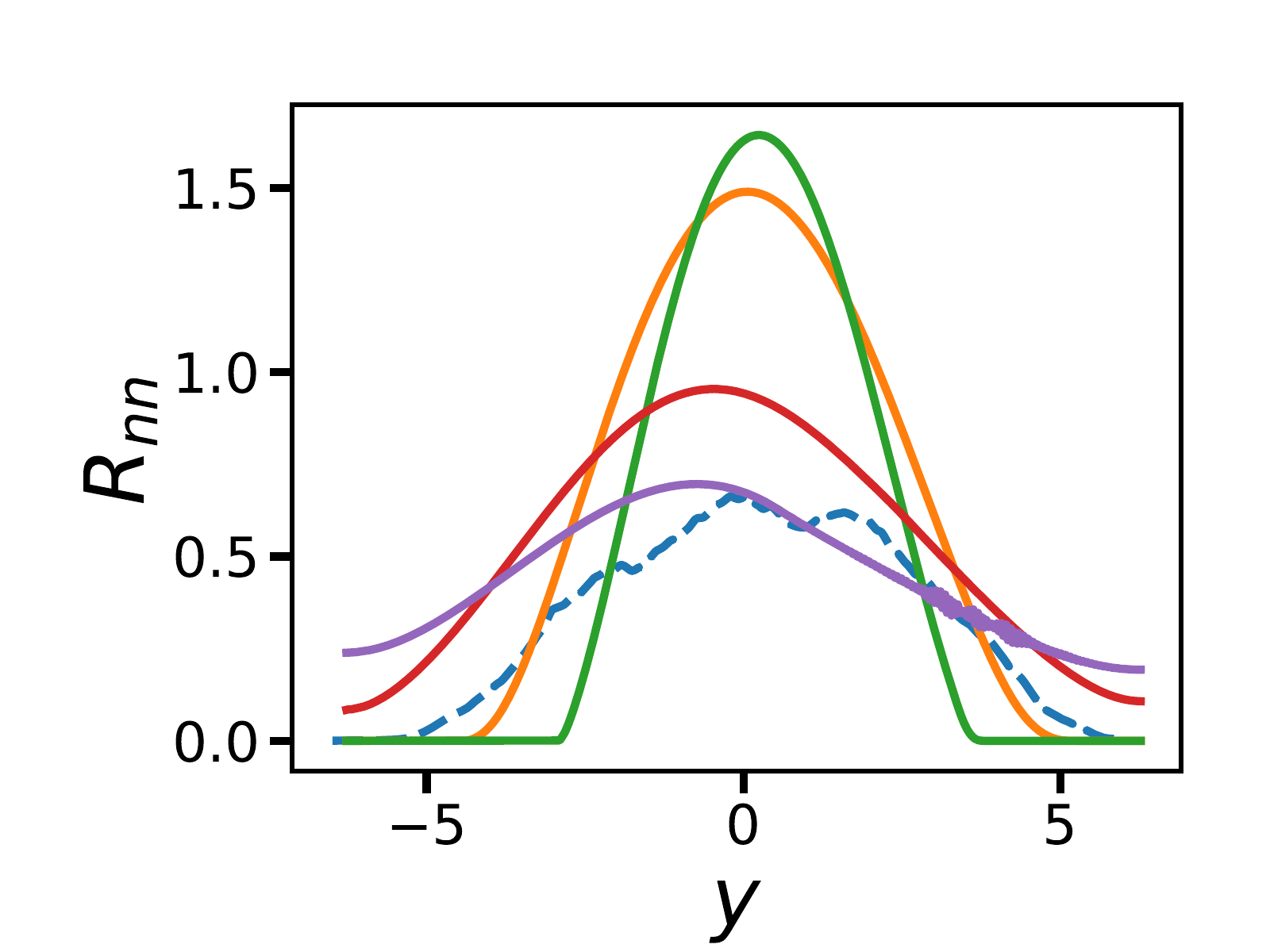}
\put(-25,95){\bf \scriptsize (c)}
\caption{Profiles of (a) $b(y,\tau)$, (b) $a_y(y,\tau)$ (in \blue{${\rm cm \ s^{-1}}$}) and (c) $R_{nn}(y,\tau)$ (in \blue{${\rm g \ cm^{-1} \ s^{-2}}$}) at time $\tau=5.3$ for the MOBILE data ({\tt R1}, blue dashed line) and that from the LWN model calculations ({\tt T1} orange, {\tt T2} green, {\tt T3} red, {\tt T4} purple).
\label{mix_prof_crp}}
\end{figure*}

In Figs.~\ref{mix_prof_crp}(a)-(c) we present mean profiles of $b(y,\tau)$, $a_y(y,\tau)$ and $R_{nn}(y,\tau)$ at $\tau = 5.3$ for the runs {\tt T1}--{\tt T4}. At $\tau=5.3$ the mixing layer evolution is in the quadratic growth regime of the mix-layer evolution. At this time mixing between the fluids is already well developed, and since there is no source term in the $\hat{b}$ equation (Eq.~\eqref{main_b}), $b(y,\tau)$ profiles, which are increasing underpredicted for increasing $C_d$, continue to decay with time. 

Figure ~\ref{mix_prof_crp}(b) shows that the rounded-top or ``dome-like'' shape of $a_y(y,\tau)$ gradually becomes broader as we increase $C_d$ but in so doing, the magnitudes of $a_y(y,\tau)$ away from the center-line, are increasingly underestimated by the LWN model. However, the spread of the profiles is fairly close to the MOBILE predictions. 
The profiles for $R_{nn}(y,\tau)$ (Figs.~\ref{mix_prof_crp}(c)) also broaden as $C_d$ is increased, an their peak value decreases at the center-line. 

\begin{figure*}[h!]
\includegraphics[width=.45\linewidth]{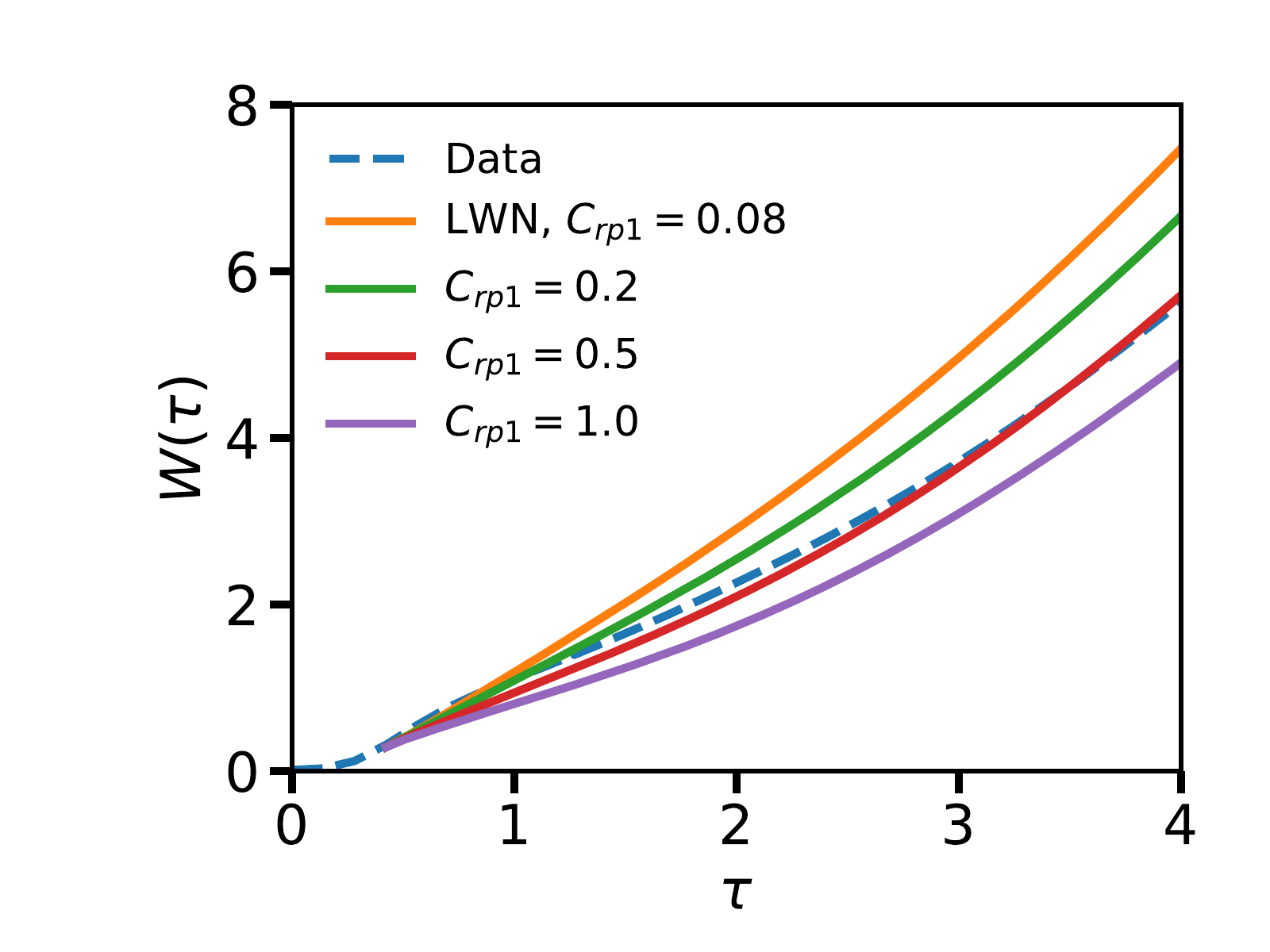}
\put(-30,45){\bf \scriptsize (a)}
\includegraphics[width=.45\linewidth]{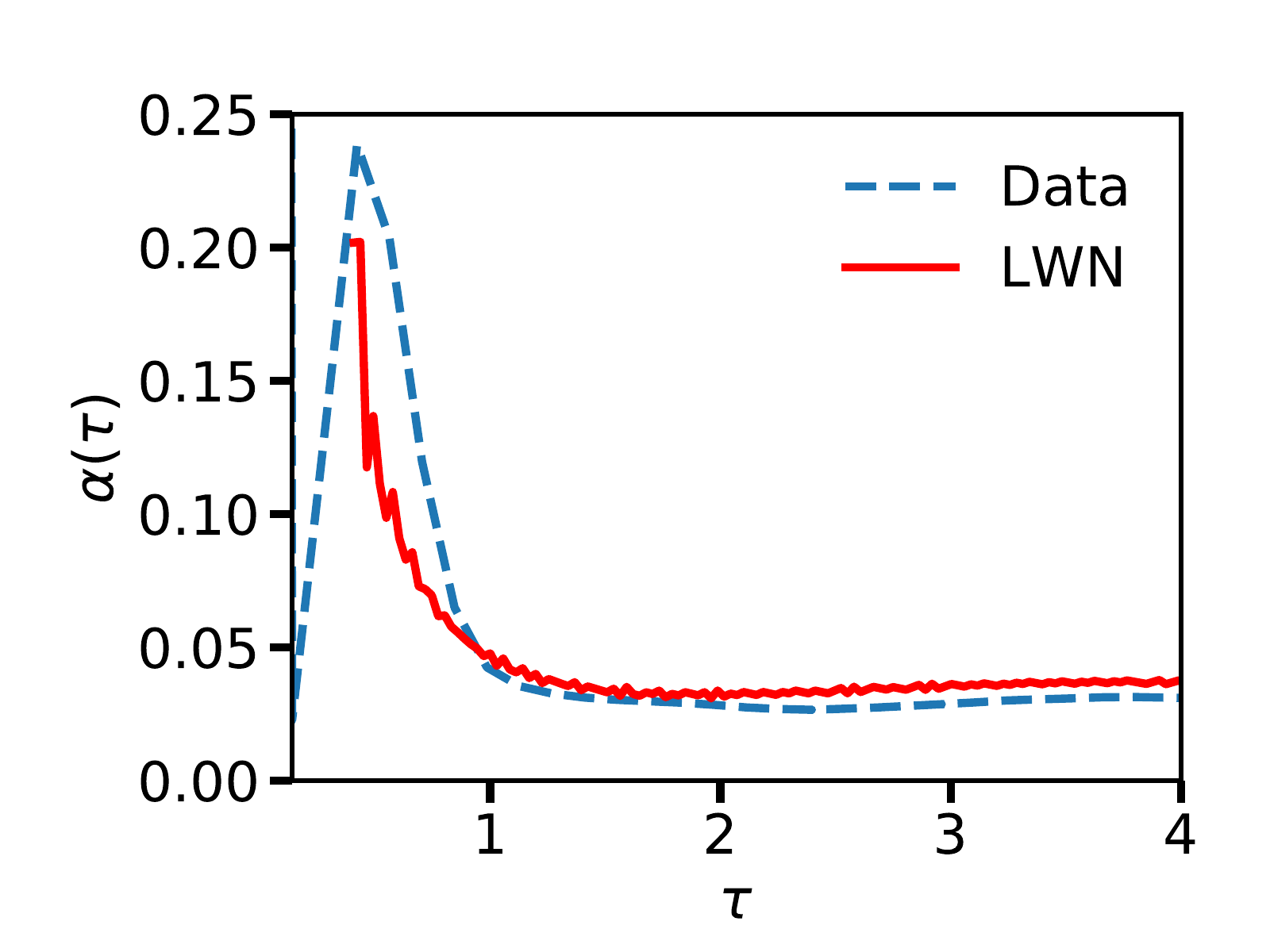}
\put(-30,45){\bf \scriptsize (b)}

\includegraphics[width=.32\linewidth]{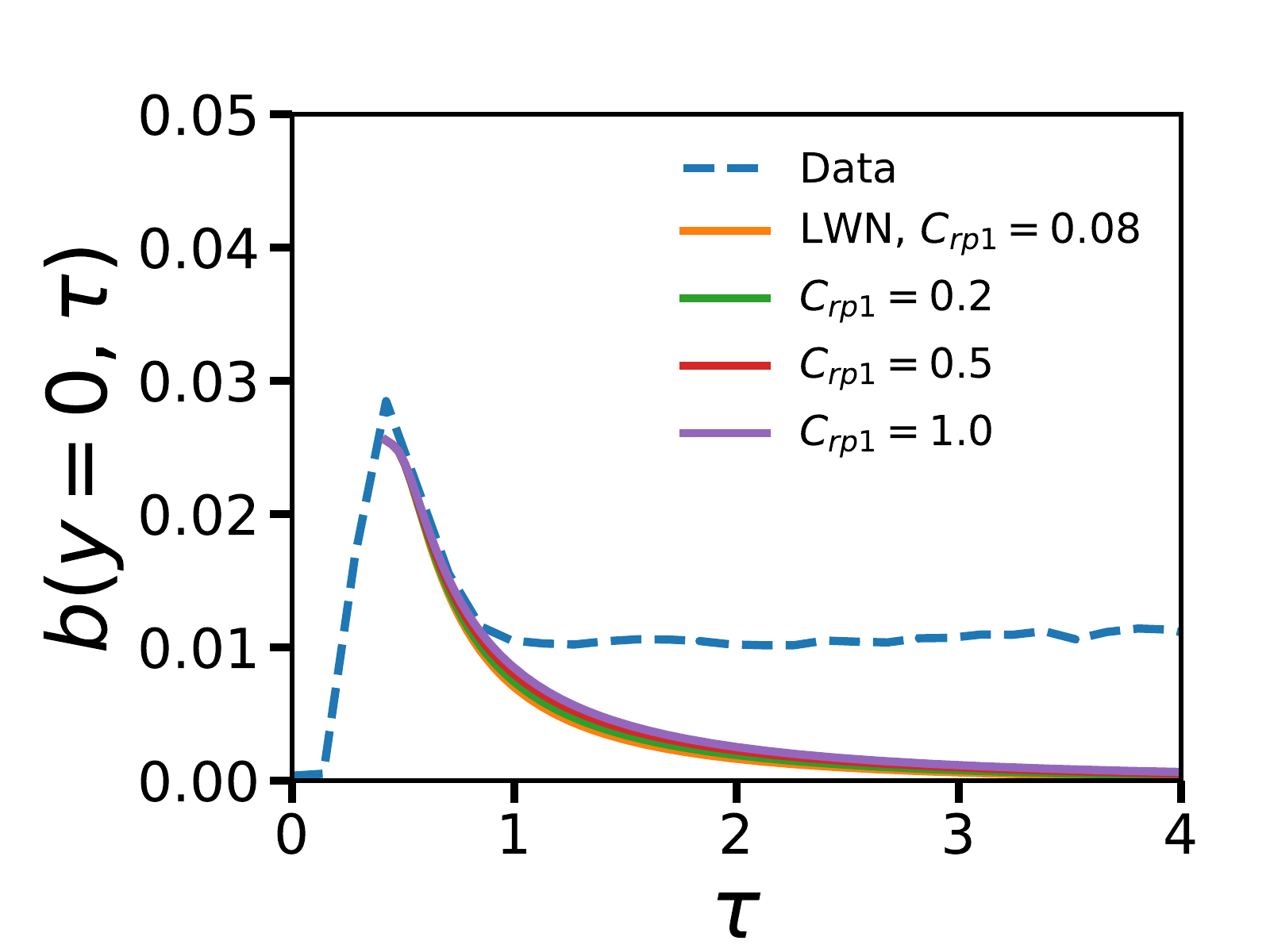}
\put(-115,90){\bf \scriptsize (c)}
\includegraphics[width=.32\linewidth]{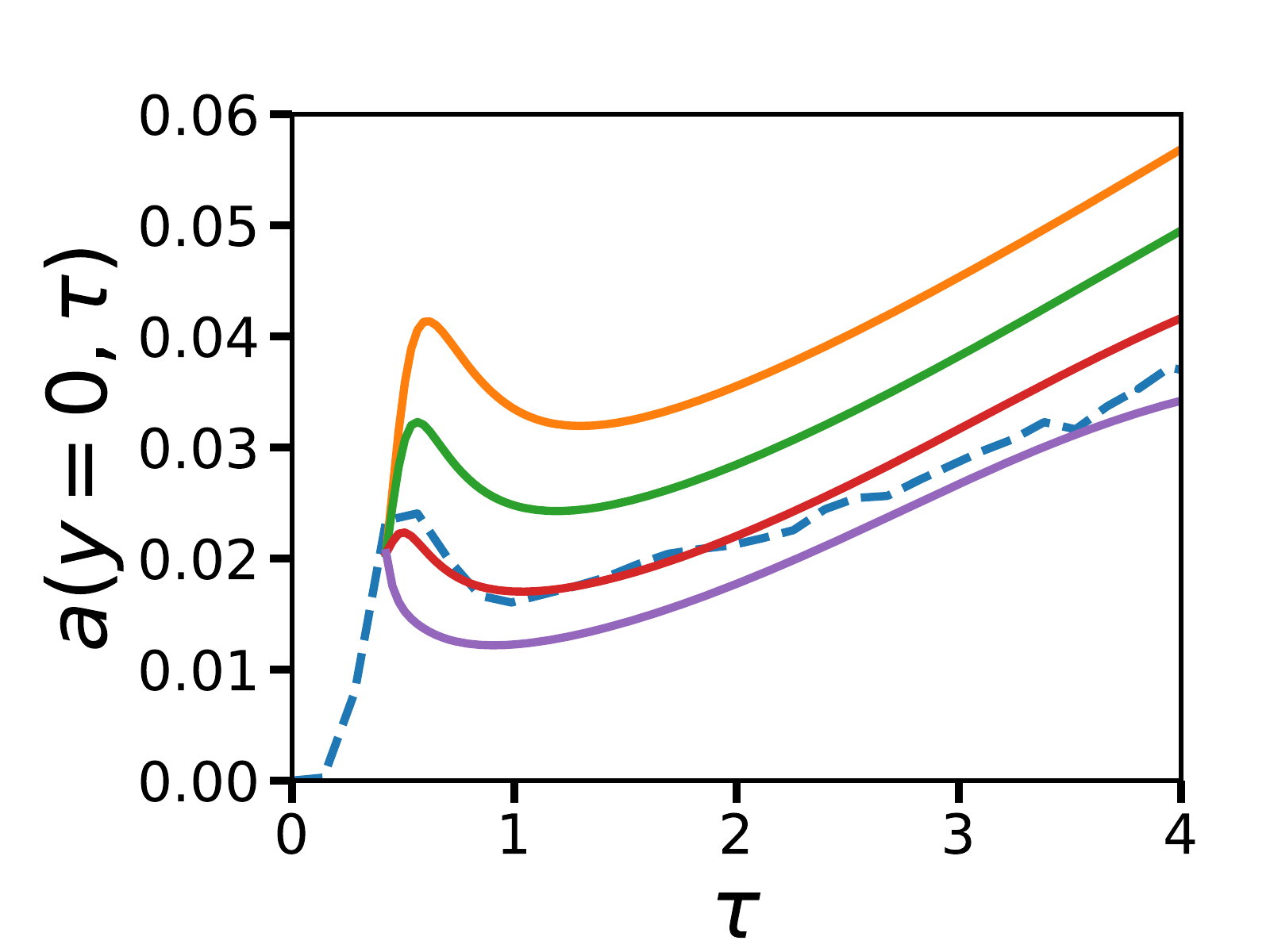}
\put(-115,92){\bf \scriptsize (d)}
\includegraphics[width=.32\linewidth]{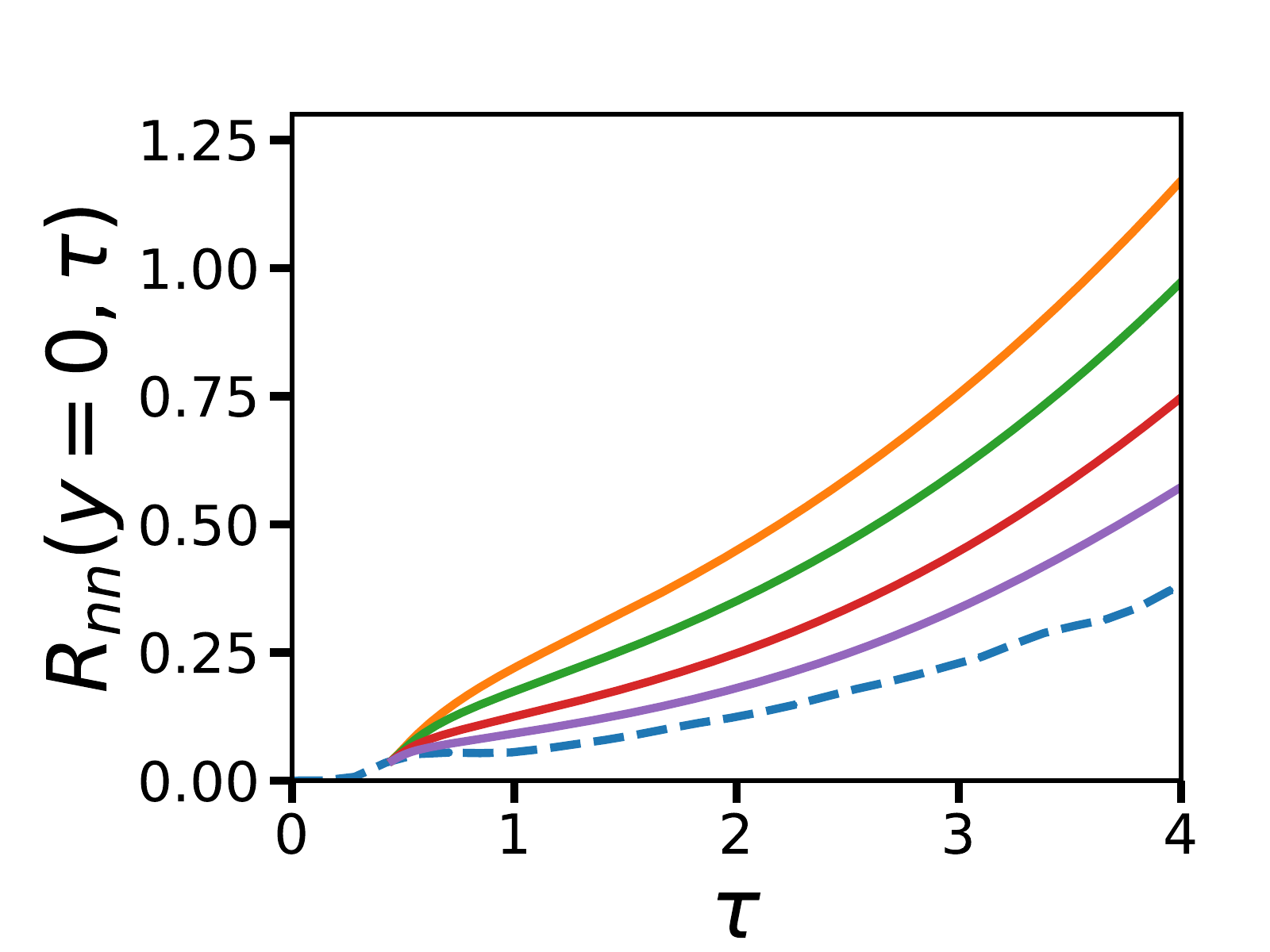}
\put(-115,92){\bf \scriptsize (e)}
\caption{(a) Mix-width (in \blue{${\rm cm}$}) from the MOBILE data ({\tt R1}, blue dashed line) and that from the LWN model calculations ({\tt T1} orange, {\tt T2} green, {\tt T3} red, {\tt T4} purple) (b) comparison of mix--width growth rate $\alpha(\tau)$ for the case $C_{rp1}=0.5$ in (a); (c) $b(y=0,\tau)$; (d) $a_y(y=0,\tau)$ (in \blue{${\rm cm \ s^{-1} }$}); (e) $R_{nn}(y=0,\tau)$ (in \blue{${\rm g \ cm^{-1} \ s^{-2}}$}).}
\label{mix_final_comp_high}
\end{figure*}
Based on the studies so far, $C_d=0.5$ yields the optimum agreement between LWN and the data with respect to the mix-layer width and individual profiles of $a_y(y,\tau)$ and $R_{nn}(y,\tau)$. We carry out further investigations by fixing this value of $C_d$ and jointly varying $C_{rp1}$ and $C_{rp2}$, ({\tt T5}-{\tt T7} in Table \ref{table2}).
We see in Fig.~\ref{mix_final_comp_high}(a) that as we increase $C_{rp1}$ and $C_{rp2}$, the rate of growth of mix-layer becomes slower eventually 
under-predicting the growth relative to the data for $C_{rp1} > 0.5$.
In Fig.\ref{mix_final_comp_high}(b), we report a comparison of the growth rate $\alpha$ calculated according to~\cite{cabot2006reynolds},  $\alpha=\frac{\dot{W}^2}{4AgW}$, where $W$ is the width of the mixing layer. $\alpha$ for the LWN run is calculated for the coefficient set {\tt T7} (see Table~\ref{table2}) with $C_{rp1}=0.5$.  
In Fig.~\ref{mix_final_comp_high}(d) we compare the evolution of $a_y(y=0,\tau)$ for the different values of $C_{rp1}$ and $C_{rp2}$. As expected, the growth of $a_y(y=0,\tau)$ is slower as the drag coefficients $C_{rp1}$ and $C_{rp2}$ are increased. Since $\hat{a}_y(y,k)$ provides the principal driving force to $\hat{R}_{nn}(y,k)$ through the pressure-gradient term, $\hat{R}_{nn}(y,k)$ growth rate also decreases as $C_{rp1}$ and $C_{rp2}$ are increased (Fig.~\ref{mix_final_comp_high}(e)). However, slower growth of $\hat{R}_{nn}(y,k)$ increases the turbulence timescale $\Theta$ and thus $\hat{b}(y,k)$ decays slower as $C_{rp1}$ and $C_{rp2}$ are increased (Fig.~\ref{mix_final_comp_high}(c)). 

\begin{figure*}[h!]
\includegraphics[width=.32\linewidth]{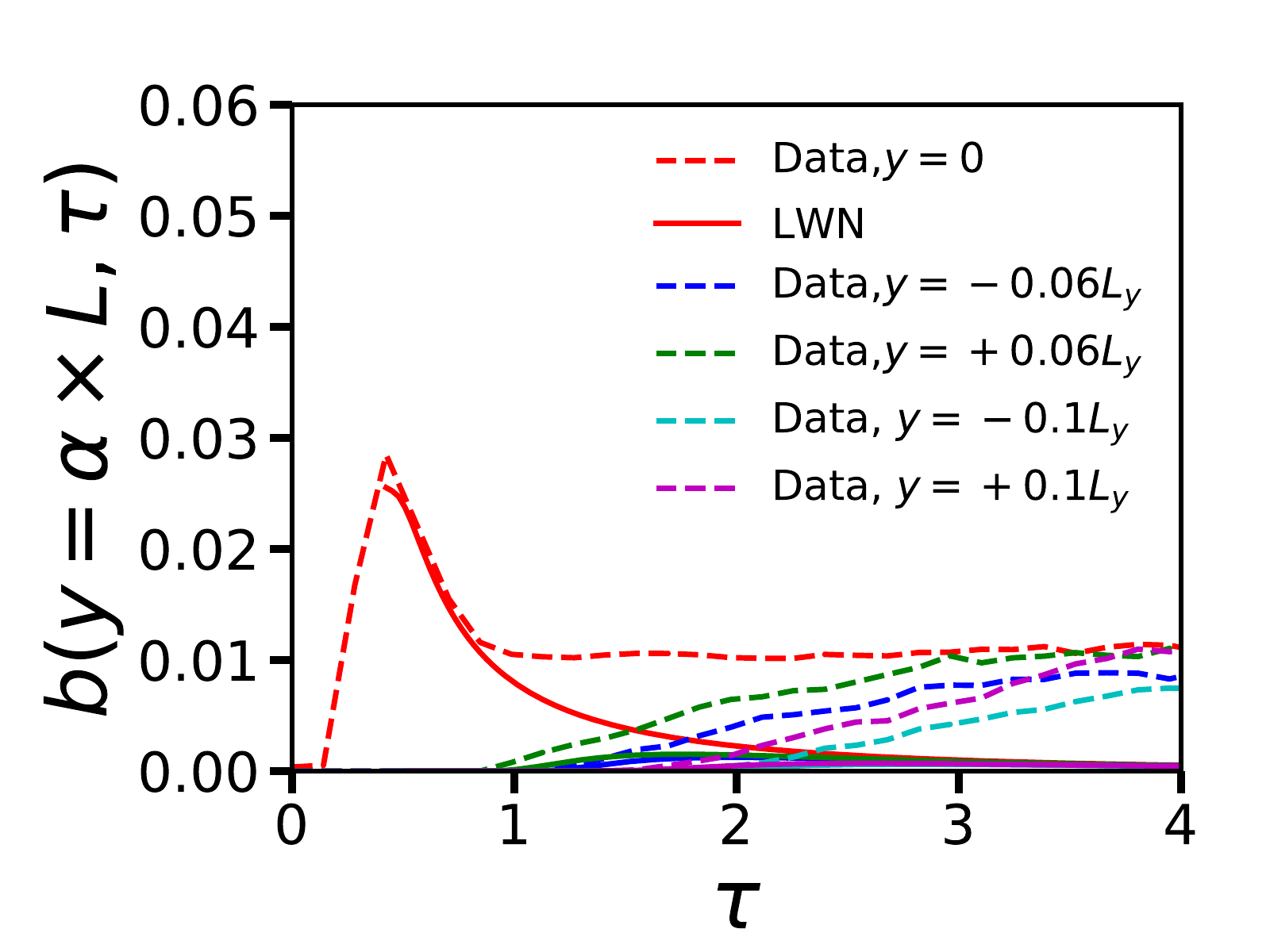}
\put(-115,90){\bf \scriptsize (a)}
\includegraphics[width=.32\linewidth]{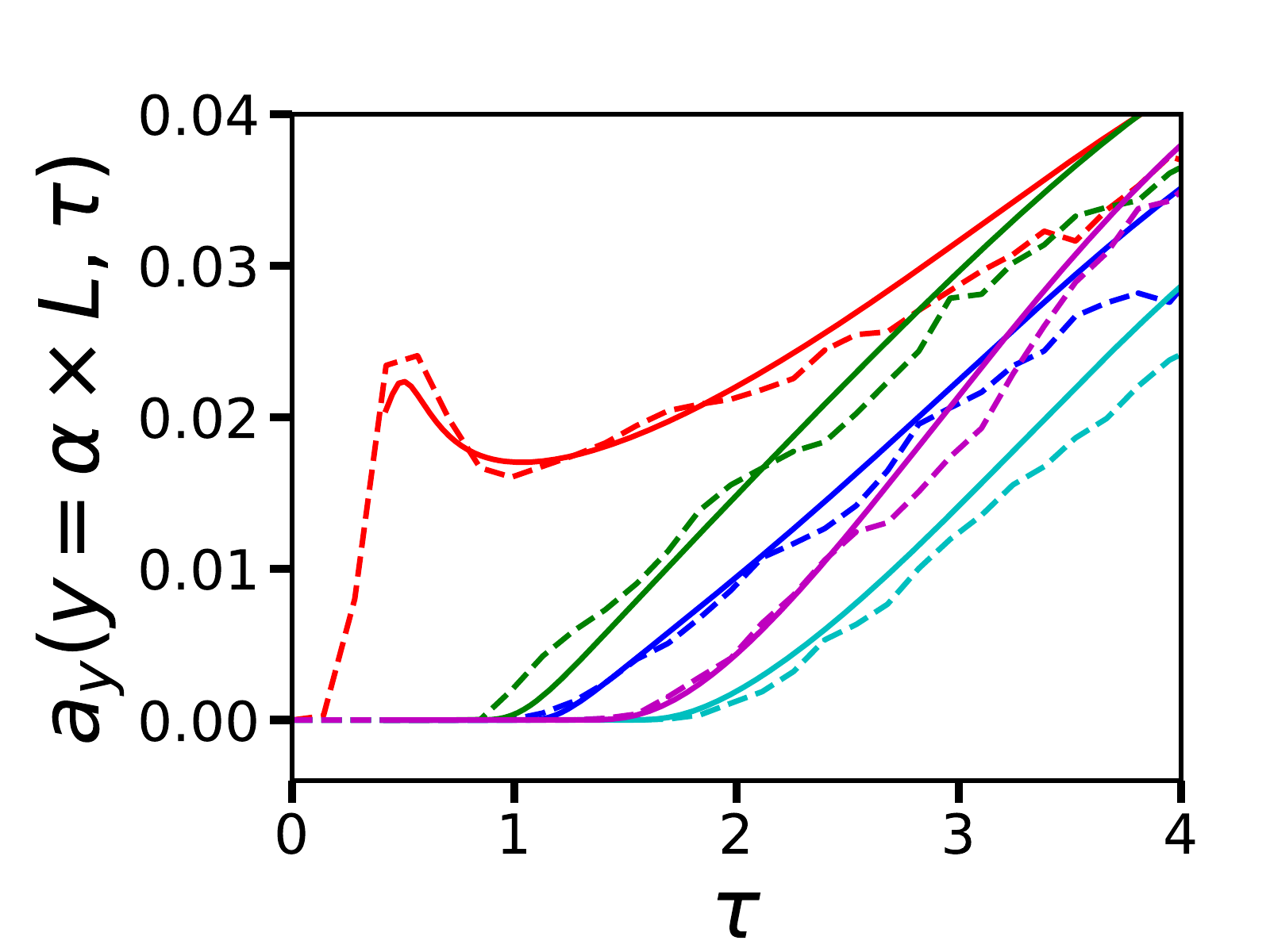}
\put(-115,92){\bf \scriptsize (b)}
\includegraphics[width=.32\linewidth]{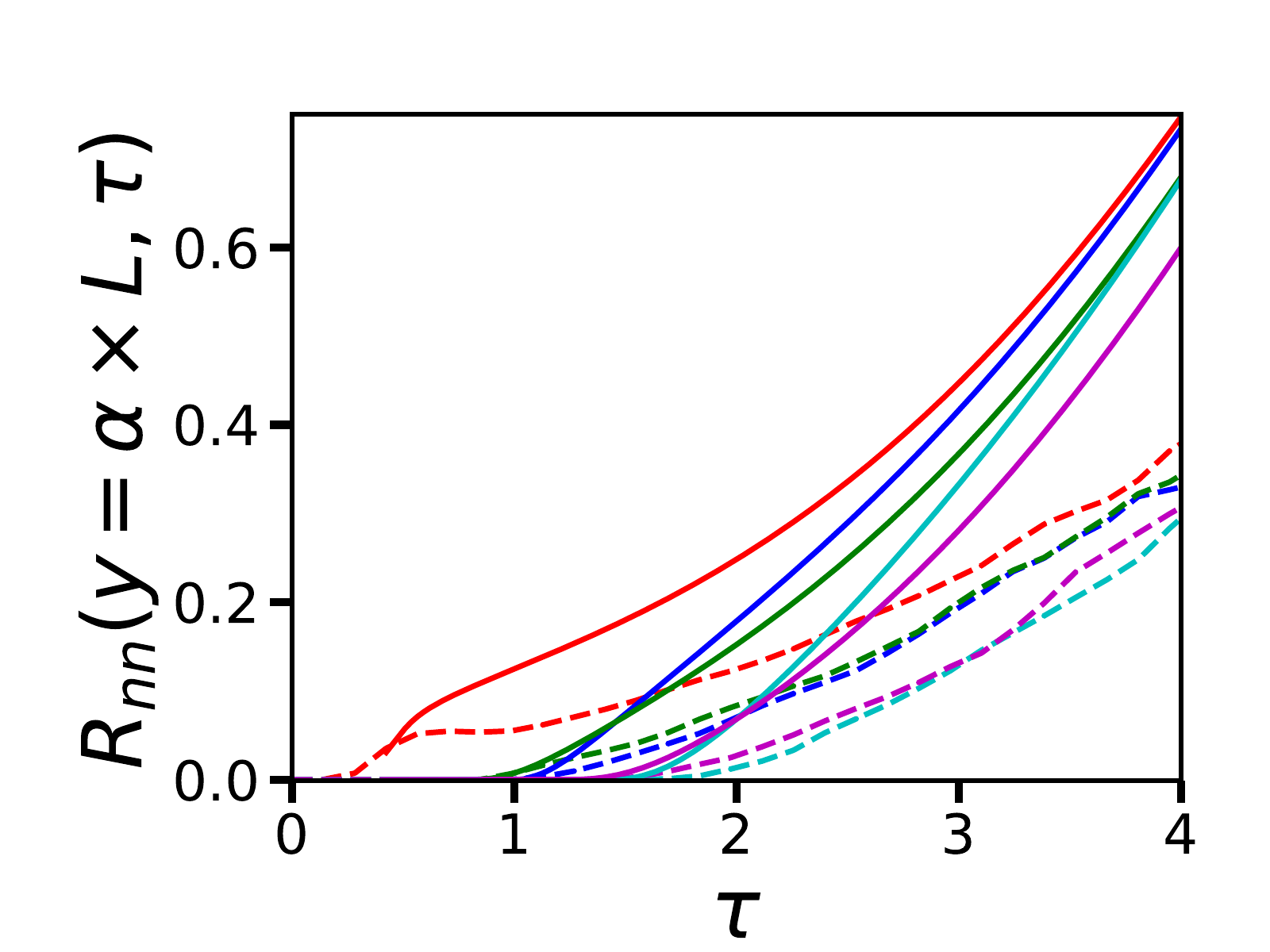}
\put(-115,92){\bf \scriptsize (c)}
\caption{
Results of comparison between MOBILE calculation ({\tt R1}, dashed lines) and that from the LWN model calculation ({\tt T7}, solid lines) at center-plane, and at distances of 6\%$L_y$ and 10\%$L_y$ from the center. Plots of (a) $b(y,\tau)$ (b) $a_y(y,\tau)$ (in \blue{${\rm cm \ s^{-1} }$}) and (c)$R_{nn}(y,\tau)$ (in \blue{${\rm g \ cm^{-1} \ s^{-2}}$}).}
\label{mix_fin}
\end{figure*}
We thus arrive at a reasonable `best' (though not rigorously optimized) set of coefficients  $C_d=0.5$, and $C_{rp1},C_{rp2}=0.5$ with run T7 in Table~\ref{table2}. With this choice we show that indeed the mixing evolution away from the center-line is also captured quite well. Figures ~\ref{mix_fin}(a), (b), and (c) we show, respectively, the evolution with time of $b$, $a_y$, and $R_{nn}$ at various horizontal planes in the domain, $y=0$, $y=\pm 0.06 L_y$ and $y=\pm 0.1L_y$. In particular $a_y(y=0,\tau)$ agrees very well with the simulation. Both $R_{yy}$ (not shown) and $R_{nn}$ growths are somewhat overpredicted, although predictions of $R_{yy}$ fare better than $R_{nn}$. The LWN underpredicts $b(y=0,\tau)$ across each of these planes, presumably due to the lack of a source term, which motivates the discussion in Section \ref{source_term}.

\begin{figure*}
\includegraphics[width=.45\linewidth]{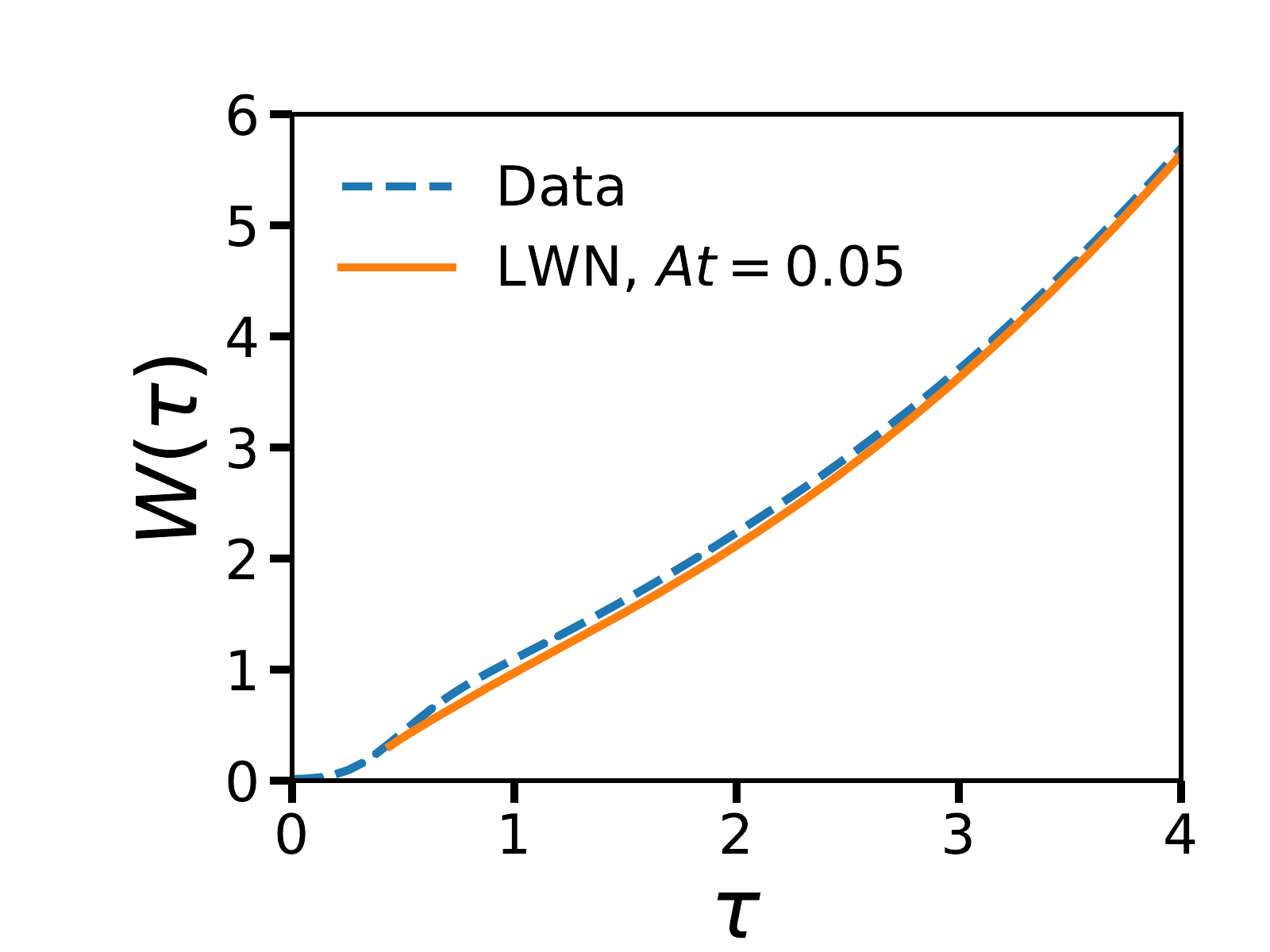}
\put(-40,40){\bf \scriptsize (a)}\\
\includegraphics[width=.32\linewidth]{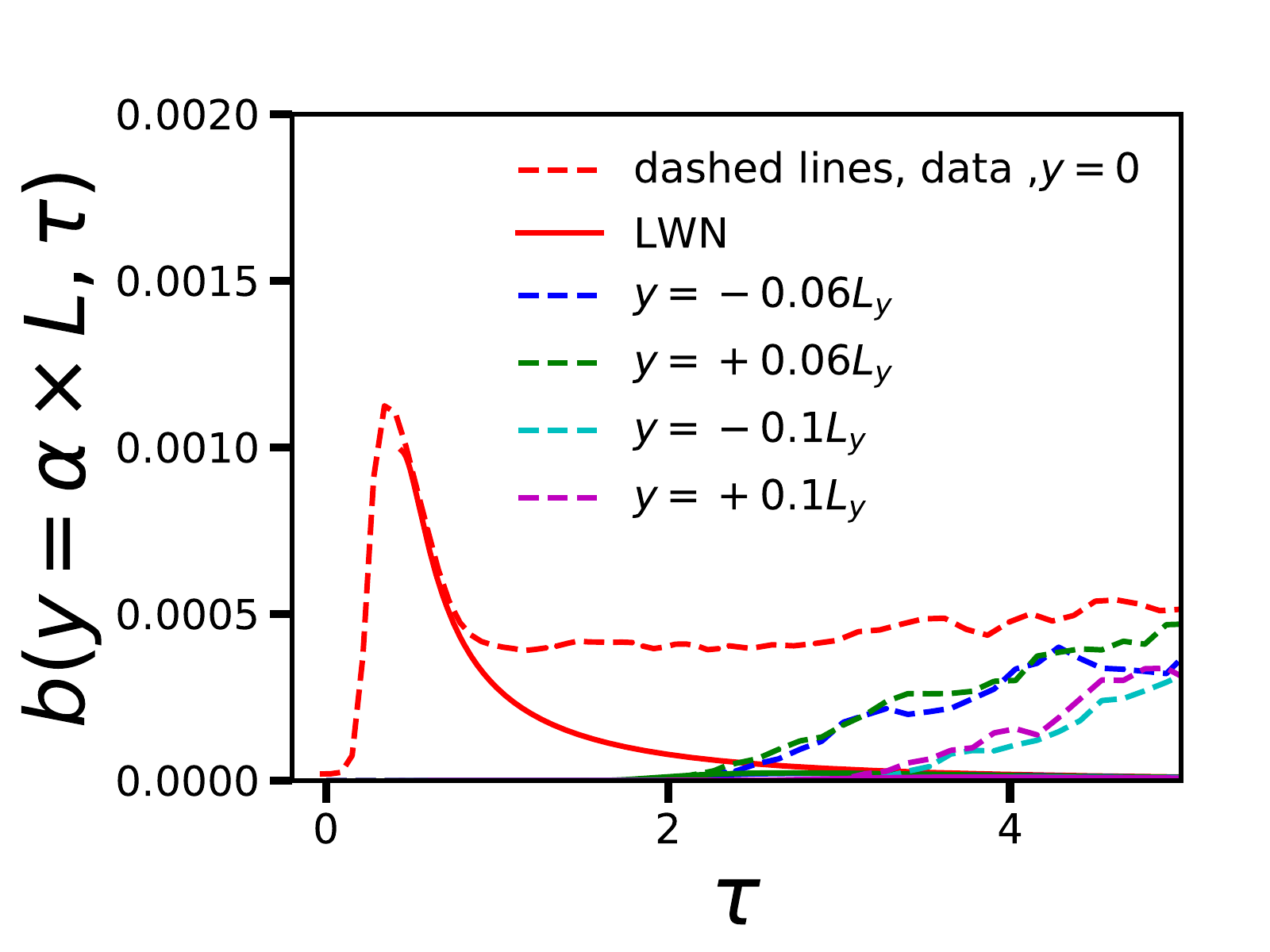}
\put(-120,100){\bf \scriptsize (b)}
\includegraphics[width=.32\linewidth]{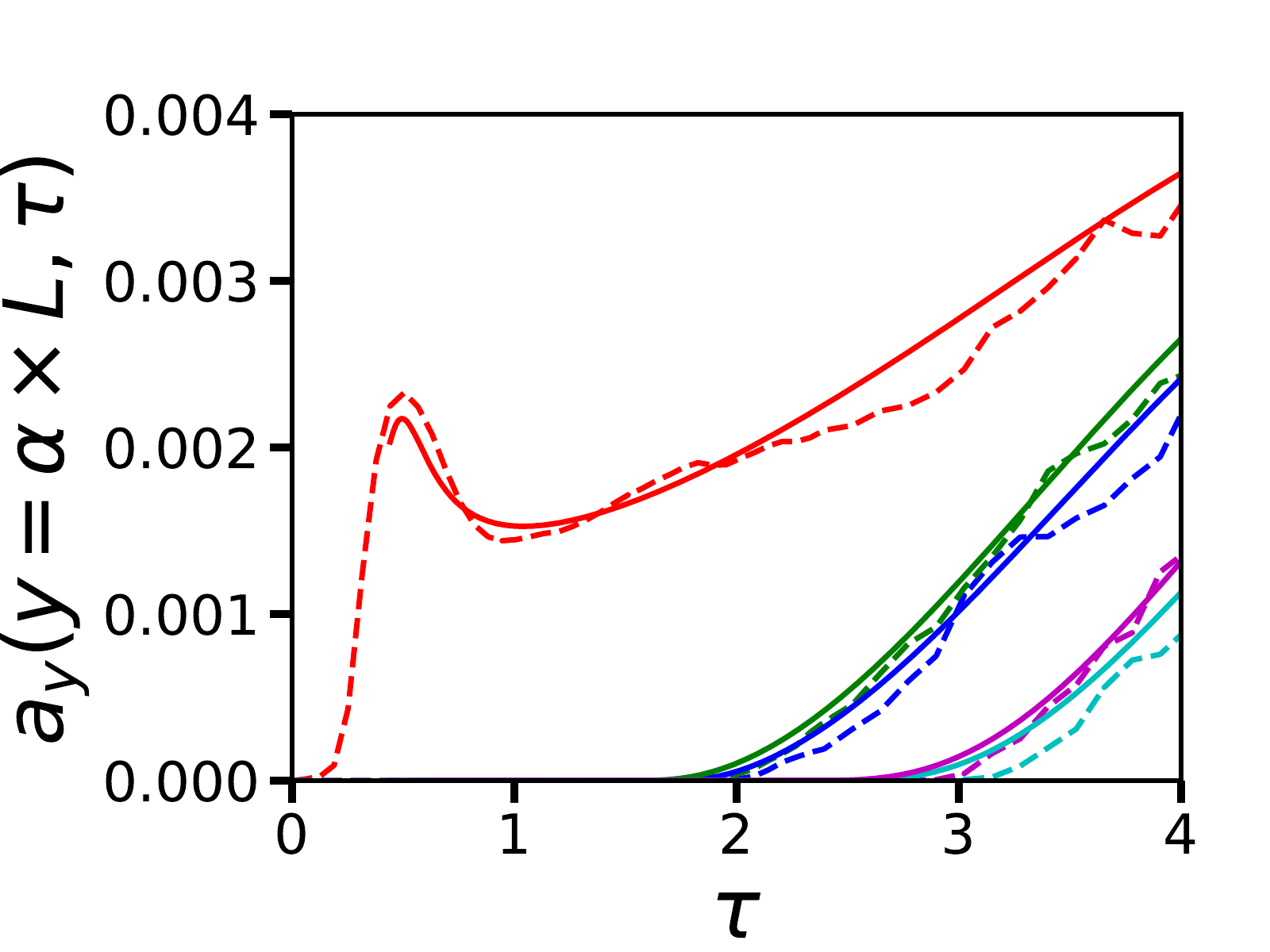}
\put(-120,100){\bf \scriptsize (c)}
\includegraphics[width=.32\linewidth]{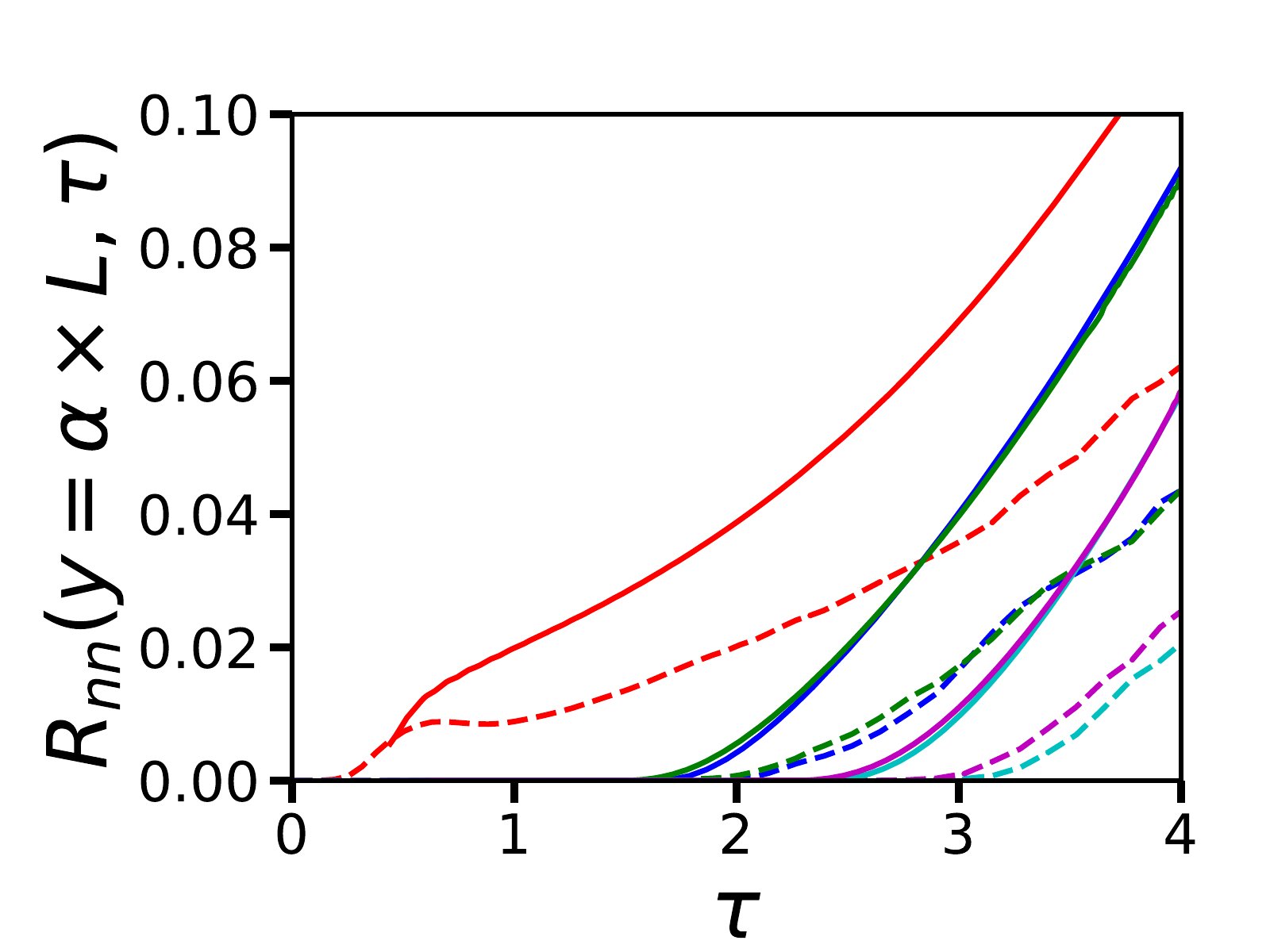}
\put(-120,100){\bf \scriptsize (d)}
\caption{Results of comparison between MOBILE data {\tt R3} (dashed line) and  LWN calculations (T7, solid lines) (a)Mix-width (in \blue{${\rm cm}$}) comparison; results at center-plan and at distances of 6\%$L_y$ and 10\%$L_y$ from the center of (b) $b(y,\tau)$ (c) $a_y(y,\tau)$ (in \blue{${\rm cm \ s^{-1} }$}) and (d) $R_{nn}(y,\tau)$ (in \blue{${\rm g \ cm^{-1} \ s^{-2}}$}).}
\label{mix_at5d-2}
\end{figure*}

With this set of coefficients roughly optimized for $A=0.25$, we carried out further calculations with the LWN model at lower values of Atwood number 0.1 and 0.05 to assess the dependence of our choices on $A$, if any. Both calculations were initialized at same non-dimensional time $\tau_0=0.42$ of the data. Figure ~\ref{mix_at5d-2} shows that the model performs just as well for $A=0.05$ flows as it does for the $A=0.25$ with no further re-tuning. The same is true for $A=0.1$ (results not shown). This is reassuring since it says that, at least at low to moderate $A$ the model does not require any change in coefficients to operate. It is also consistent with the fact that the model does not have $A$-dependent assumptions built into it.


\section{Growth and saturation of $b$ in the LWN model \label{source_term}}

\begin{figure*}
\includegraphics[width=.32\linewidth]{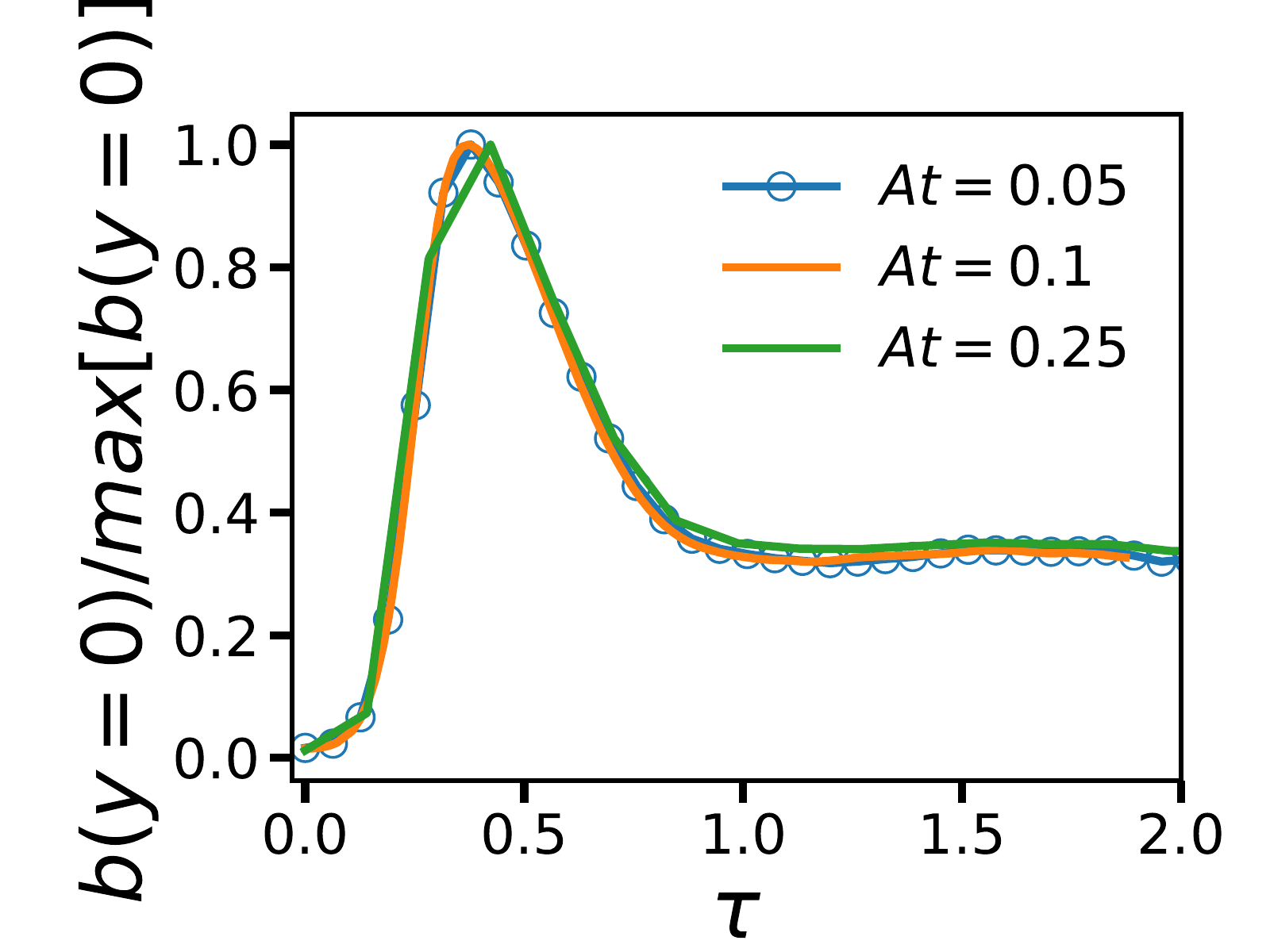}
\put(-120,90){\bf \scriptsize (a)}
\includegraphics[width=.32\linewidth]{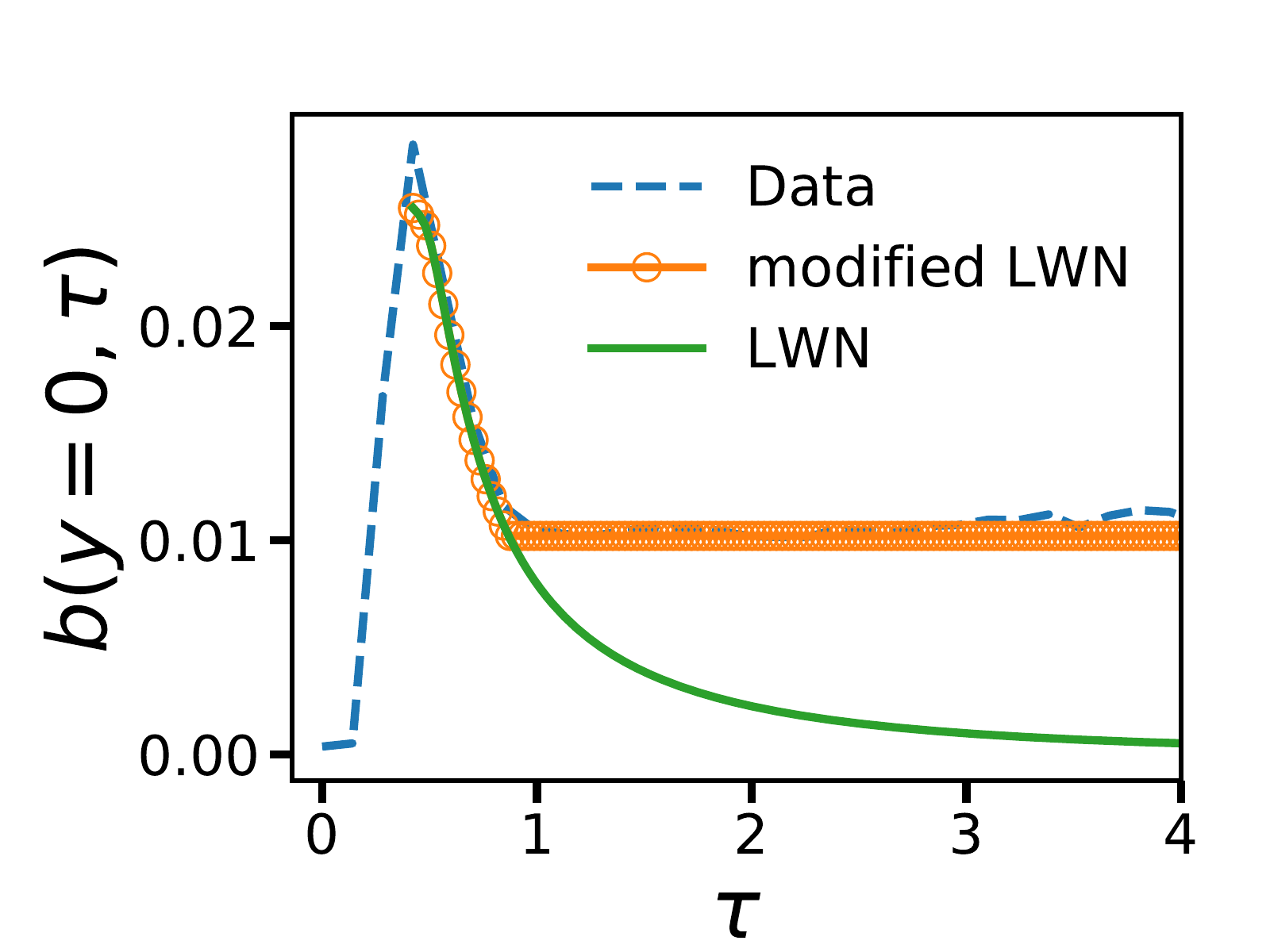}
\put(-130,90){\bf \scriptsize (b)}
\includegraphics[width=.32\linewidth]{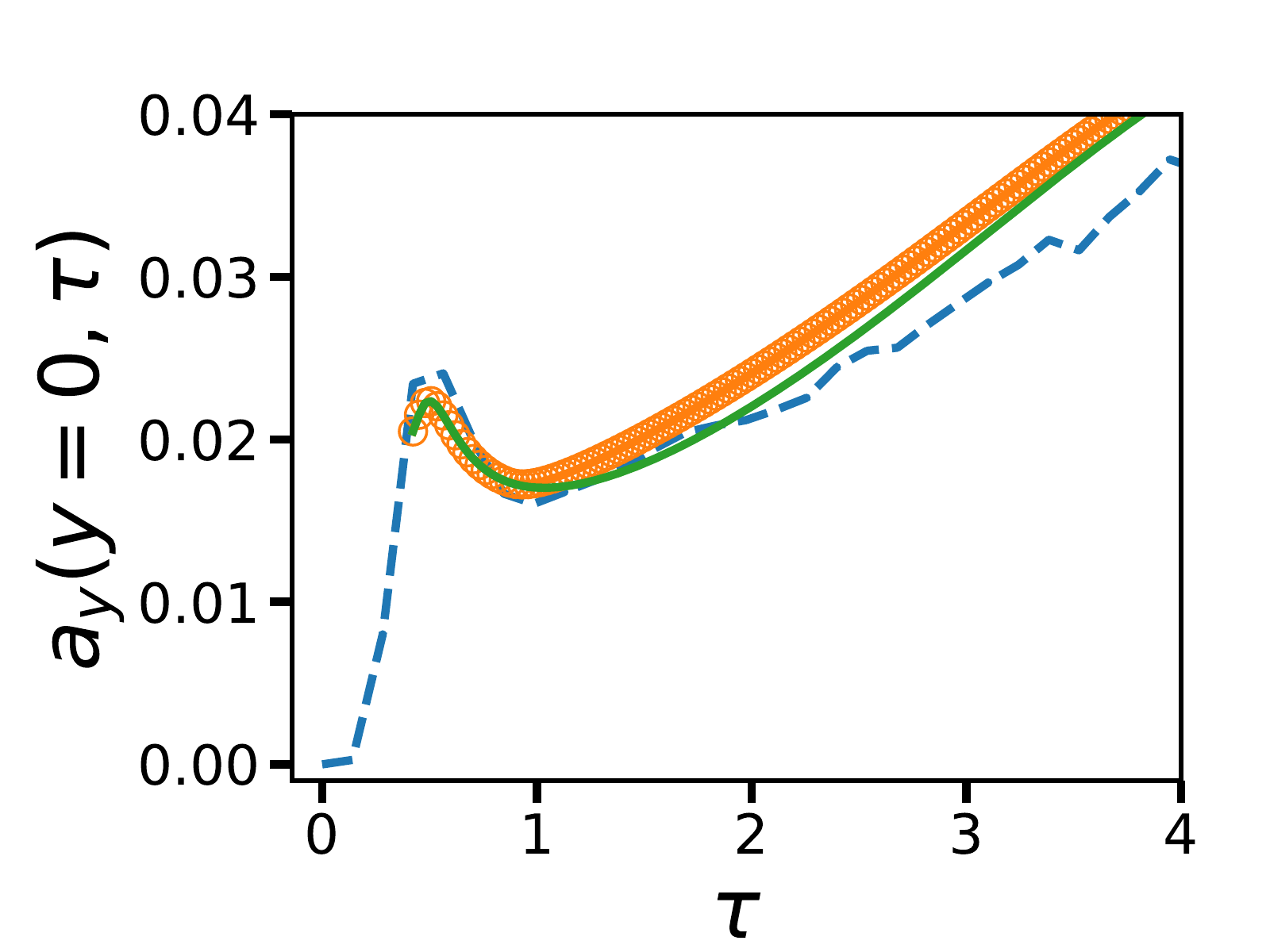}
\put(-116,90){\bf \scriptsize (c)}
\caption{(a) Plots of $\df{b(y=0)}{max[b(y=0)]}$ for different Atwood
  numbers from MOBILE data {\tt R1, R2, R3}. (b)Comparison of
  $b(y=0,\tau)$ among results from MOBILE data {\tt R1} (blue dashed
  line), modified LWN (orange line with circles), and
  LWN (green line).(b) Comparison of $a_y(y=0,\tau)$ (in \blue{${\rm cm \ s^{-1} }$}) for the
  same runs.}
\label{steady_b}
\end{figure*}
We have noted that the LWN model does not predict the correct magnitude of the $b(y,\tau)$ profile for $\tau > 1$, and we attribute the decay of $b(y,\tau)$ to the omission of a source term in equation (Eq.~\eqref{main_b}) for $b(y,\tau)$. Previous efforts on RT using spectral turbulence models~\cite{Steinkamp1999a, Steinkamp1999b,besnard1996spectral,ristorcelli2004rayleigh, cabot2006reynolds,morgan2017self}
stress the importance of maintaining the centerline $b$, i.e., $b(y=0,\tau)$ at a constant value. Indeed
\cite{Steinkamp1999a,Steinkamp1999b} introduced a kinematic source term in an {\textit {ad hoc}} manner to maintain $b(y=0,\tau)$ (see~\cite{Steinkamp1999a}). With access to detailed simulations we can solidify this notion further. In Fig. \ref{steady_b}(a) we plot the time-evolution of $\df{b(y=0)}{max[b(y=0)]}$ as obtained from MOBILE simulations and we find independence of $b(y=0)$ across the Atwood number range studied. Furthermore, simulations show that $b(y=0,\tau)$ saturates at $\df{b(y=0)}{max[b(y=0)]}\sim \df{1}{3}$. Motivated by this observation, and as a first attempt to reduce model errors in $b(y=0,\tau)$, we modified the LWN model to keep $b(y=0,\tau)$ constant at its final steady-state value observed in (Fig.~\ref{steady_b}(b)). This modified model shows much improved predictions for $b(y=\pm 0.06 L_y,\tau)$ and $b(y=\pm 0.1L_y,\tau)$, especially compared with Fig.~\ref{mix_final_comp_high}(b). This is consistent with the new data-motivated ``source'' term that we now hold constant at the centerline, and over time the $C_d$-weighted diffusion spreads the signal outwards.

With the above modifications to the model, other quantities such as $a_y(y=0,\tau)$ (see Fig.~\ref{steady_b}(c)) and
$R_{nn}(y=0,\tau)$ (not shown here), both at the centerline and outward from it, change very little. This approach is not entirely satisfactory from the point of view of the evolution equations. 

Motivated by single-point studies of variable-density RT~\cite{denissen2012implementation,bertsch2015rans,israel2009validating} we can improve our approach further by modifying the $\hat{b}(y,k,t)$ equation as follows:
\begin{widetext}
\begin{eqnarray}
\displaystyle\frac{\partial \hat{b}(y,k,t)}{\partial t}&=&-\df{2(b(y)+1)}{\overline{\rho}}\hat{a}_y\df{\partial \overline{\rho}}{\partial y}+\frac{\partial}{\partial k}\left[k\Theta^{-1}\left[-C_{b1}\hat{b}+C_{b2}k\frac{\partial \hat{b}}{\partial k}\right]\right]+C_d\frac{\partial }{\partial y}\left(\upsilon_t\frac{\partial \hat{b}}{\partial y}\right)
\label{app2}
\end{eqnarray}
\end{widetext}
where the first term on the right-hand side is the spectral extension of the form of the source term used in single-point studies~\cite{banerjee2010development, denissen2012implementation,bertsch2015rans,israel2009validating}, with $\hat{a}_y = \hat{a}_y(k,y,t)=\hat{a}_y(k,y,\tau)$ and $b(y)$ is the integrated value of $\hat{b}$ at $y$. \blue{Again, the explicit arguments are dropped for brevity}.
\begin{figure}
\includegraphics[width=.8\linewidth]{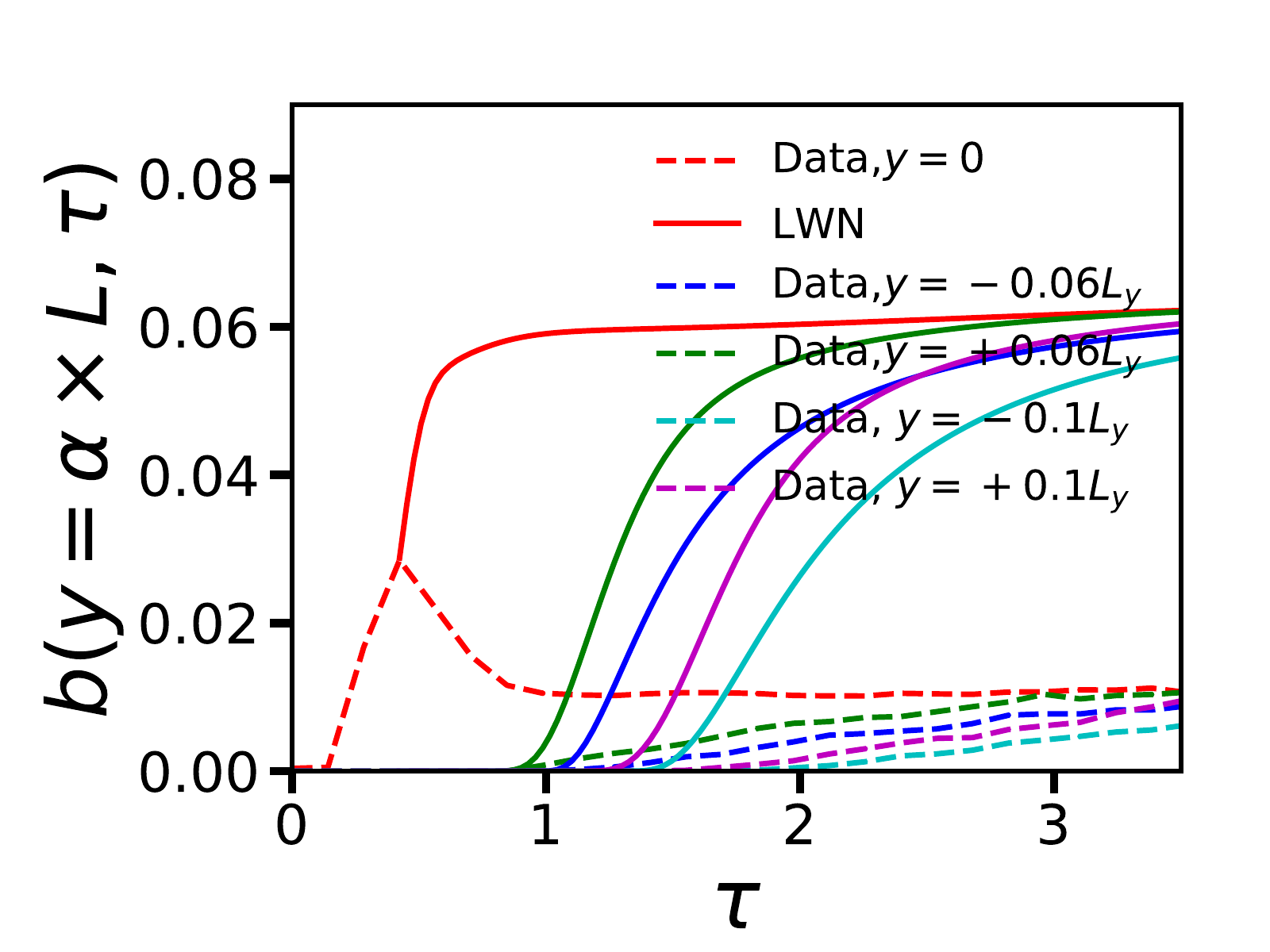}
\put(-150,120){\bf \scriptsize (a)}
\\
\includegraphics[width=.8\linewidth]{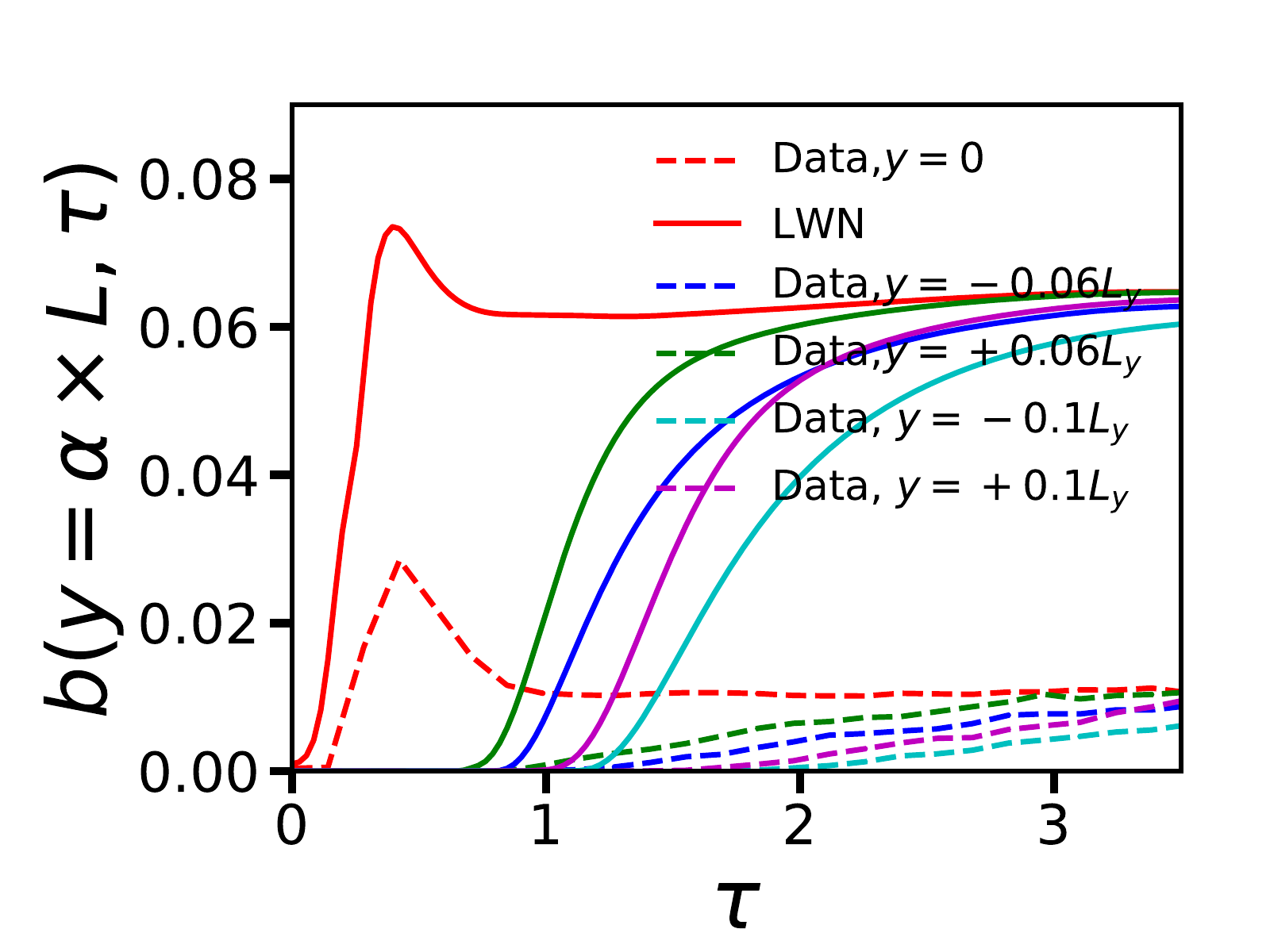}
\put(-150,120){\bf \scriptsize (b)}
\caption{The evolution of $b(y,\tau)$ at the the center-plane and at two distances away from the centerplane, computed using a source term as in Eq. (\ref{app2}), for two different start times, 
(a) $\tau = 0.42$ (run {\tt PT1, Table~\ref{table3}}) and (b) $\tau = 0.0$ (run {\tt PT2}, Table~\ref{table3}). 
}
\label{lwnb}
\end{figure}

\begin{figure}
\includegraphics[width=.8\linewidth]{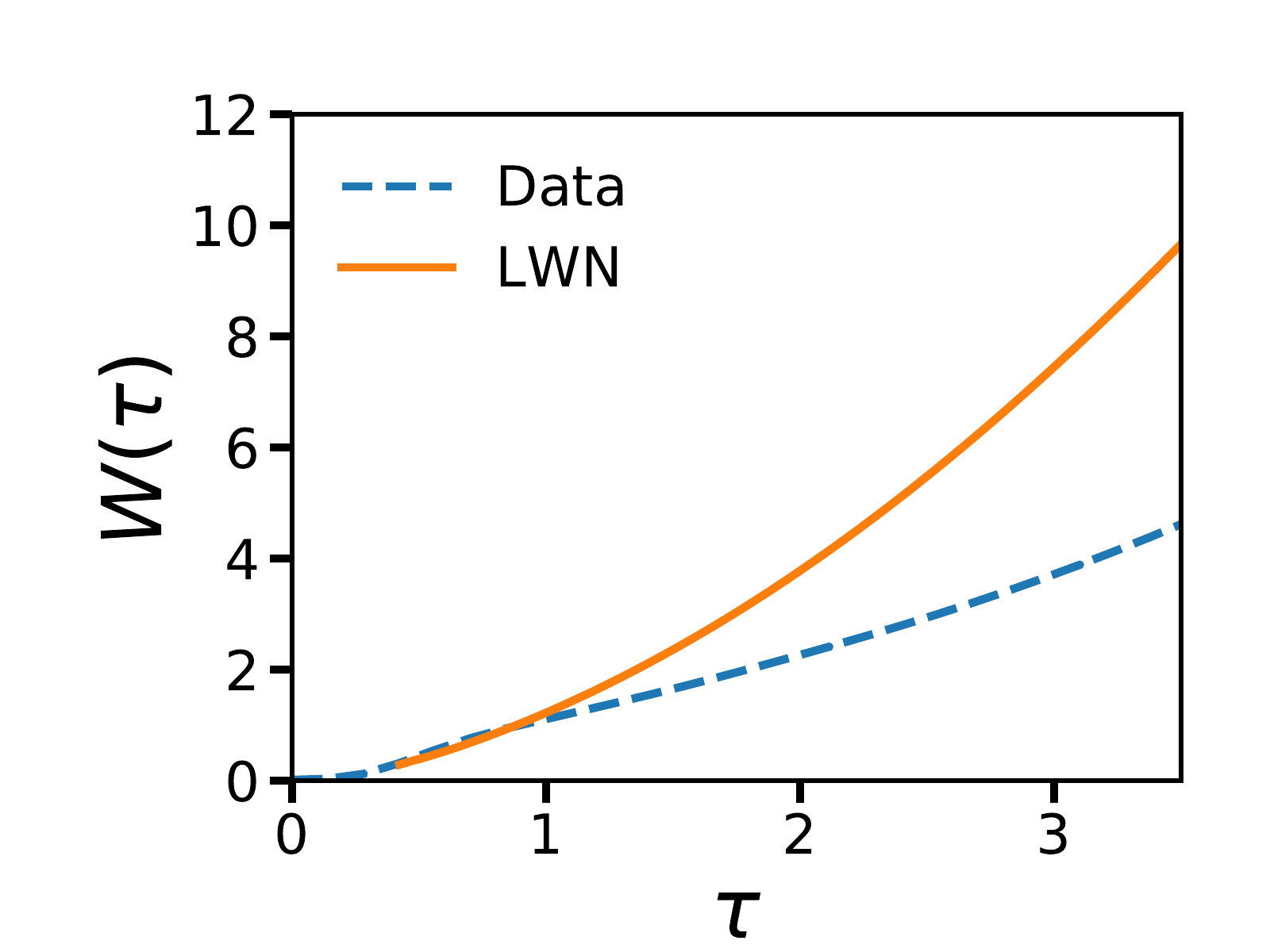}
\put(-30,120){\bf \scriptsize (a)}
\\
\includegraphics[width=.8\linewidth]{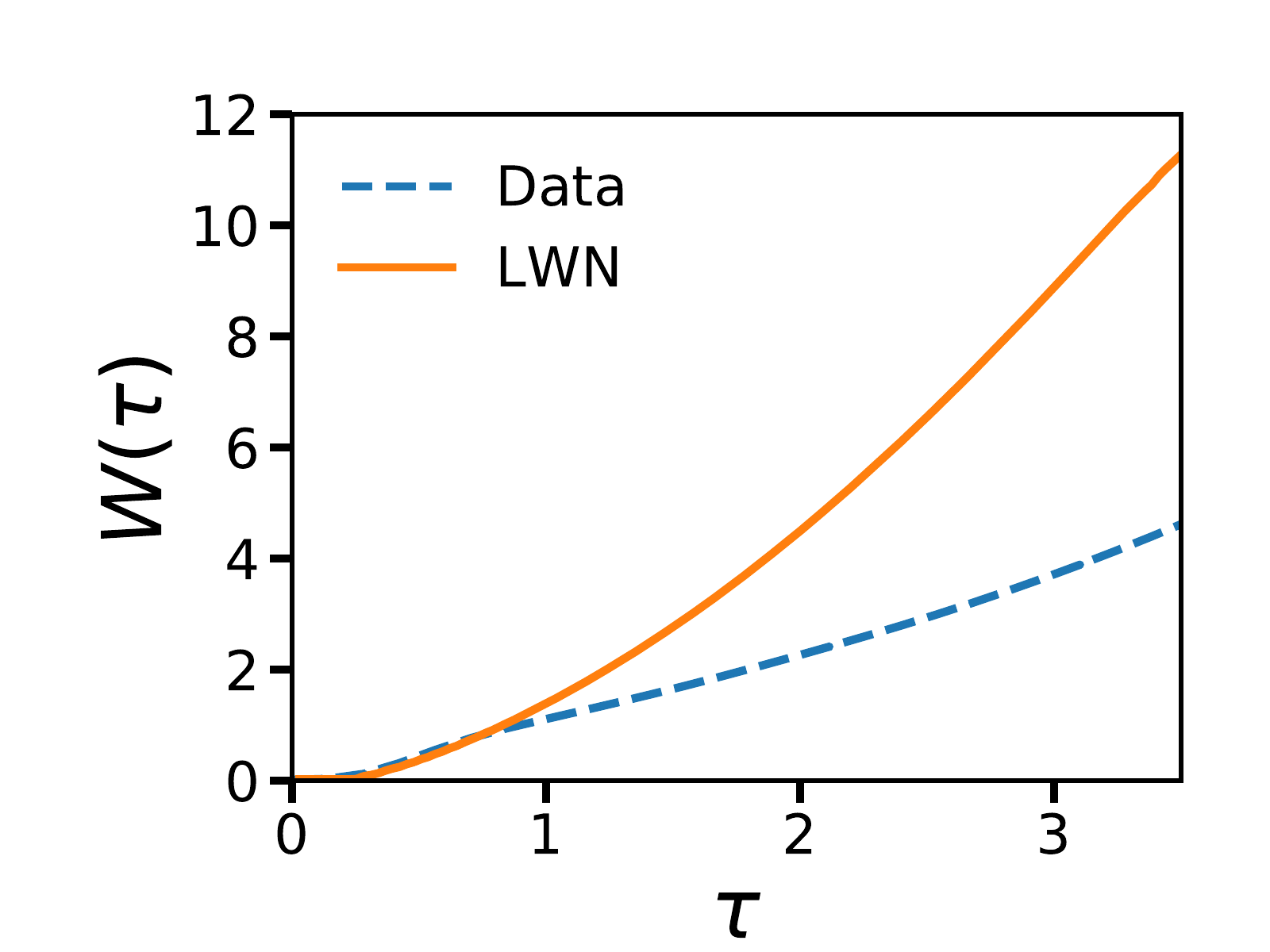}
\put(-30,120){\bf \scriptsize (b)}
\caption{The evolution of mix--layer (in \blue{${\rm cm}$}), for two different start times, (a) $\tau = 0.42$ (run {\tt PT1}, Table~\ref{table3}) and (b) $\tau = 0.0$ (run {\tt PT2}, Table~\ref{table3}). 
}
\label{lwnb_mix}
\end{figure}

With coefficients unchanged, the results of incorporating this source term on $\hat{b}$ is shown in the top two panels of Fig. \ref{lwnb} for two different start times (initial conditions). The first start time is at $\tau = 0.42$ as before and the second is at $\tau = 0$, in anticipation of better transition capture with a source term. The first thing to note is that both cases show initial growth of $b$ followed by an asymptote to roughly $0.06$. Model calculations corresponding to lower $A$ (\texttt{R2}, \texttt{R3}s) were also observed to saturate to correspondingly lower $b(y=0,\tau)$. In all cases the final value of the $b(y=0,\tau)$ was the same for a given flow irrespective of initial (start) time, but larger than for the MOBILE simulation. The mix-width evolution with the source term included is showin in Fig.~\ref{lwnb_mix} for both start times. The mix--layer as predicted by the LWN model is  overpredicted in both cases although the expected $~t^2$ behavior is recovered.  \blue{As we show below, it is possible to get a better agreement for the mix--width by re-tuning of $C_d$, $C_{rp1}$, $C_{rp2}$ and $C_{b2}$}.

\blue{
The saturation value of $b(y=0,\tau)$ achieved by the model is roughly $6$ times the value of $b(y=0,\tau)$ achieved by the MOBILE calculation. {\sout{As discussed in \cite{youngs2009application} 
there are two mechanisms that decide the saturation value of $b$. The first being the entrainment of immiscible unmixed fluid into the mix layer and the second being the molecular diffusion of the two fluids into each other. The model only captures the former, we do not have a molecular diffusion mechanism in place. The MOBILE simulation results are not purely immiscible due to numerical diffusion. Therefore it is not surprising that the final value of $b$ is overpredicted by the model. The impact on other quantities of interest such as mix-width (see Fig. \ref{lwnb_mix}), $a_y$ and $R_{nn}$, of the overprediction in the final $b$ is therefore not unexpected.}}
When the spatial diffusion coefficient $C_d$ is small, there is a tendency of $b(y=0,\tau)$ to relax to the ``configurational'' or the no--mix value of $b$.  The configurational or ``no--mix'' value of $b(y=0,\tau)$ is defined as~\cite{steinkamp1996spectral,Steinkamp1999b}
    $b=\df{\alpha_1\alpha_2(\rho_1-\rho_2)^2}{\rho_1\rho_2}$, where $\alpha_1$ and $\alpha_2$ are the volume fractions of the light and heavy fluids respectively. $b(y=0,\tau)$ attains the configurational or ``no--mix'' value at $\alpha_1=\alpha_2=0.5$ (the configurational $b=0.067$ for $\rho_1=1.0$ and $\rho_2=1.667$ as in our case) during late times. Similar spectral turbulence models (\cite{steinkamp1996spectral,Steinkamp1999b}) have also observed this same tendency to relax towards the configurational value of $b(y=0,\tau)$.
The MOBILE simulation results are not purely immiscible. 
\sout{due to numerical diffusion}. }
Therefore it is not surprising that the final value of $b(y=0,\tau)$ is overpredicted by the model (Fig.~\ref{lwnb}). The impact on other quantities of interest such as mix-width (see Fig. \ref{lwnb_mix}), $a_y(y=0,\tau)$ and $R_{nn}(y=0,\tau)$, of the overprediction in the final $b(y=0,\tau)$ is therefore not unexpected.\blue{ The coefficients for these runs are provided in Table~\ref{table3} (runs {\tt PT1} and {\tt PT2}).}

\begin{figure*}
\includegraphics[width=.45\linewidth]{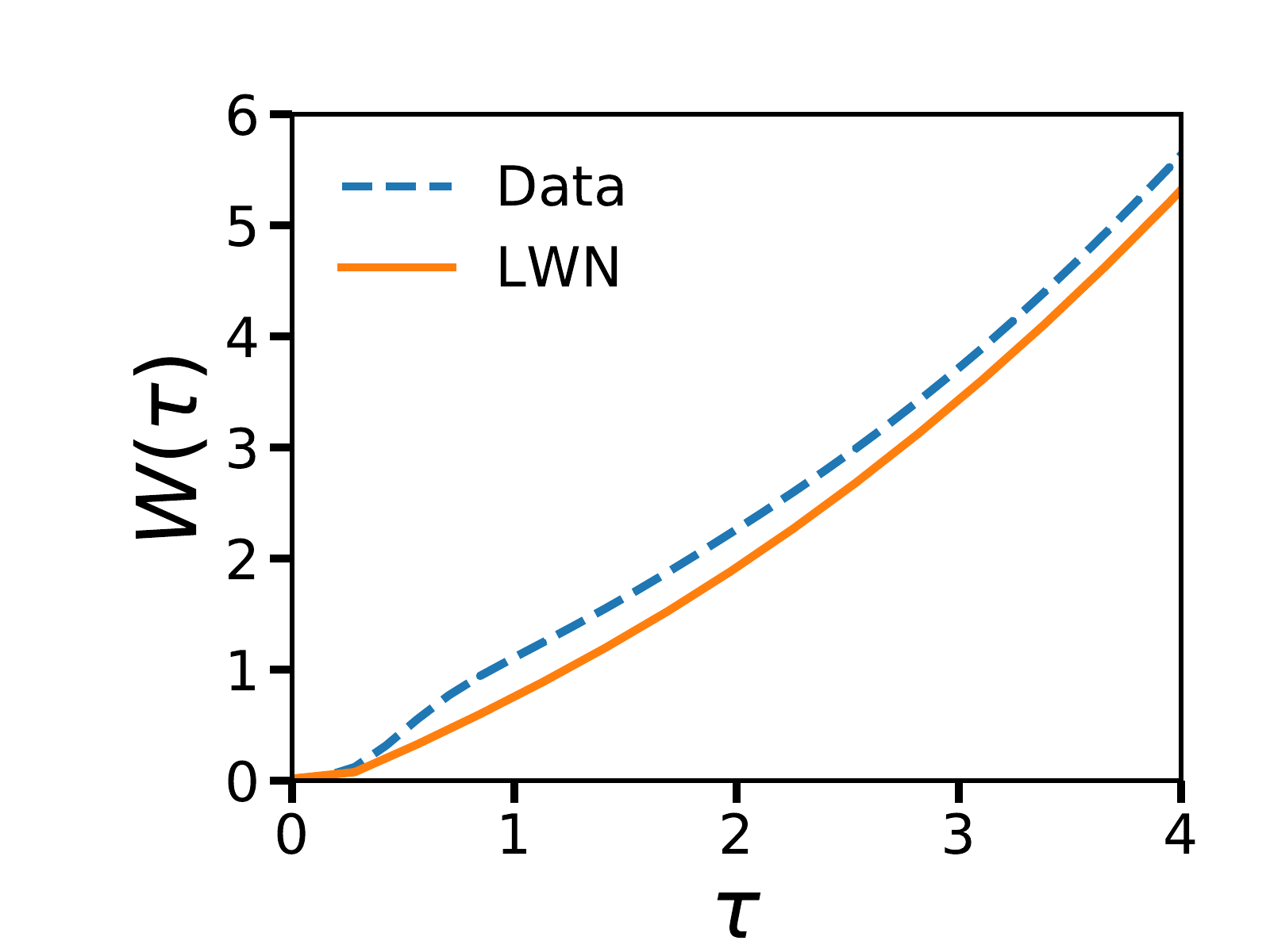}
\put(-40,40){\bf \scriptsize (a)}\\
\includegraphics[width=.32\linewidth]{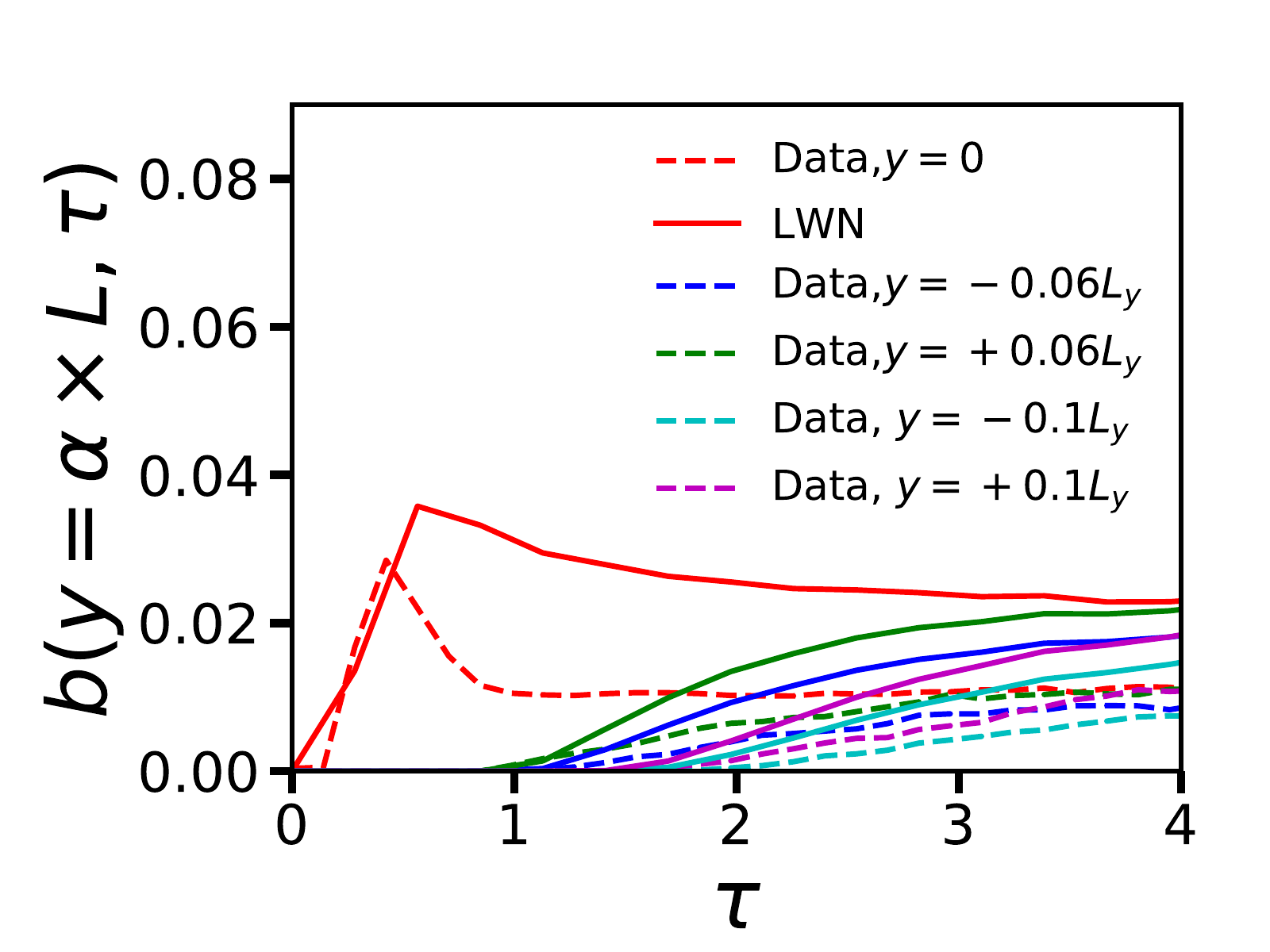}
\put(-120,100){\bf \scriptsize (b)}
\includegraphics[width=.32\linewidth]{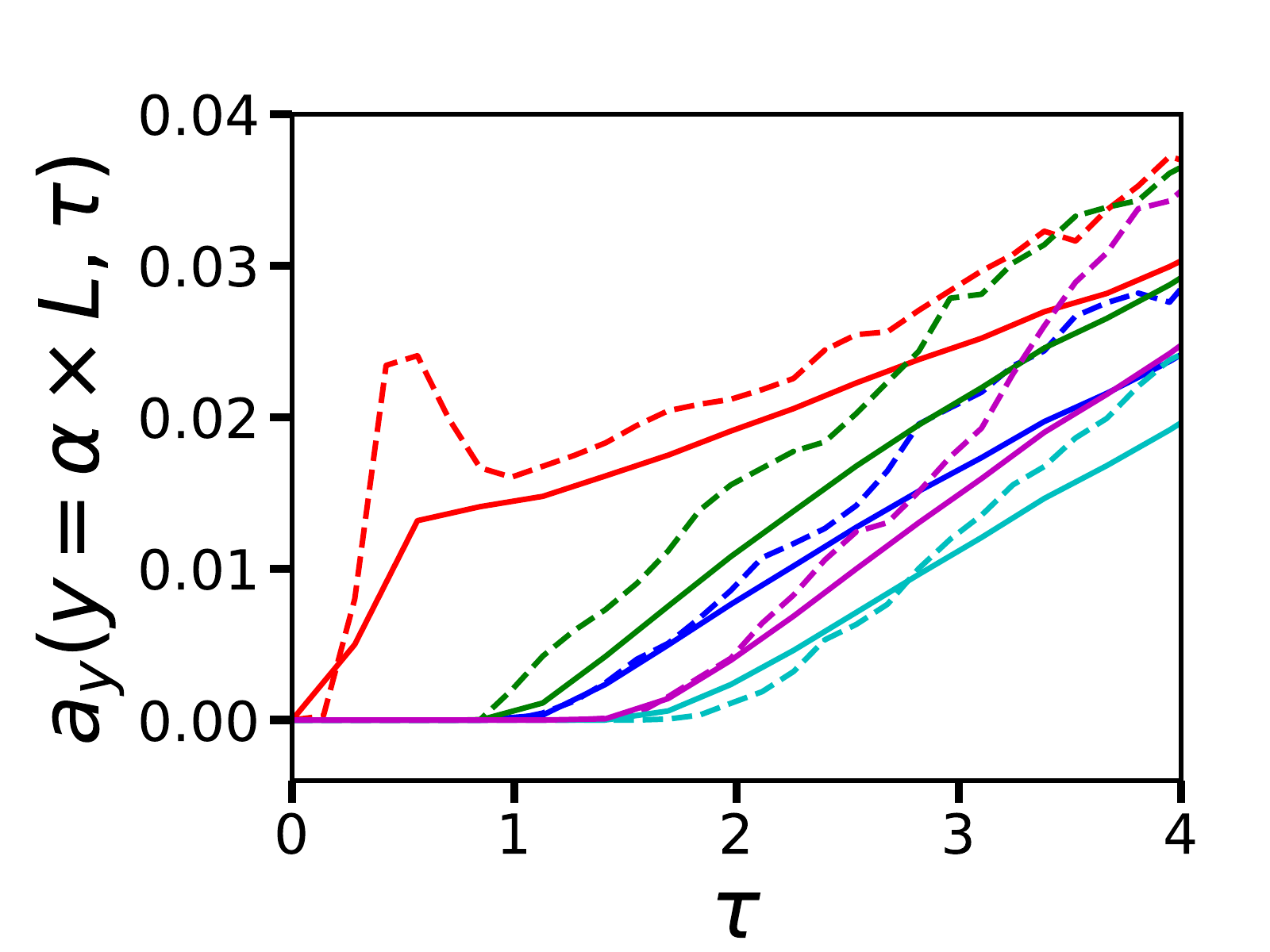}
\put(-120,100){\bf \scriptsize (c)}
\includegraphics[width=.32\linewidth]{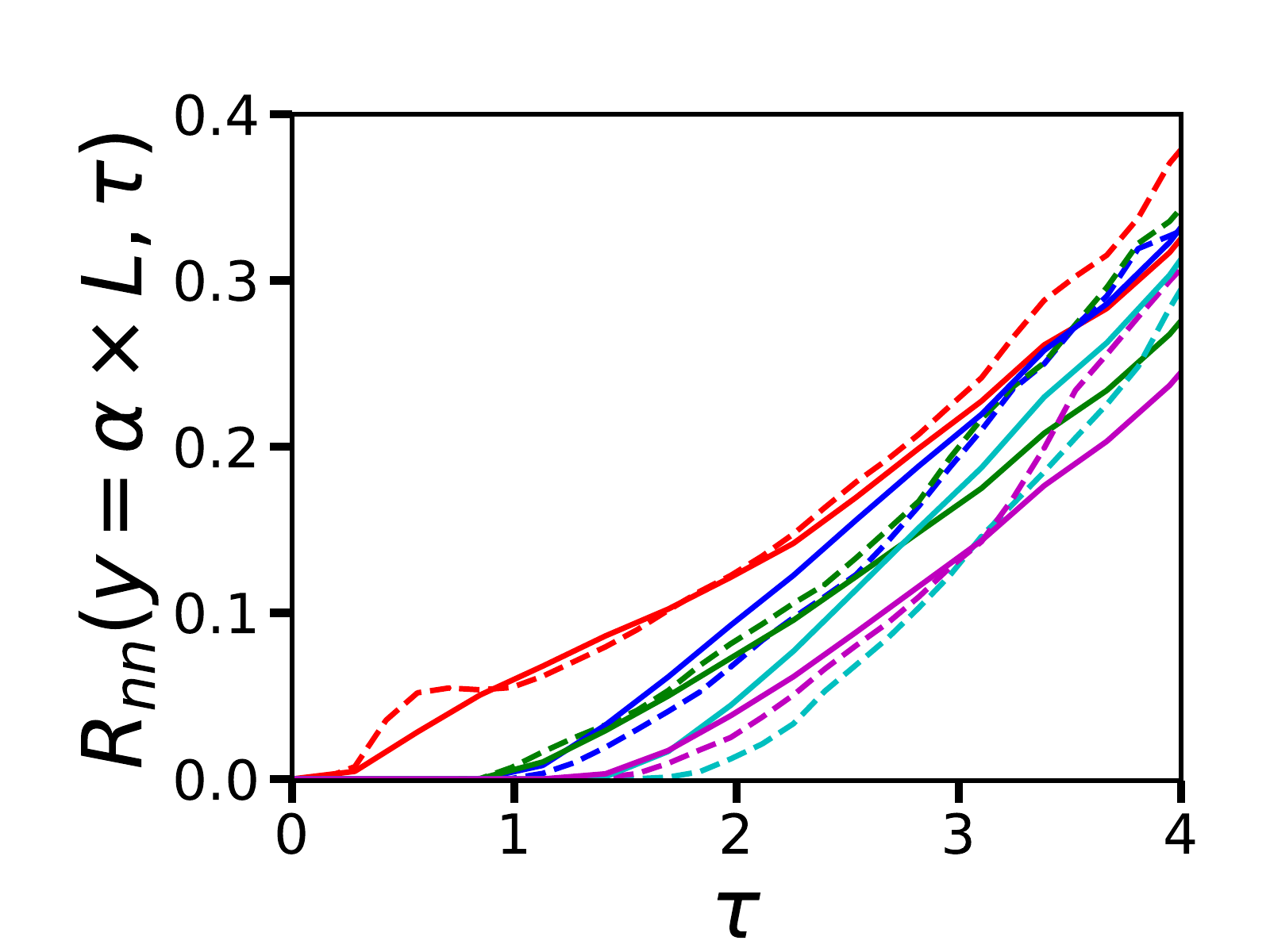}
\put(-120,100){\bf \scriptsize (d)}
\caption{Results of comparison between MOBILE data {\tt R1} (dashed line) and  LWN calculations (run {\tt PT3} in Table~\ref{table3}, solid lines) (a)Mix-width (in ${\rm cm}$) comparison; results at center-plane and at distances of 6\%$L_y$ and 10\%$L_y$ from the center of (b) $b(y,\tau)$ (c) $a_y(y,\tau)$ (in \blue{${\rm cm \ s^{-1} }$}) and (d) $R_{nn}(y,\tau)$ (in \blue{${\rm g \ cm^{-1} \ s^{-2}}$}).}
\label{mix_at25d-1_prod_rescaled}
\end{figure*}

\blue{
With further tuning of the spatial diffusion coefficient ($C_d$), the spectral transfer coefficient of $\hat{b}(y,k,\tau)$, i.e., $C_{b2}$, and the drag coefficients of $\hat{a}_y(y,k,\tau)$, i.e., $C_{rp1}$ and $C_{rp2}$, we arrive at a much improved comparison for the different metrics under study (Fig.~\ref{mix_at25d-1_prod_rescaled}). The coefficients are given in Table~\ref{table3} (run {\tt PT3}). As is shown in Fig.~\ref{mix_at25d-1_prod_rescaled}(a), the trend of the mix-width evolution is captured by the LWN model. The centerline $b(y,\tau)$ (Fig.~\ref{mix_at25d-1_prod_rescaled}(b)) remains overestimated, but the qualitative behavior is captured. Similarly, the evolution of $a_y(y,\tau)$ (Fig.~\ref{mix_at25d-1_prod_rescaled}(c)) and $R_{nn}(y,\tau)$ (Fig.~\ref{mix_at25d-1_prod_rescaled}(d)) are captured by the LWN model. 
}

\begin{table}
{
\begin{tabular}{|l|l|l|l|l|l|l|l|}
\hline
run &$A$ & $\tau_0$ &$C_{b1}$ & $C_{b2}$ &$C_d$&$C_{rp1}$ & $C_{rp2}$ \\
\hline
{\tt PT1}&$0.25$&$0.42$ &$0.12$ &$0.06$&$0.5$&$0.5$ & $0.5$\\
\hline
{\tt PT2}&$0.25$&$0.0$ &$0.12$ &$0.06$&$0.5$&$0.5$ & $0.5$\\
\hline
{\tt PT3}&$0.25$&$0.0$ &$0.12$ &$0.12$&$2.0$&$2.0$ & $2.0$\\
\hline
\end{tabular}
}
\caption{\blue{Table summarizing comparison study of LWN model against the MOBILE data (run {\tt R1}).}} 
\label{table3} 
\end{table}

\blue{
While further tuning of coefficients could optimize among the main metrics, the important thing to note with this source term for $\hat{b}(y,k,\tau)$ 
is that it successfully captures the evolution of the mix--width (Fig.~\ref{mix_at25d-1_prod_rescaled}(a)), the qualitative evolution of $b(y,\tau)$ (Fig.~\ref{mix_at25d-1_prod_rescaled}(b)), $a_y(y,\tau)$ (Fig.~\ref{mix_at25d-1_prod_rescaled}(c)), and   $R_{nn}(y,\tau)$ (Fig.~\ref{mix_at25d-1_prod_rescaled}(d)),  both at the center-plane and elsewhere in the mix-layer.
\sout{timing of the peak of $b$ at the center line, and the qualitative behavior of a ramp-up, peak and decay to an asymptotically constant value.}
}

\section{~\label{conclusions}Discussion and conclusions}
In this study of Rayleigh-Taylor instability, we have compared the LWN model with the results from implicit large-eddy simulations using MOBILE. 
Comparisons of plane-averaged quantities and the time-evolution of their spatial distribution have been made of mix-layer width, the time-evolutions of the specific volume and density fluctuation correlation 
$b(y,\tau)$, the mass-flux velocity $a_y(y,\tau)$, and the trace of the Reynolds stress $R_{nn}(y,\tau)$. Although the LWN model has previously been applied to homogeneous turbulent problems, 
this is the first study using initial conditions comprising high wave-number, narrow-band density interface perturbations that are a pre-requisite for the study of the classical Rayleigh-Taylor instability. 
Following on from the homogeneous, isotropic variable-density two-point model presented in our previous work~\cite{pal2018two}, the enhancements to the model presented in this paper provide the most compact feasible representation of behaviors in inhomogeneous turbulence.

We separated our study into two parts in order to attempt to isolate the physics associated with the main model capabilities.  The first part was an assessment of the minimal augmentation of the model required to capture inhomogeneous mixing. This yielded excellent outcomes for the mix width and mass flux velocity $a_y$ over a modest range of $A$ without additional tuning. Of note is the sensitivity to the spatial diffusion coefficient, $C_d$, in predicting the correct profile of the turbulent mass flux velocity $a_y(y,\tau)$. When $C_d$ is small, the dominant processes governing the turbulence are the inertial range scale-to-scale transfers of $\hat{R}_{nn},\hat{R}_{yy},\hat{a}_y,\hat{b}$ that in the LWN model are represented in $k$-space, and the baroclinic drive due to the pressure and density gradients (see Eqs.~\ref{main_rnn}--\ref{main_b}). None of these terms directly induce spreading of the mean $a_y(y,\tau)$ profiles, and thus we find (see Fig.~\ref{mix_prof_crp}(b)) that it acquires a rounded-top or ``dome-like'' shape. Inevitably, larger values of $C_d$ make spatial diffusion more rapid, producing smoother distributions in space. 
\begin{figure}
	\includegraphics[width=.7\linewidth]{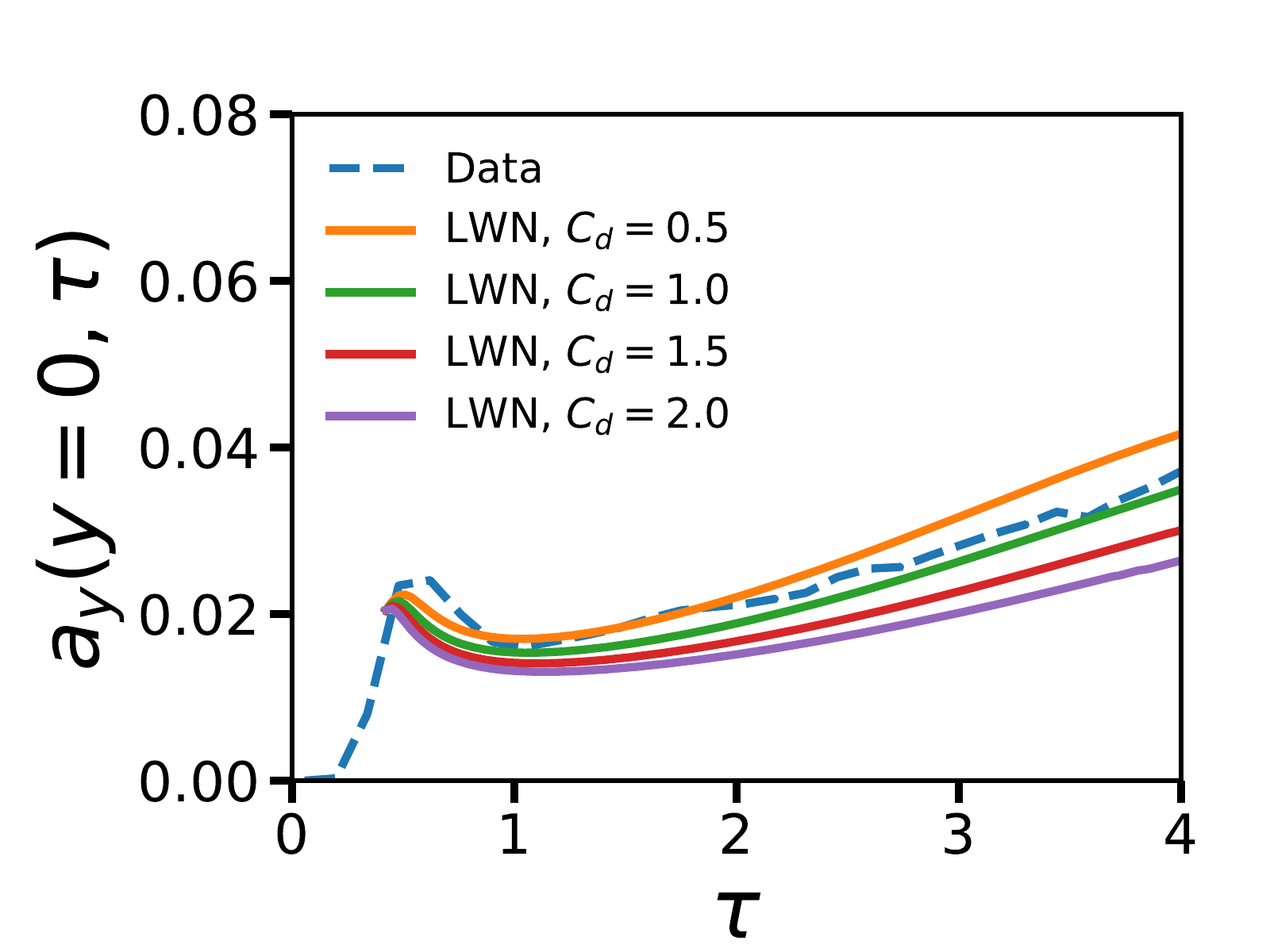}
	\put(-25,100){\bf \scriptsize (a)}\\
	\includegraphics[width=.7\linewidth]{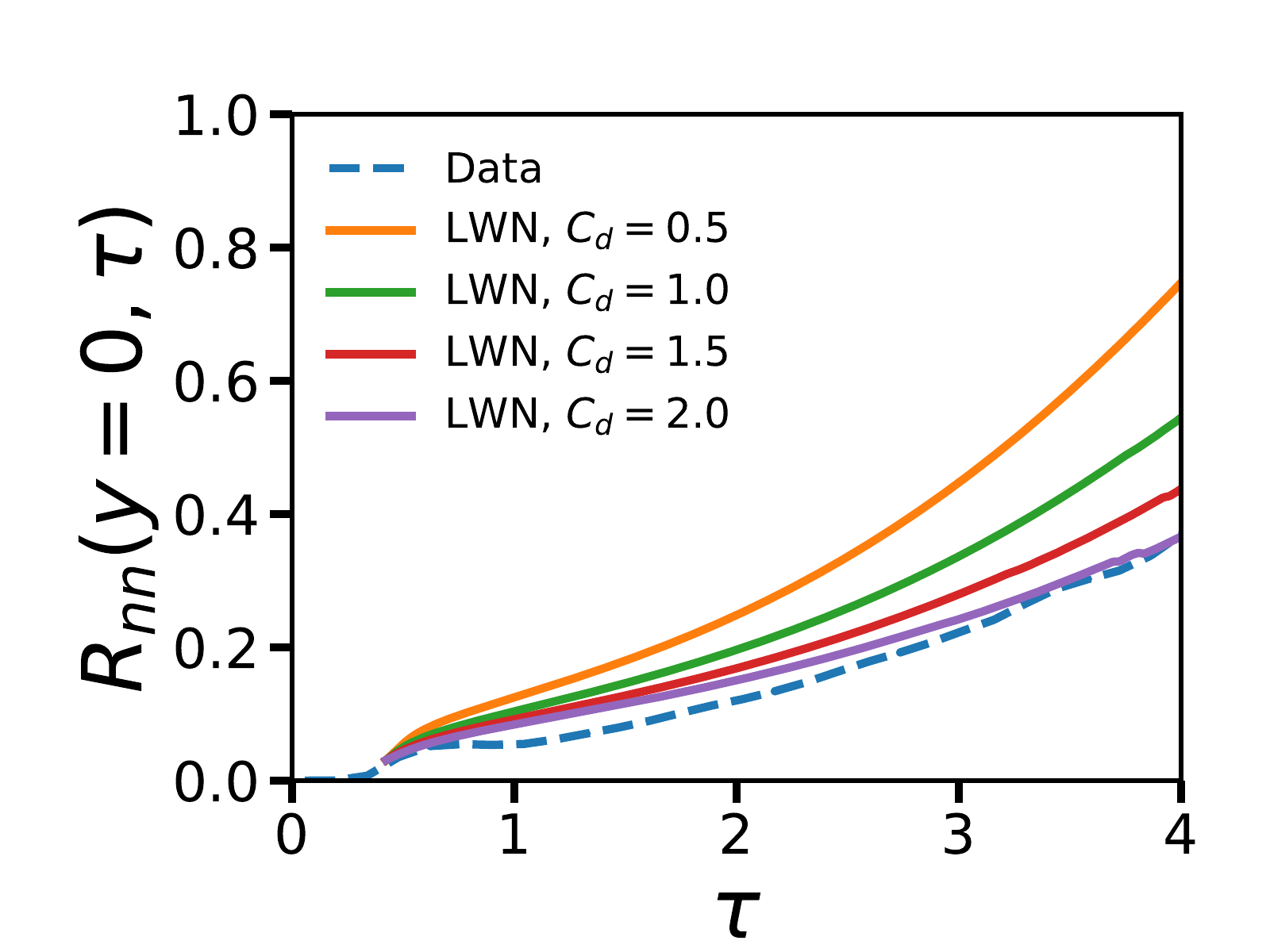}
	\put(-25,100){\bf \scriptsize (b)}
	\caption{(a) Comparison of $a_y(y=0,\tau)$ (in \blue{${\rm cm \ s^{-1} }$}) among results from
          MOBILE data (blue dashed line), and LWN runs with
          $C_d=0.5$ (orange line), $C_d=1.0$ (green line), $C_d=1.5$
          (red line) and $C_d=2.0$ (purple line). All other
          coefficients are same as in run {\tt T7}.(b) Comparison of
          $R_{nn}(y=0,\tau)$ (in \blue{${\rm g \ cm^{-1} \ s^{-2}}$}) for these runs.}
	\label{rnn_agree}
\end{figure}

The simplest LWN model yields larger magnitudes of $R_{nn}(y=0,\tau)$ (see Fig.~\ref{mix_fin}(d)) compared to MOBILE simulation. Careful optimization of the $C_d$ coefficient gives a somewhat better agreement on this parameter, but then $a_y(y=0,\tau)$ becomes under-predicted (see Figs.~\ref{rnn_agree}(a) and (b)). While it may be possible to select a separate $C_d$ for each variable, or modify the Leith model~\cite{rubinstein2017leith}, such fine-tuning lies outside the scope of the present paper. 

Another important factor that controls the quality of comparison of the variable $R_{nn}$ is the selection of a suitable return-to-isotropy coefficient, $C_m$. As we mention in Section~\ref{implement}, $C_m$ in Eq.~\eqref{main_ryy} governs the distribution of energy between the components of the stress tensor $R_{ij}$ and in setting $C_m=1.0$ we fix $R_{yy}(y,\tau)\sim 0.4 R_{nn}(y,\tau)$ at all times. The quality of comparison between model and simulation is rather better for $R_{yy}$ than for $R_{nn}$. This is because a close match for the mass flux velocity $a_y$ results in a better prediction of $R_{yy}$. However, we find in the simulations that $R_{yy} \sim c R_{nn}$, where $c$ is a constant between $0.6$ and $0.8$ depending on the flow regime, and so energy is preferentially contained in $R_{yy}$. In contrast, the model energy is approximately equidistributed. As discussed in~\cite{bragg2017model}, bias in the Reynolds stress is known to be a feature in the development of anisotropic flows and so its appearance here is consistent with the problem configuration. In principle, a biased distribution of energy amongst the Reynolds stress components is sensitive to $C_m$, but there are intrinsic limitations in the spectral model with respect to anisotropic flows, for which more detailed study has been proposed in~\cite{rubinstein2017leith,clark2018generation}.
In the asymptotic case, $a_y \sim \tau$ and $R_{nn} \sim \tau^2$, which we do observe in the present study.

The LWN model as implemented in the first part of the study did not predict the correct magnitude of the $b(y,\tau)$ profile for $\tau > 1$, and we attributed the decay of $b(y,\tau)$ to the omission of a source term in equation (Eq.~\eqref{main_b}) for $b(y,\tau)$. Since there is no term in the $\hat{b}(y,k)$ equation (Eq.~\eqref{main_b}) for growth of $\hat{b}(y,k)$, it must decay with time, reducing the strength of production $\df{\hat{b}(y,k)}{\overline \rho}\df{\partial \overline p}{\partial y}$ of $\hat{a}_y(y,k)$, and decreasing its growth rate. On the contrary we note that for $0.42 < \tau < 1$, $b(y,\tau)$ agrees well very well with simulation results, and the growth of $a_y(y,\tau)$ in the model is commensurate with the growth of $a_y(y,\tau)$ in the simulation through $\tau \sim 4$ and later. This apparent contradiction may be attributed to the additional spectral production term for $\hat{a}_y(y,k)$ in (Eq.~\eqref{main_a}), $\left(\df{\hat{R}_{yy}(k)}{\overline{\rho}^2}\df{\partial\overline{\rho}}{\partial y}\right)$. In the nonlinear flow regime $\tau > 1$, the mixing between the two fluids becomes important, and the $k$-space transport and drag terms in $\hat{a}_y(y,k)$ (Eq.~\eqref{main_a}) offer a balance with the production term. Thus $a_y(y,\tau)$, $R_{nn}(y,\tau)$ and the evolution of the mix-layer width are captured reasonably well even though $b(y,\tau)$ is under-predicted at this later stage. 
A similar mechanism in a single-point formulation is described in \cite{livescu2009cfdns}.

The second part of the study was an effort to improve the model with a suitable source term for $\hat{b}$ which would help sustain an asymptotic value and also capture 
early transitional regimes. An accurate prediction of $b(y,\tau)$ is thought to be essential for any study involving variable-density flow, and as discussed in~\cite{ristorcelli2004rayleigh, cabot2006reynolds}, $b(y,\tau)$ should achieve an asymptotic steady-state. Our study in section~\ref{source_term} shows that LWN is indeed capable of capturing the early and asymptotic evolution of $b(y,\tau)$, when a suitable kinematic source term is used in the model. \sout{The question of how to assess \textit{a priori} the asymptotic value of configurational $b$ for miscible flows remains open.} \blue{The choice and inclusion of kinematic source term, as well as careful tuning of the spatial diffusion ($C_d$) and the drag ($C_{rp1}, C_{rp2}$) coefficients has enabled the LWN model to capture the evolution of the mix--layer, $a_y(y,\tau)$, $R_{nn}(y,\tau)$ and $b(y,\tau)$. \sout{suggests further tuning and optimization of the coefficients may be performed, which we leave for future specific applications.}}

The current version of the source term is a simple extrapolation from the single-point form which integrates to the latter for consistency. The source term could in principle be made more sophisticated with better spectral characteristics \cite{steinkamp1996spectral,canfieldlwn}. 
However this involves a convolution which is not straightforward to compute or implement in a practical model so we leave this for future research.

Our approach to this analysis, of separating two components of the model, reveals that 
it remains difficult to achieve equal fidelity simultaneously across all variables. 
Without the source term for $\hat{b}$ we were able to obtain good quantitative values for mass flux velocity and the mix width by starting at not too early times. With the source term, additional mixing physics was captured at early times and for the final state of $b$, as was the Reynolds stress with adequate diffusion, \blue{but some aspects the previously adequate results were compromised, for example the growth of $a_y$}. Further improvement might be had with additional calibration, depending on the user priorities and the flow to be modeled. However it remains the case that this is a second-order statistical model which has intrinsic limitations relative to the full turbulent dynamics, even if it is an improvement in some respects over single-point models.
While the goal of this paper is not to provide an optimal or universal set of coefficients, we have nevertheless shown that the elements of this model show significant promise in capturing the key physics of inhomogeneous Rayleigh-Taylor mixing.


\section{Acknowledgements}
\blue{We thank the anonymous reviewers for motivating much of the effort described in Section 5.} The authors thank Timothy T. Clark (University of New Mexico) for useful discussions. NP, IB and SK were funded by the Mix and Burn project under the Physics and Engineering Models  program of the DOE Advanced Simulation and Computing program. Work at LANL was performed under the auspices of Triad National Security, LLC which operates Los Alamos National Laboratory under Contract No. 89233218CNA000001 with the U.S. Department of Energy/National Nuclear Security Administration. 

\bibliography{var}
\end{document}